%% file: main.tex
\documentclass[preprint,12pt]{elsarticle}

\usepackage{bm}
\usepackage{color}

\usepackage{amssymb,amsmath}
\usepackage{amsthm}
\theoremstyle{definition}
\newtheorem{dfn}{Definition}

\usepackage{lineno}
\modulolinenumbers[5]
\usepackage{url}
\usepackage[subrefformat=parens]{subcaption}
\journal{Acta Astronautica}

\begin{document}

\begin{frontmatter}



\title{Design of low-energy transfers in cislunar space \\ using sequences of lobe dynamics}


\author[inst1]{Naoki Hiraiwa\corref{cor1}\fnref{fn1}} \ead{hiraiwa.naoki.978@gmail.com}
\cortext[cor1]{Corresponding author.}
\fntext[fn1]{Currently at Faculty of Environment and Information Science, Yokohama National University, 79-7 Tokiwadai, Hodogaya-ku, Yokohama, 240-8501, Kanagawa, Japan.}
\author[inst1]{Mai Bando} \ead{mbando@aero.kyushu-u.ac.jp}
\author[inst2]{Yuzuru Sato}
\ead{ysato@math.sci.hokudai.ac.jp}
\author[inst1]{Shinji Hokamoto} \ead{hokamoto@aero.kyushu-u.ac.jp}
\affiliation[inst1]{organization={Department of Aeronautics and Astronautics, Kyushu University},%Department and Organization
            addressline={744 Motooka, Nishu-ku}, 
            city={Fukuoka},
            postcode={819-0395}, 
            state={Fukuoka},
            country={Japan}}
\affiliation[inst2]{organization={RIES-MSC / Department of Mathematics, Hokkaido University},
            addressline={N12 W7 Kita-ku}, 
            city={Sapporo},
            postcode={060-0812}, 
            state={Hokkaido},
            country={Japan}}
            
\begin{abstract}
Dynamical structures in the circular restricted three-body problem (CR3BP) are fundamental for designing low-energy transfers, as they aid in analyzing phase space transport and designing desirable trajectories.
One of these dynamical structures, lobe dynamics, can be exploited to achieve local chaotic transport around celestial bodies.
This study proposes and fully validates a systematic method for designing low-energy transfers by combining multiple sequences of lobe dynamics, building upon the authors' prior preliminary framework.
A graph-based framework is developed to explore possible transfer paths between departure and arrival orbits, reducing the complexity of the combinatorial optimization problem for fuel-efficient transfer design.
Using this graph, low-energy transfer trajectories are constructed by connecting chaotic orbits within lobes.
The resulting optimal trajectory in the Earth--Moon CR3BP is then converted into an optimal transfer in the bicircular restricted four-body problem via multiple shooting.
This transfer is compared with existing optimal solutions to demonstrate the effectiveness of the proposed method.
\end{abstract}

%% Research highlights (Up to 85 characters, including spaces, per bullet point)
\begin{highlights}
\item A new low-energy trajectory design method based on lobe dynamics is proposed.
\item The proposed method systematically combines the sequences of multiple lobe dynamics.
\item The application to the fuel-optimal Earth--Moon transfer is demonstrated.
\end{highlights}


\end{frontmatter}

% \linenumbers

\input{1_introduction}
\input{2_preliminaries}
\input{3_lobe_dynamics}
\input{4_trajectory_design}
\input{5_application}
\input{6_extension}
\input{7_conclusion}

\section*{Acknowledgments}
The first author thanks the support of Japan Society for the Promotion of Science (JSPS) KAKENHI Grant No. JP 23KJ1692. The third author was supported by JSPS Grant-in-Aid for Scientific Research (B), JP No. 21H01002.

\section*{Declaration of Generative AI and AI-assisted technologies in the writing process}
During the preparation of this work, the authors used Grammarly and ChatGPT 4.1 in order to improve readability.
After using this tool, the authors reviewed and edited the content as needed and take full responsibility for the content of the publication.

\bibliographystyle{elsarticle-num}
\biboptions{sort&compress}
\bibliography{reference}

\end{document}

%% file: 1_introduction.tex
\section{Introduction}
Space exploration in cislunar space has attracted increasing interest due to the Artemis program~\cite{creech2022artemis} and the growing number of small satellite missions~\cite{Kopacz2020small,Turan2022autonomous}.
Cislunar missions have different requirements for fuel consumption and time of flight, which directly affect trajectory design.
Recent missions such as Chang'e-3~\cite{liu2015change,zuo2021china}, Artemis 1~\cite{batcha2020artemis,eckman2023trajectory}, and Chandrayaan-3~\cite{mathavaraj2024chandrayaan} used fast transfer trajectories that require more fuel.
In contrast, missions like Hiten~\cite{uesugi1991japanese,uesugi1996results}, SMART-1~\cite{racca2002smart,foing2018smart}, GRAIL~\cite{zuber2013gravity}, CAPSTONE~\cite{gardner2023capstone}, and SLIM~\cite{ishida2025vision,kitamura2025trajectory,ueda2025onorbit} adopted fuel-efficient transfers with longer time of flight.
A key challenge in trajectory design is to find a variety of transfer options that meet specific mission criteria~\cite{howell2012application}.
This is difficult because multi-body environments like cislunar space are highly sensitive to initial conditions, with spacecraft influenced by multiple celestial bodies simultaneously.
Trajectory design in such environments must harness the chaotic dynamics~\cite{yamaguti1994towards,jaeger2004harnessing}.

Trajectory design methods in the literature fall into two main categories: numerical and geometrical. 
Numerical approaches (e.g., Refs.~\cite{russell2007primer,topputo2013optimal,oshima2017global,grossi2024optimal}) identify transfer trajectories by solving optimization problems. 
By evaluating a large set of initial conditions, these methods can find a wide range of solutions for fuel consumption and time of flight, often yielding Pareto-optimal sets. 
Their drawbacks include the need for accurate initial guesses and high computational cost. 
Geometrical or dynamical systems approaches (e.g., Refs.~\cite{koon2000heteroclinic,koon2002constructing,gomez2004connecting,McCarthy2023four,henry2024fully,hiraiwa2024designing,scheuerle2024energy}) use the natural dynamics described by dynamical systems theory to generate low-energy transfers. 
These approaches are limited to planar problems and often result in longer transfer times.
The Genesis mission~\cite{lo2001genesis,burnett2019future} was the first to demonstrate the practical value of this approach by designing its nominal trajectory using modern dynamical systems theory. 
Both approaches are valuable and complementary, but further development of the geometrical approach is crucial to address the key challenge in trajectory design. 
The geometrical approach identifies desirable trajectories by analyzing phase space transport, while the numerical approach focuses on local dynamics near given points. 
Although the numerical approach may imply underlying dynamical structures, the geometrical approach is generally more effective at providing a variety of trajectories with specific properties.

The geometrical approach produces low-energy transfers by exploiting chaotic transport in natural flows, which is described by the theories of tube dynamics and lobe dynamics (e.g., Refs.~\cite{koon2004geometric,koon2011dynamical,ren2012two}).
Both theories characterize the transport structures formed by invariant manifolds associated with periodic orbits in the planar circular restricted three-body problem (CR3BP)~\cite{szebehely1967theories}, but from different perspectives.
Tube dynamics describes transport as a continuous flow along invariant manifolds, whereas lobe dynamics represents it as discrete transfer across manifold intersections.
In astrodynamics, these two transport theories can be applied to interpret dynamics at different scales and in different regions, as detailed below.

Tube dynamics~\cite{koon2000heteroclinic,marsden2006new} describes global transport between two celestial bodies based on the stable and unstable manifolds of libration point orbits~\cite{koon2004geometric}, especially around $L_1$ or $L_2$ in the planar CR3BP.
These manifolds are called \textit{Conley--McGehee tubes} or simply \textit{tubes}~\cite{marsden2006new}.
The tubes around $L_1$ and $L_2$ provide a framework for understanding the dynamics in the planar CR3BP, as shown in Ref.~\cite{koon2002constructing}.
For example, the natural orbit transitions of Jupiter comets are described using tube dynamics in the Sun--Jupiter system~\cite{koon2001resonance}.
Tube dynamics also explains the mechanism of ballistic lunar transfers through Earth--Moon $L_2$ and enables their systematic design~\cite{koon2001low}.
One limitation of tube dynamics is that tubes associated with $L_1$ Lyapunov orbits do not facilitate departure from the primary, as they move away from it in the CR3BP.
Another limitation is its limited applicability to interplanetary transfers in the Solar System~\cite{ren2012two}.
For example, tubes in the Sun--Earth and Sun--Mars planar CR3BP do not naturally intersect~\cite{ren2012two,mingotti2011earth}.

Lobe dynamics describes finer transport between regions around a single celestial body~\cite{koon2004geometric}.
This mechanism complements tube dynamics by revealing pathways that connect the vicinity of the primary to tubes, which is difficult to achieve using tube dynamics alone.
The theory of lobe dynamics was first developed to analyze fluid mixing in two-\hspace{0pt}dimensional inviscid incompressible flows~\cite{rom1990analytical} and then extended to two-dimensional, area- and orientation-\hspace{0pt}preserving maps~\cite{rom1990tranport}.
In such maps, segments of the stable and unstable manifolds associated with periodic points may form small enclosed regions called \textit{lobes}, which are key elements of transport by lobe dynamics.
Within the theory of transport in dynamical systems~\cite{meiss1992symplectic,meiss2015thirty}, this idea has been used to analyze the chaotic transport of phase space volume between regions in various fields, such as fluid dynamics~\cite{camassa1991chaotic,duan1997lagrangian,malhotra1998geometric,watanabe2021experimental,speetjens2022linear}, geophysical flows~\cite{wiggins2005dynamical,raynal2006lobe,duToit2010horseshoes}, celestial mechanics~\cite{ren2012two,oshima2015jumping}, and chemical reaction dynamics~\cite{krajnak2020phase,katsanikas2020phase}.
Computational tools for lobe area and transport rate~\cite{dellnitz2005transport,naik2017computational} have been developed, as the application of lobe dynamics is limited by the computational difficulty of identifying lobes numerically~\cite{koon2004geometric}.
In spacecraft trajectory design, lobe dynamics typically leverages the stable and unstable manifolds of resonant orbits.
Previous studies~\cite{ross2004application,grover2009designing} have developed trajectory design methods using multiple gravity assists of Jupiter's moons by numerically exploiting lobe dynamics in the Jovian system.
Furthermore, Refs.~\cite{odashima2016design,odashima2017design} have used the lobe dynamics of the $3$:$1$ resonant orbit to design transfers to the Moon in the Earth--Moon system.
In summary, lobe dynamics has been investigated numerically for moon-to-moon transfers or analytically for trajectory design around a single resonant orbit in the literature.
These studies have not yet provided a simple interpretation of phase space transport for trajectory design based on lobe dynamics.

Unlike the previous studies~\cite{ross2004application,grover2009designing,odashima2016design,odashima2017design}, this study proposes a new method that systematically incorporates the dynamical structures of lobes associated with multiple resonant orbits into trajectory design.
Earlier work by the authors' group~\cite{hiraiwa2024designing} provided the first theoretical investigation into the combination of multiple lobe dynamics in the standard map~\cite{chirikov1979universal} via a graph-based framework.
That Letter~\cite{hiraiwa2024designing} also included a preliminary application to the planar CR3BP.
This paper fully develops and validates a trajectory design method based on this approach.
The proposed method is applied to construct Earth--Moon transfers in both the CR3BP and the bicircular restricted four-body problem (BCR4BP)~\cite{huang1960very,boudad2022trajectory}.
The optimal transfer obtained in the BCR4BP is compared with existing solutions to demonstrate the effectiveness of the proposed method.

%% file: 2_preliminaries.tex
\section{Preliminaries}\label{sec:preliminaries}
This section introduces fundamental knowledge related to this study.
The Earth--Moon CR3BP is mainly employed in this paper, while the Sun--Earth--Moon BCR4BP is used in Section~\ref{sec:extension} to compare the results with those in the literature.
Note that lobe dynamics is described separately in the next section.

\subsection{Circular restricted three-body problem (CR3BP)}
\begin{figure}[t]
    \centering
    \includegraphics[width=0.45\columnwidth]{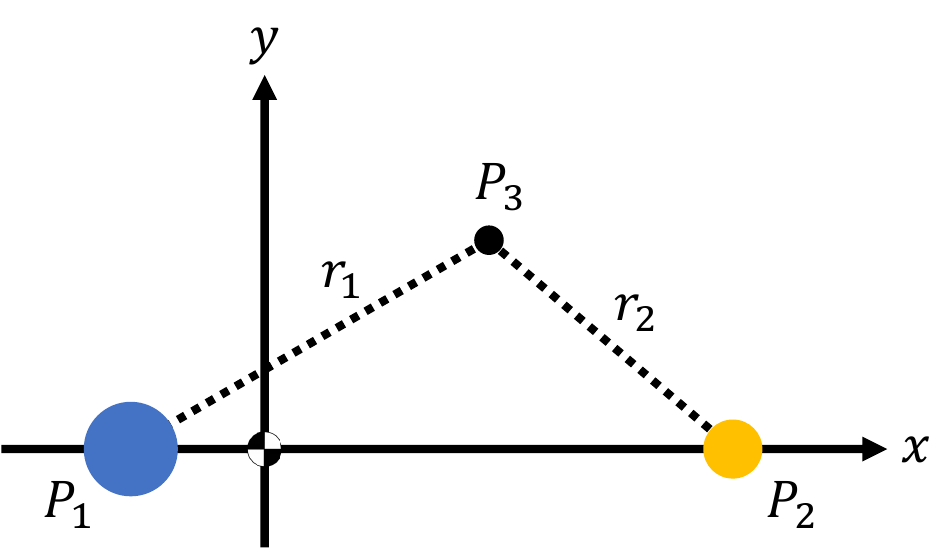}
    \caption{Coordinate system in CR3BP (Rotating frame).}
    \label{fig:CR3BP}
\end{figure}
This subsection describes key features of the CR3BP, whose coordinate system is shown in Fig.~\ref{fig:CR3BP}.
This problem considers two celestial bodies, $P_1$ and $P_2$, and a spacecraft $P_3$, all treated as point masses.
Let $m_1$, $m_2$ and $m_3$ denote the masses of $P_1$, $P_2$ and $P_3$, respectively, with the assumption that $m_1 > m_2 \gg m_3$.
This condition means $P_3$ has no effect on the motion of $P_1$ and $P_2$.
Another assumption in the CR3BP is that $P_1$ and $P_2$ move in circular orbits around their barycenter.
Under these assumptions, the equations of motion of $P_3$ are expressed in a non-dimensional form with a pseudo-potential function $U$ as follows:  
\begin{equation}
    \left\{\,
    \begin{aligned}
	\ddot{x} - 2\dot{y} &= \frac{\partial U}{\partial x}\\
	\ddot{y} + 2\dot{x} &= \frac{\partial U}{\partial y}\\
	\ddot{z} &= \frac{\partial U}{\partial z}
    \end{aligned}
    \right.\,,
    \label{eq:EoM}
\end{equation}
\begin{align}
    U &= \frac{1}{2}\left( x^2 + y^2 \right) + \frac{1 - \mu}{r_1} + \frac{\mu}{r_2} + \frac{1}{2}\mu\left(1 - \mu\right),\\
    r_1 &= \sqrt{ (x + \mu)^2 + y^2 + z^2 },\\
    r_2 &= \sqrt{ \left\{ x  - \left(1 - \mu \right)\right\}^2 + y^2 + z^2},
\end{align}
where $\mu = m_2 / (m_1 + m_2)$ is the non-dimensional mass ratio of the primaries, and $\mu = 1.21509\times10^{-2}$ in the Earth--Moon system.
Note that the relatively large $\mu$ value in the Earth--Moon system leads to highly chaotic dynamics, posing challenges in trajectory design.

The CR3BP has several characteristics useful for trajectory design.
First, the CR3BP has five libration points satisfying $\nabla U = \bm{0}$, denoted as $L_i$ ($i = 1,2,\cdots,5$).
The dynamics around these points have been well investigated in the literature (e.g., Refs.~\cite{doedel2007elemental,qiao2024cislunar,chen2025stable}) because each libration point $L_i$ has families of periodic orbits.
These periodic orbits and their stable and unstable manifolds are useful for analyzing the dynamics and designing trajectories.
In addition, this model possesses one integral of motion, called the Jacobi constant $C_J$, which is defined as 
\begin{equation}
    C_J = 2U - \left(\dot{x}^2 + \dot{y}^2 + \dot{z}^2\right). \label{eq:jacobi}
\end{equation}
Equation~\eqref{eq:jacobi} can be rewritten as
\begin{equation}
    2U - C_J = \dot{x}^2 + \dot{y}^2 + \dot{z}^2 \geq 0.
\end{equation}
This relationship shows that the condition $2U - C_J = 0$ defines the boundary of possible motion, known as the zero-velocity surface.
In the planar problem, the zero-velocity surface becomes a closed curve called the zero-velocity curve.
The zero-velocity curve with a given Jacobi constant divides position space into realms of possible motion~\cite{koon2011dynamical}.
The realm around the primary is referred to as the \textit{interior realm}, and the realm around the secondary is called the \textit{moon realm}.
The realm outside both of the primaries is called the \textit{exterior realm}.
These realms can be connected through the necks around $L_1$ and $L_2$.

\subsection{Resonant orbit}
Mean-motion resonances or orbital resonances play an essential role in describing the dynamics of small bodies or spacecraft~\cite{pan2022review,rawat2025cislunar}.
In the Earth--Moon CR3BP, a $p:q$ resonant orbit ($p, q \in \mathbb{N}$) about the Earth in resonance with the Moon is defined as an orbit whose period in the inertial frame, $T_3^I$, is related to the sidereal period of the Moon, $T_2$, as $p T_3^I \simeq q T_2$.
In other words, the resonance ratio $p:q$ indicates that a spacecraft rotates around the Earth $p$ times in approximately the same time required for the Moon to complete $q$ revolutions~\cite{Vaquero2014design}.
If $p > q$, resonant orbits have a period less than that of the Moon and are called interior resonant orbits; if $p < q$, they are called exterior resonant orbits~\cite{anderson2016broad}. 
Since $T_3^I = 2\pi a^{3/2} / \sqrt{1 - \mu}$ and $T_2 = 2\pi$ in non-dimensional units, the semi-major axis $a$ of the resonance orbit is expressed as
\begin{equation}
    a \simeq \left(\frac{q}{p}\right)^{2/3} \left(1 - \mu\right)^{1/3}. \label{eq:resonant_a}
\end{equation}
\begin{figure}[t]
    \centering
    \begin{minipage}{0.45\columnwidth}
        \centering
        \includegraphics[width=\columnwidth]{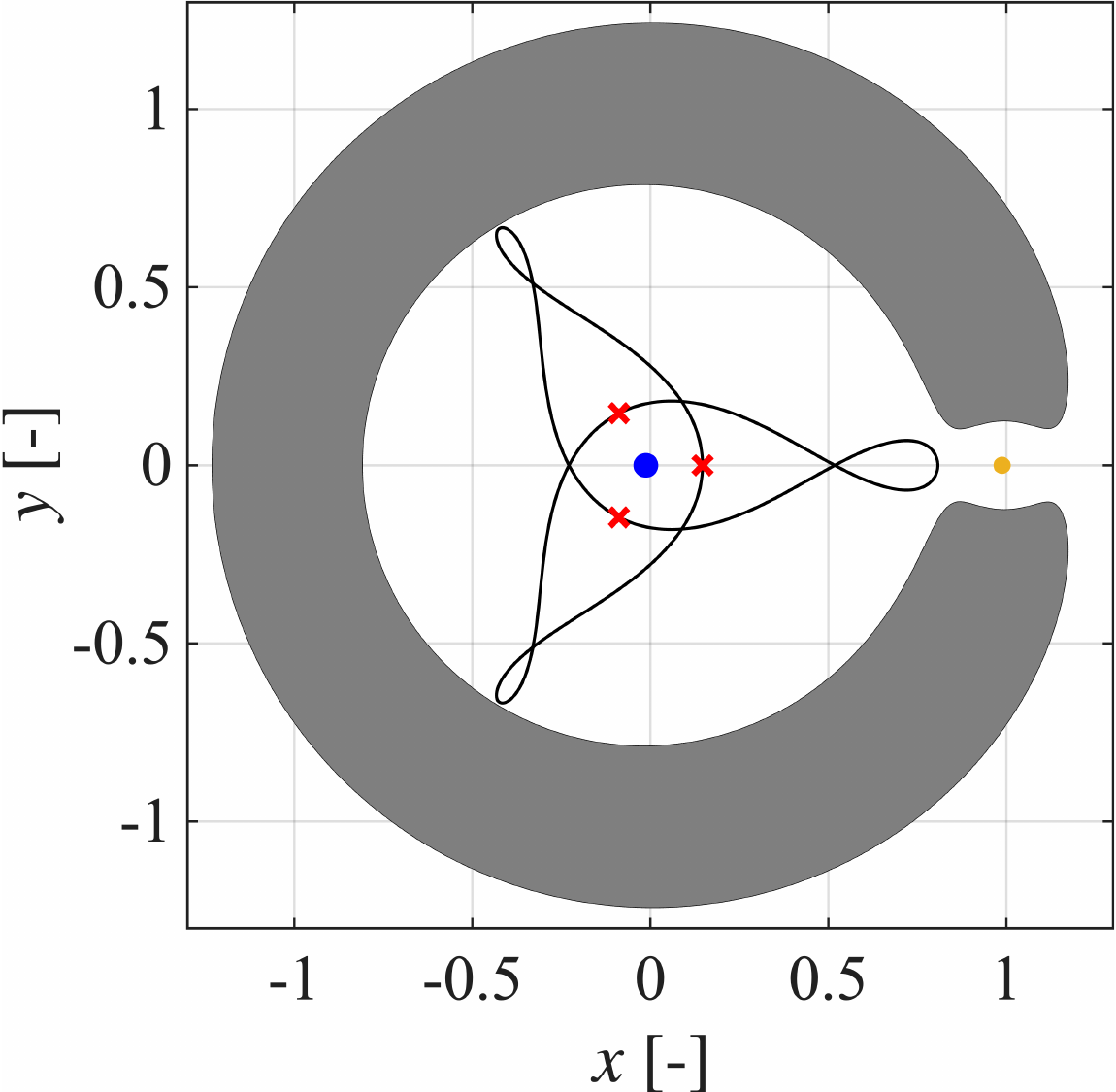}
        \subcaption{In the $x$-$y$ plane.} \label{fig:resonant_orbit_xy}
    \end{minipage}\\
    \vspace{5mm}
    \begin{minipage}{0.45\columnwidth}
        \centering
        \includegraphics[width=\columnwidth]{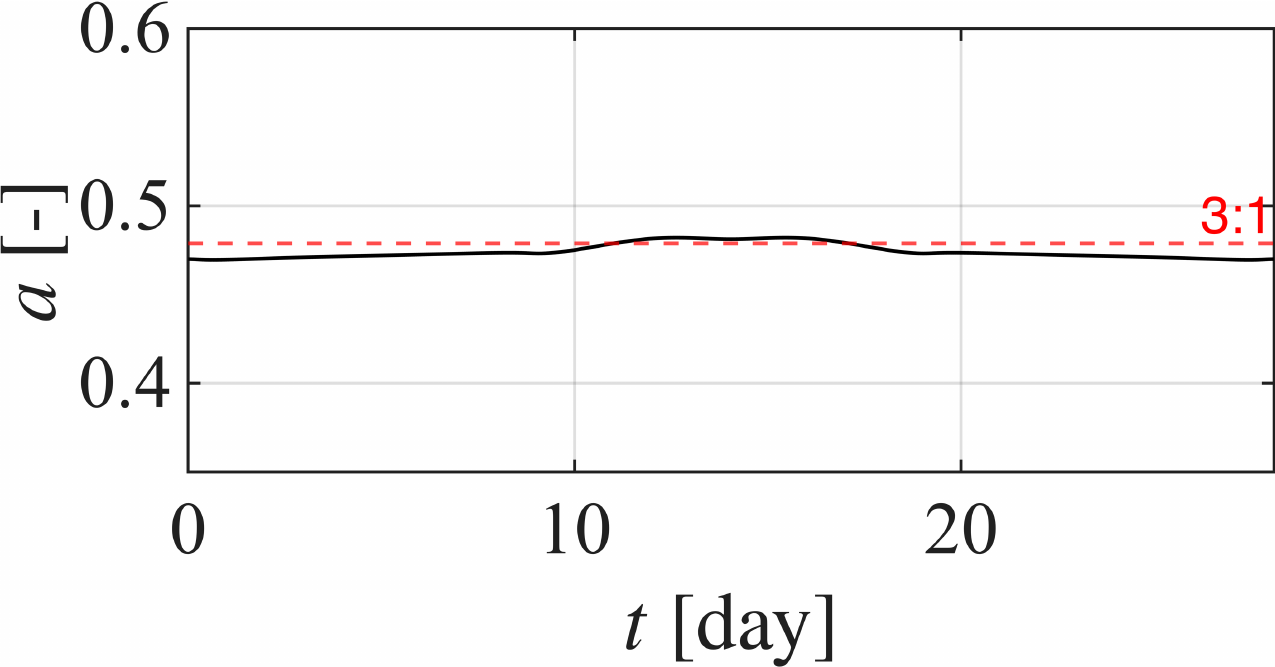}
        \subcaption{Time history of the semi-major axis along the orbit.} \label{fig:resonant_orbit_a}
    \end{minipage}%
    \caption{Example of a resonant orbit (the $3$:$1$ unstable resonant orbit when $C_J = 3.16$) and its semi-major axis.} \label{fig:resonant_orbit}
\end{figure}%
Figure~\ref{fig:resonant_orbit} illustrates a $3$:$1$ unstable resonant orbit and the time history of its semi-major axis.
The red crosses indicate periapses of this resonant orbit, and the blue and yellow dots represent the Earth and Moon, respectively.
The gray region is the forbidden realm where $2U - C_J < 0$.
As shown in Fig.~\ref{fig:resonant_orbit_a}, the semi-major axis roughly satisfies Eq.~\eqref{eq:resonant_a}, as represented by the red dashed line.
The perturbation of the semi-major axis comes from the gravitational influence of the Moon.
Resonant orbits have planar~\cite{Arenstorf1963existence} and spatial~\cite{Vaquero2014design} families, and quasi-periodic tori may exist around these orbits~\cite{Bonasera2023computing}.
These orbits are useful in trajectory design~\cite{anderson2021tour,canales2022transfer}, as has been demonstrated by the extended mission of IBEX~\cite{dichmann2013dynamics} and the TESS mission~\cite{parker2018transiting}.

\begin{figure}[!b]
    \centering
    \begin{minipage}{0.45\columnwidth}
        \centering
        \includegraphics[width=\columnwidth]{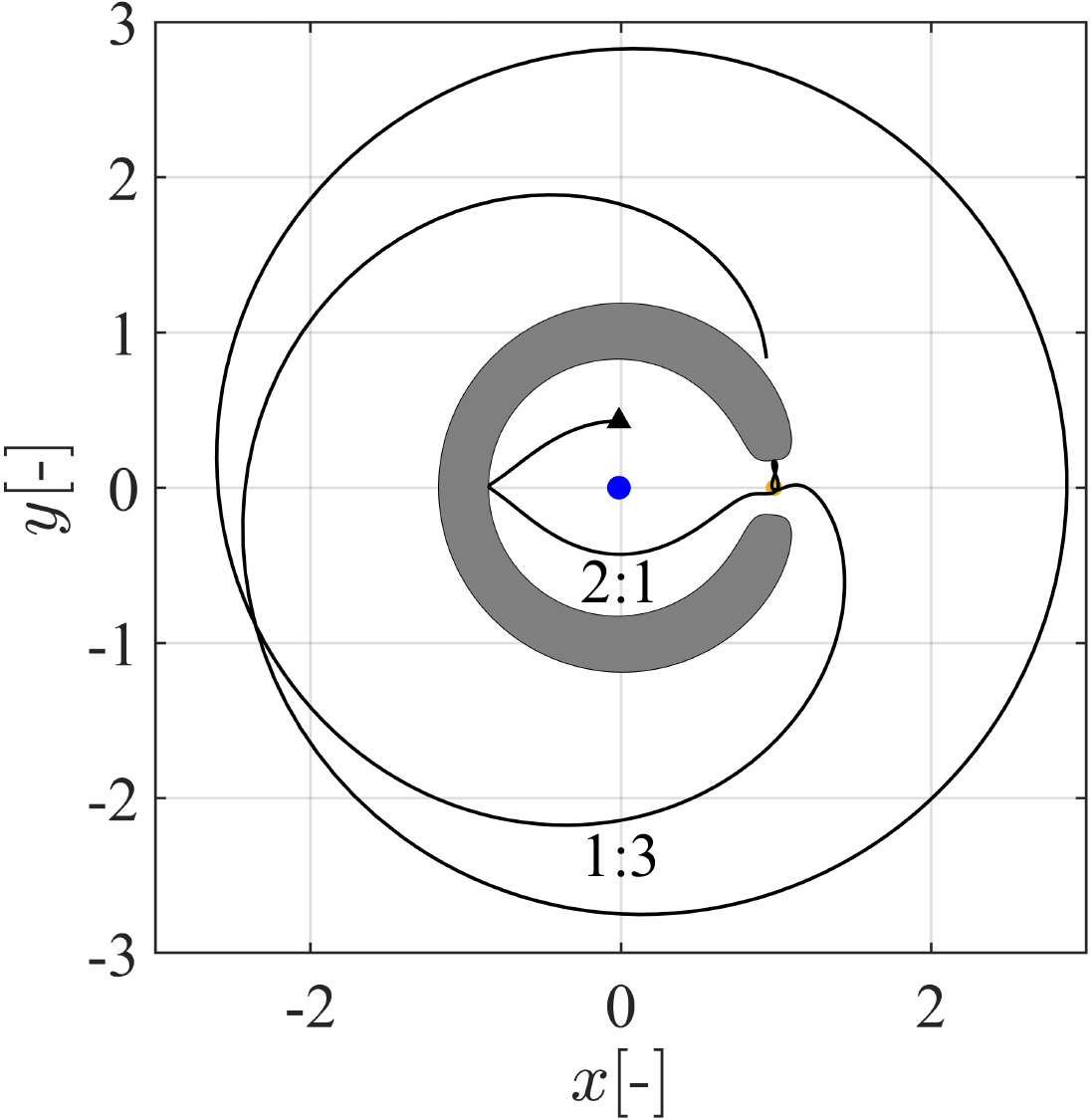}
        \subcaption{In the rotating frame.} \label{fig:resonance_hop_rotating}
    \end{minipage}
    \hspace{5mm}
    \begin{minipage}{0.45\columnwidth}
        \centering
        \includegraphics[width=\columnwidth]{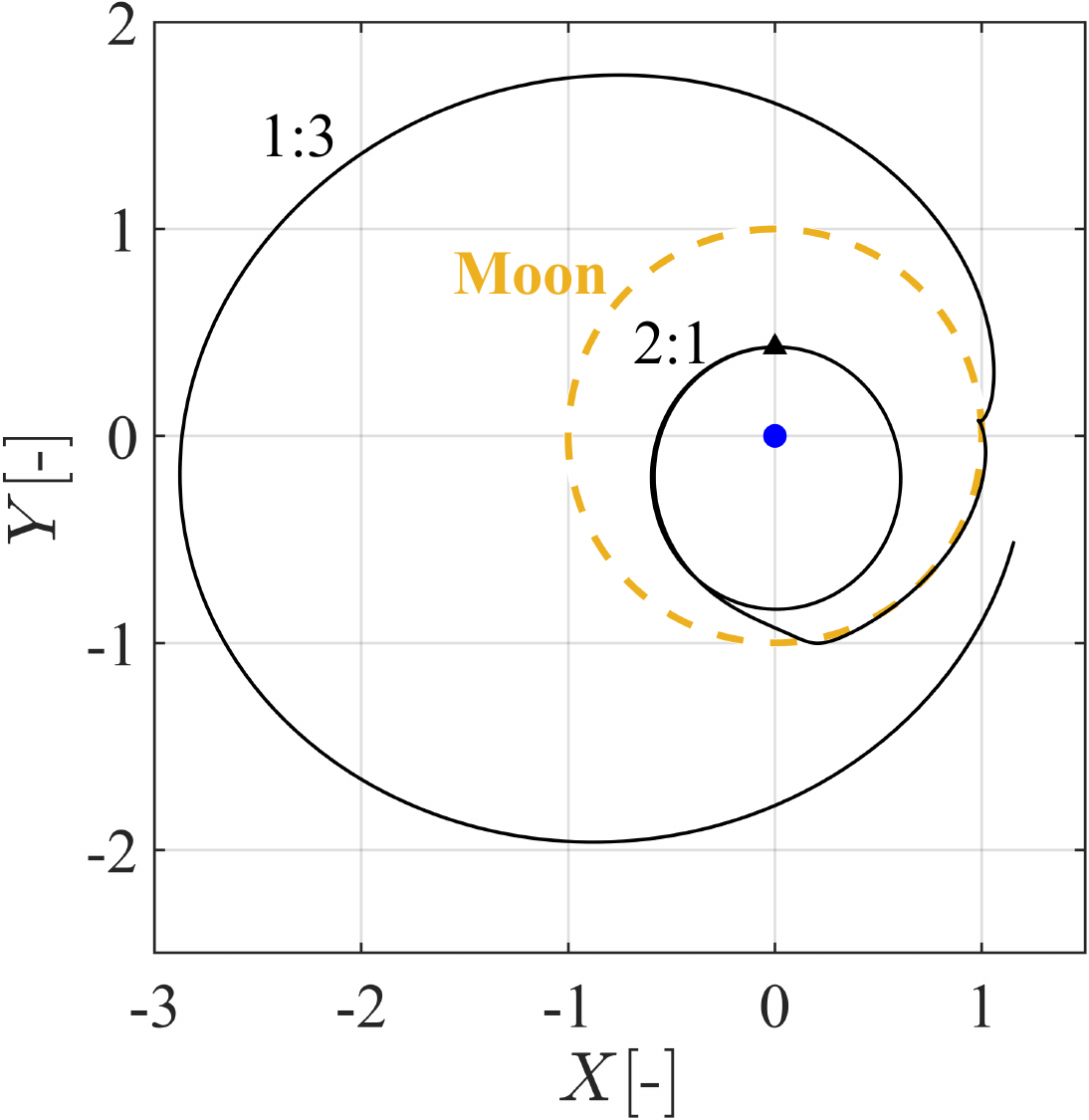}
        \subcaption{In the inertial frame.} \label{fig:resonance_hop_inertial}
    \end{minipage}
    \caption{Example of resonance transition in the Earth-Moon CR3BP~\cite{belbruno2008resonance}.}
    \label{fig:resonance_hop}
\end{figure}

Interesting transitions between orbits with different resonance ratios, called resonance transitions~\cite{koon2000heteroclinic,koon2001resonance,belbruno2008resonance}, have been observed for short-period comets such as Oterma.
Resonance transitions, also referred to as resonant hopping~\cite{belbruno1997resonance,lantoine2011optimization} or resonant gravity assists~\cite{ross2003design,ross2007multi}, are useful in trajectory design, especially for a tour design in multi-moon systems~\cite{campagnola2014jovian,campagnola2019tour}.
In this phenomenon, weak capture at the Moon plays a significant role~\cite{belbruno2008resonance}, where the Kepler energy around the Moon is non-positive in multi-body dynamics.
Figure~\ref{fig:resonance_hop} illustrates the resonance transition from $2$:$1$ to $1$:$3$ in the rotating frame and in the Earth-centered inertial frame.
The region where weak capture occurs is defined as the weak stability boundary (WSB), which was first introduced by Belbruno~\cite{belbruno1987lunar} and has been extended in Ref.~\cite{belbruno2008resonance}.
The WSB can be regarded as a more general and precise estimation of the sphere of influence in multi-body dynamics~\cite{belbruno1993sun}.

\subsection{Periapsis Poincar{\'e} map}
This study focuses on lobes divided by the stable and unstable manifolds of resonant orbits. 
The geometry of these manifolds is complex in the position space, and therefore periapsis Poincar{\'e} maps are utilized to understand the dynamics.
The periapsis Poincar{\'e} map is a Poincar{\'e} map whose surface of section lies in periapsis passage.
This idea was first introduced for astrodynamics in Ref.~\cite{villac2003escaping}.
One of its advantages is enabling the projection of dynamical structures onto position space, which is applied to design transfer trajectories~\cite{scheuerle2024energy,ikeda2023design}.
This study extracts dynamical structures related to resonances by numerically constructing a periapsis Poincar{\'e} map in the Delaunay elements~\cite{vallado2001fundamentals}. 
These elements are canonical and suitable for analyzing lobe dynamics.

\begin{figure}[!t]
    \centering
    \includegraphics[width=0.45\columnwidth]{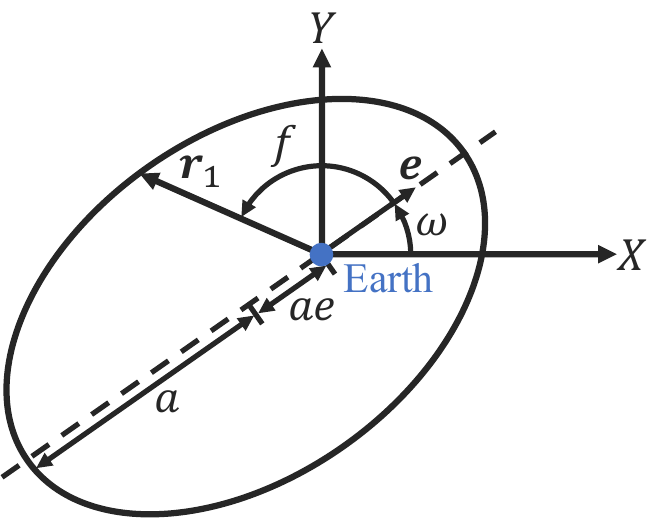}
    \caption{Classical orbital elements for an elliptic orbit in the temporary inertial frame (when $X\dot{Y} - Y\dot{X} > 0$).} \label{fig:inertial_frame}
\end{figure}

Periapses are calculated with respect to the Earth (i.e., $\dot{r}_1 = 0$ and $\ddot{r}_1 > 0$) throughout this paper.
The following coordinate transformation is necessary to obtain periapsis Poincar{\'e} maps in the Delaunay elements. 
First, the temporary inertial frame, the $X$-$Y$ frame, is defined so that its axes coincide with those of the rotating $x$-$y$ frame, and its origin is located at the Earth. 
Specifically, the relationship between the $X$-$Y$ frame and the $x$-$y$ plane is expressed as
\begin{equation}
    \begin{bmatrix} X \\ Y \\ \dot{X} \\ \dot{Y} \end{bmatrix} = \begin{bmatrix} 1 & 0 & 0 & 0 \\ 0 & 1 & 0 & 0 \\ 0 & -1 & 1 & 0 \\ 1 & 0 & 0 & 1 \end{bmatrix} \begin{bmatrix} x+\mu \\ y \\ \dot{x} \\ \dot{y} \end{bmatrix}.
\end{equation}
On this inertial frame, classical orbital elements for the planar problem (the semi-major axis $a$, eccentricity $e$, argument of periapsis $\omega$, and true anomaly $f$) are defined such that
\begin{align}
    a &= \frac{\left(1-\mu\right)r_1}{2\left(1-\mu\right) - r_1 V^2}, & & \\
    \bm{e} &= - \frac{\left(\bm{r}_1 \times \bm{V}\right) \times \bm{V}}{1-\mu} - \frac{\bm{r}_1}{r_1}, & e &= \|\bm{e}\|,
\end{align}
$\omega$ is the angle from the $X$ axis to the $\bm{e}$ vector, and $f$ is the angle from the $\bm{e}$ vector to the $\bm{r}_1$ vector, where
\begin{align}
    \bm{r}_1 &= \begin{bmatrix} X \\ Y \\ 0 \end{bmatrix}, & r_1 &= \|\bm{r}_1\|, &
    \bm{V} &= \begin{bmatrix} \dot{X} \\ \dot{Y} \\ 0 \end{bmatrix}, & V &= \|\bm{V}\|.
\end{align}
Figure~\ref{fig:inertial_frame} illustrates these orbital elements on the $X$-$Y$ frame.
Note that this transformation assumes that the inclination of the elliptic orbit is zero and the angular momentum is positive, i.e., $X\dot{Y} - Y\dot{X} > 0$.

The orbital elements $\left(a,\,e,\,\omega,\,f\right)$ are converted to the Delaunay elements $\left(l_d,\,g_d,\,L_d,\,G_d\right)$ as follows:
\begin{align}
    l_d &= M = E - e\sin E, \label{eq:def_ld}\\
    g_d &= \omega, \\
    L_d &= \sqrt{\left(1 - \mu\right)a}, \\
    G_d &= \sqrt{\left(1 - \mu\right)a\left(1 - e^2\right)},
\end{align}
where $G_d$ is identical to the magnitude of the angular momentum $h = |X\dot{Y} - Y\dot{X}|$, $M$ is the mean anomaly, and $E$ is the eccentric anomaly satisfying
\begin{equation}
    E = 2\tan^{-1}\left(\sqrt{\frac{1 - e}{1 + e}}\tan\frac{f}{2}\right). \label{eq:trans_f2e}
\end{equation}
Periapsis passage corresponds to $f = 0$, which is equal to $l_d = 0$ from Eqs.~\eqref{eq:def_ld} and~\eqref{eq:trans_f2e} when $0 < e < 1$.

\begin{figure}[!t]
    \centering
    \includegraphics[width=0.5\columnwidth]{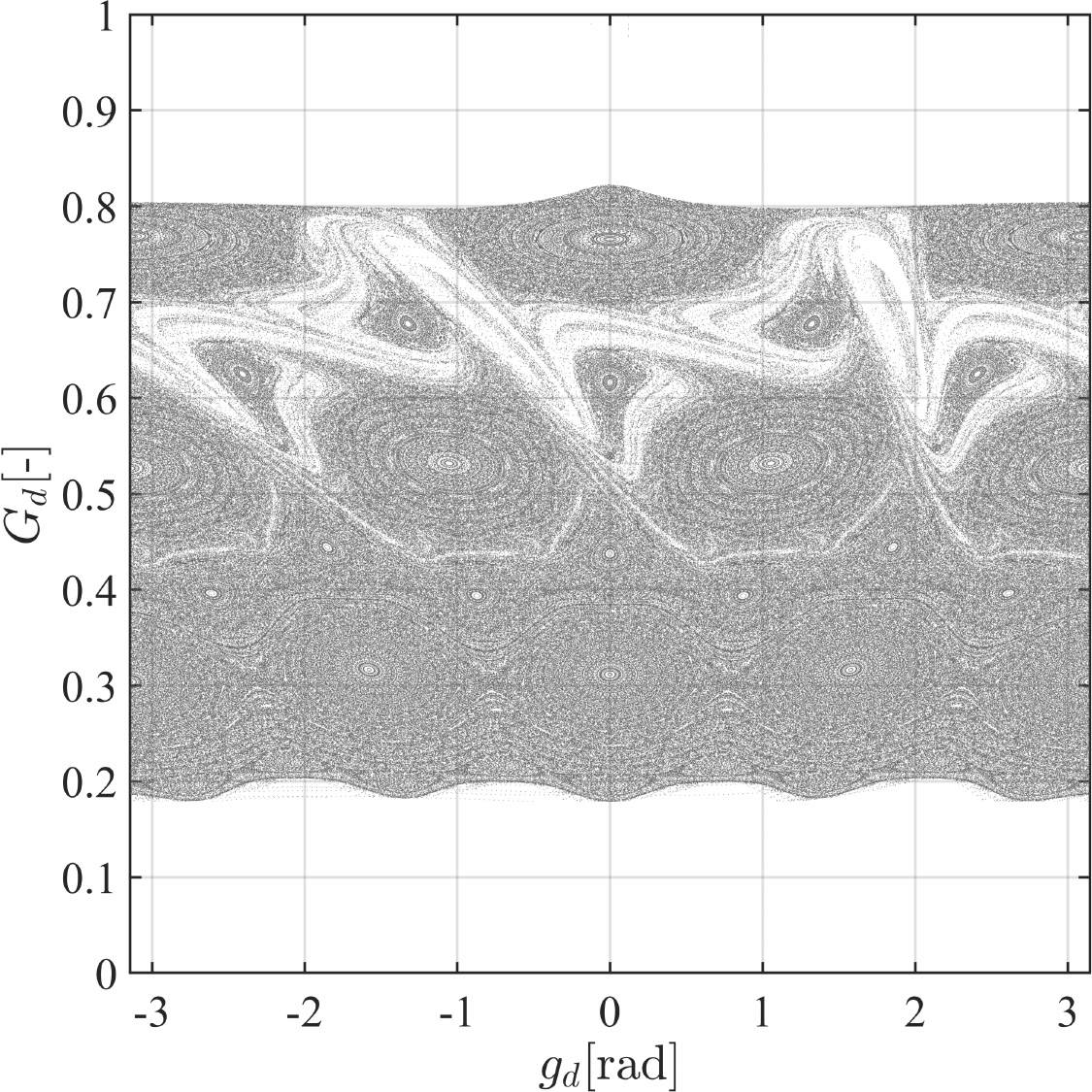}
    \caption{Periapsis passage points on the Poincar{\'e} section for the Earth-Moon system ($l_d = 0$,\,$X\dot{Y} - Y\dot{X} > 0$,\,and $C_J = 3.16$).} \label{fig:poincare_map}
\end{figure}

In this paper, the periapsis Poincar{\'e} map is defined as $l_d = 0\,(X\dot{Y} - Y\dot{X} > 0)$ in the canonical $g_d$-$G_d$ plane and illustrated in Fig.~\ref{fig:poincare_map}.
The four-dimensional phase space in the planar CR3BP is reduced to two-dimensional Poincar{\'e} maps by taking a surface of section (periapsis passage for Fig.~\ref{fig:poincare_map}) and fixing the Jacobi constant ($C_J = 3.16$ for this case).
On this map, the phase space can be divided into \textit{resonance regions}~\cite{mackay1987resonances,koon2011dynamical} and \textit{chaotic zones}.
The former includes quasi-periodic and periodic orbits, while the latter consists of chaotic orbits.

\subsection{Bicircular restricted four-body problem (BCR4BP)}
\begin{figure}[t]
    \centering
    \includegraphics[width=0.45\columnwidth]{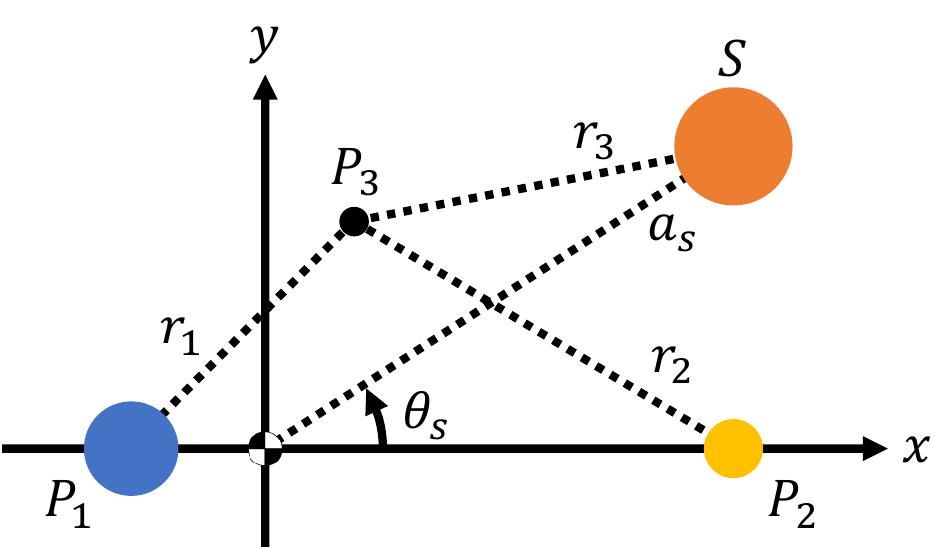}
    \caption{Coordinate system in BCR4BP (Earth--Moon rotating frame).}
    \label{fig:BCR4BP}
\end{figure}
This subsection briefly introduces the BCR4BP, one of the models for the Sun--Earth--Moon system.
The BCR4BP is the simplest model for this four-body system, assuming that the Sun $S$ moves around the barycenter of the Earth--Moon rotating frame in a circular orbit.
Its coordinate system is shown in Fig.~\ref{fig:BCR4BP}.
The Sun is often assumed to be on the Earth--Moon orbital plane. 
The position of the Sun is then written as
\begin{align}
    x_s &= a_s\cos\theta_s\,, & y_s &= a_s\sin\theta_s\,, & z_s &= 0,
\end{align}
where $a_s$ is the non-dimensional distance between the Sun and the barycenter of the Earth--Moon system ($3.89173\times10^2$~[-]), and $\theta_s$ is the phase angle of the Sun, given by
\begin{equation}
    \theta_s = \theta_s^\ast + \omega_s t.
\end{equation}
Note that $\theta_s^\ast$ is the initial value of $\theta_s$, $\omega_s$ is the non-dimensional angular velocity of the Sun, defined as  $\omega_s = \sqrt{\left(1 + m_s\right) / a_s^3} - 1$, and $m_s$ is the non-dimensional mass of the Sun ($3.28914\times10^5$~[-]).
Thus, the BCR4BP is time-dependent, and the Earth--Moon system is periodically perturbed by the Sun.

Under these assumptions, the non-dimensional equations of motion of $P_3$ are expressed with a pseudo-potential function $\Upsilon$ based on Eq.~\eqref{eq:EoM} as follows:
\begin{equation}
    \left\{\,
    \begin{aligned}
        \ddot{x} - 2\dot{y} &= \frac{\partial \Upsilon}{\partial x}\\
        \ddot{y} + 2\dot{x} &= \frac{\partial \Upsilon}{\partial y}\\
        \ddot{z} &= \frac{\partial \Upsilon}{\partial z}
    \end{aligned}
    \right.\,,
    \label{eq:EoM_BCR4BP}
\end{equation}
\begin{align}
    \Upsilon &= \frac{1}{2}\left( x^2 + y^2 \right) + \frac{1 - \mu}{r_1} + \frac{\mu}{r_2} + \frac{m_s}{r_s} - \frac{m_s}{a_s^3}\left(x_s x + y_s y + z_s z\right),\\
    r_s &= \sqrt{ (x - x_s)^2 + (y - y_s)^2 + (z - z_s)^2 }.
\end{align}
When $m_s \to 0$, Eq.~\eqref{eq:EoM} of the CR3BP is recovered from Eq.~\eqref{eq:EoM_BCR4BP} of the BCR4BP.

%% file: 3_lobe_dynamics.tex
\section{Lobe dynamics}\label{sec:lobe_dynamics}
This section introduces the idea of extracting lobes suitable for guiding a spacecraft in chaotic zones~\cite{hiraiwa2024designing}.
In the planar CR3BP, the stable and unstable manifolds of a resonant orbit may form regions called lobes.
Lobe dynamics describes the chaotic transport of phase space volume between lobes within a single realm.
A detailed explanation of lobe dynamics for trajectory design is given below.

\subsection{Description of lobe dynamics for trajectory design}
\begin{figure}[!b]
    \centering
    \includegraphics[width=0.5\columnwidth]{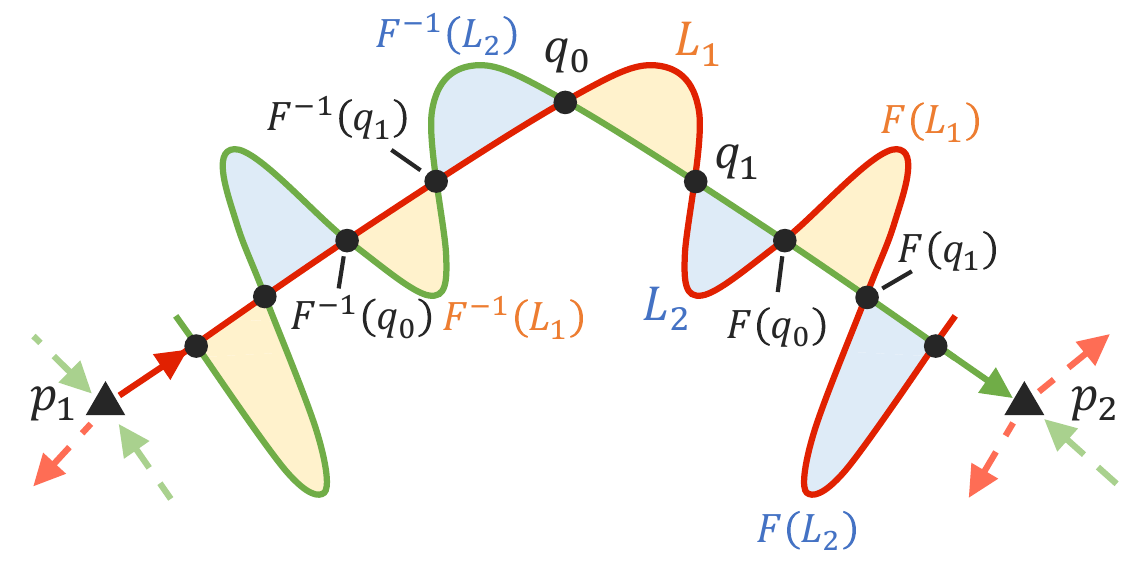}
    \caption{General description of lobes.} \label{fig:lobe_def}
\end{figure}%
Consider hyperbolic fixed points ($p_1$ and $p_2$) on a two-dimensional, area- and orientation-\hspace{0pt}preserving map $F$ (see Fig.~\ref{fig:lobe_def}).
Examples of $F$ include the standard map and a Poincar{\'e} map in the planar CR3BP.
On the map $F$, it is assumed that an unstable manifold of $p_1$, $W^{u}_{p_1}$, and a stable manifold of $p_2$, $W^{s}_{p_2}$, form a heteroclinic tangle (or a homoclinic tangle when $p_1 = p_2$).
Heteroclinic and homoclinic points are defined as follows:
\begin{dfn}[Heteroclinic/\hspace{0pt}homoclinic point]
    A point $q_i$ on the map $F$ is called a \textit{heteroclinic point} if $q_i \in W^{u}_{p_1} \cap W^{s}_{p_2}$.
    If $p_1 = p_2$, then $q_i$ is called a \textit{homoclinic point}.
\end{dfn}

To describe lobes and lobe dynamics, primary intersection points are defined as follows:
\begin{dfn}[Primary intersection point, pip~\cite{rom1990tranport}]
    Consider a heteroclinic/\hspace{0pt}homoclinic point $q_i \in W^{u}_{p_1} \cap W^{s}_{p_2}$.
    The segment of $W^{u}_{p_1}$ from $X$ to $Y$ and the segment of $W^{s}_{p_2}$ from $X$ to $Y$ are denoted as $U[X, Y]$ and $S[X, Y]$, respectively.
    Then, $q_i$ is called a \textit{primary intersection point} if $U[p_1, q_i]$ and $S[p_2, q_i]$ intersect only in $q_i$ (and possibly in $p_1$ when $q_i$ is a homoclinic point).
    In Fig.~\ref{fig:lobe_def}, all the black dots represent pips.
\end{dfn}
\noindent Lobes are then defined as follows:
\begin{dfn}[Lobe~\cite{rom1990tranport}]
    Let $q_0$ and $q_1$ be adjacent pips, i.e., no other pips exist on $U[q_0, q_1]$ and $S[q_0, q_1]$.
    A \textit{lobe} is defined as a region bounded by $U[q_0, q_1]$ and $S[q_0, q_1]$.
\end{dfn}
\noindent The transport structure of lobes, called \textit{lobe dynamics}, is governed by the map $F$.
For example, the lobe bounded by $U[q_0, q_1]$ and $S[q_0, q_1]$ is mapped to the lobe bounded by $U[F(q_0), F(q_1)]$ and $S[F(q_0), F(q_1)]$. 
Specific lobes moving into/\hspace{0pt}out of resonance regions under one mapping of $F$ (e.g., $F^{-1}(L_2)$ and $F^{-1}(L_1)$ in Fig.~\ref{fig:lobe_def}) are called \textit{turnstiles}~\cite{mackay1984transport,meiss1992symplectic,meiss2015thirty}, which govern the transport between resonance regions and chaotic zones.

A series of lobes mapped by $F$ is called a \textit{lobe sequence}, which is introduced in Ref.~\cite{hiraiwa2024designing} and defined as follows:
\begin{dfn}[Lobe sequence~\cite{hiraiwa2024designing}]
    Assume that a lobe $L$ exists on the map $F$.
    A sequence
    \begin{equation*}
        \{\cdots, F^{-n}(L), \cdots, F^{-1}(L), L, F(L), \cdots, F^{n}(L), \cdots\}\,,
    \end{equation*}
    for $\forall n$ is called a lobe sequence $\Lambda(F, L)$.
\end{dfn}
\noindent In Fig.~\ref{fig:lobe_def}, the stable and unstable manifolds form two lobe sequences
\begin{equation*}
    \{\cdots, F^{-1}(L_1), L_1, F(L_1), \cdots\} \,\,\mathrm{(a\;sequence\;of\;the\;yellow\;lobes),}
\end{equation*}
and
\begin{equation*}
    \{\cdots, F^{-1}(L_2), L_2, F(L_2), \cdots\} \,\,\mathrm{(a\;sequence\;of\;the\;blue\;lobes).}
\end{equation*}
The number of lobe sequences depends on pairs of manifolds dividing the phase space into lobes.

\begin{figure}[!b]
    \centering
    \includegraphics[width=0.3\columnwidth]{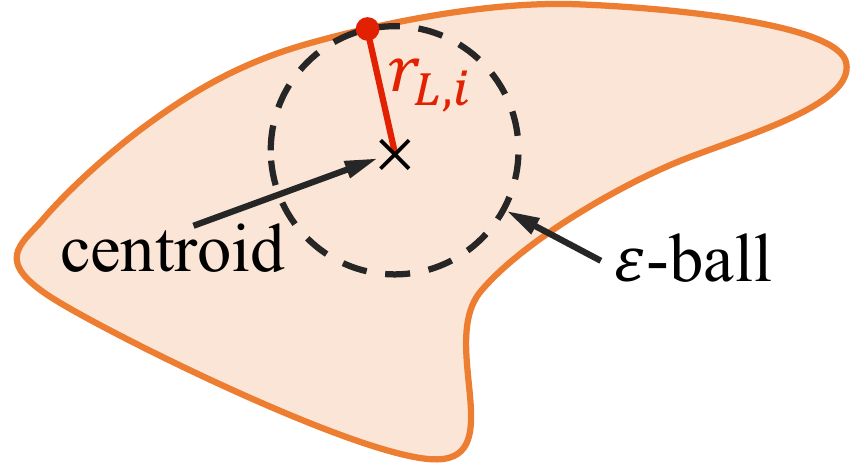}
    \caption{Illustration of the radius of a lobe for computation.} \label{fig:radius_lobe}
\end{figure}

It is difficult to visualize the entire picture of a lobe sequence because the accurate and extended propagation of the stable and unstable manifolds is necessary to detect finer lobes in the sequence.
In this study, this numerical difficulty is avoided by leveraging \textit{effective lobes}~\cite{hiraiwa2024designing}.
First, the \textit{radius of a lobe} is defined as follows: 
\begin{dfn}[Radius of a lobe~\cite{hiraiwa2024designing}]
    The $\varepsilon$-ball is an open ball whose radius is $\varepsilon \in \mathbb{R} \ (\varepsilon > 0)$, defined as: 
    \begin{equation*}
        B_\varepsilon\left(\bm{c}\right) := \{\bm{\xi} \in \mathbb{R}^2 : \|\bm{\xi} - \bm{c}\| < \varepsilon \}\,,
    \end{equation*}
    where $\bm{\xi}$ is the state on the map $F$, and $\bm{c} \in \mathbb{R}^2$ is the center of this ball.
    The radius of a lobe is the maximum radius of $\varepsilon$-ball contained within a lobe region $L_i$, defined as:
    \begin{equation*}
        r_{L, i} := \max_{\bm{c} \in L_i,\,B_\varepsilon\left(\bm{c}\right) \subset L_i} \varepsilon\,.
    \end{equation*}
\end{dfn}
\noindent For simplicity, the radius of a lobe $r_{L, i}$ is calculated as the minimum distance between the centroid of a lobe and its boundary (see Fig.~\ref{fig:radius_lobe}).
Now, the \textit{effective lobes} and \textit{effective lobe sequences} are defined using a threshold radius $r_L^\ast$ for robust transfer design.
\begin{dfn}[Effective lobe~\cite{hiraiwa2024designing}]
    An effective lobe is defined as a lobe whose radius satisfies the condition
    \begin{equation*}
        r_{L, i} > r_L^\ast\,,
    \end{equation*}
    where $r_L^\ast$ is the threshold value for the radius of lobes.
\end{dfn}
\begin{dfn}[Effective lobe sequence~\cite{hiraiwa2024designing}]
    A finite lobe sequence $\Lambda(F, L, r_L^\ast) \subset \Lambda(F, L)$ in which all lobes are effective lobes satisfying $r_{L, i} > r_L^\ast$ and are selected in order from the sequence $\Lambda(F, L)$ is called an \textit{effective lobe sequence} with respect to $r_L^\ast$.
\end{dfn}

To investigate the transport structure between lobe sequences, this study assumes that the transfers between lobes (region-to-region transfer) are represented by transfers between the centroids of the lobes (point-to-point transfer).
Under this assumption, effective lobes allow errors whose size is less than $r_L^\ast$ (e.g., control errors and/or orbit determination errors).
Despite such errors, spacecraft will remain within effective lobes, and the transfer between effective lobes becomes robust.

\subsection{Lobe sequences in the CR3BP}\label{sec:lobe_sequence}
This subsection describes effective lobe sequences in the planar Earth--Moon CR3BP.
The $7$:$2$ and $3$:$1$ unstable resonant orbits are selected for trajectory design because the corresponding resonance regions are relatively large in Fig.~\ref{fig:poincare_map} and their lobe dynamics is considered to enhance phase space transport.
The initial guesses for these resonant orbits can be obtained by a grid search on the Poincar{\'e} map, and the exact initial conditions are acquired through the differential correction method~\cite{howell1984three}.
By propagating the stable and unstable manifolds of the obtained resonant orbits, heteroclinic tangles on the Poincar{\'e} map are revealed.
Based on the heteroclinic tangles, one can find the initial guesses for pips.
Pips are precisely identified by parameterizing the manifolds and calculating their crossing points.
The lobes are determined by the adjacent pips and the manifolds connecting them.
Thus, parameterizing the manifolds makes it easier to detect lobe boundaries.
Once a lobe in a certain lobe sequence is found, the whole lobe sequence can be revealed by propagating the boundary of the lobe forward and backward in time.

\begin{figure}[!t]
    \centering
    \includegraphics[width=\columnwidth]{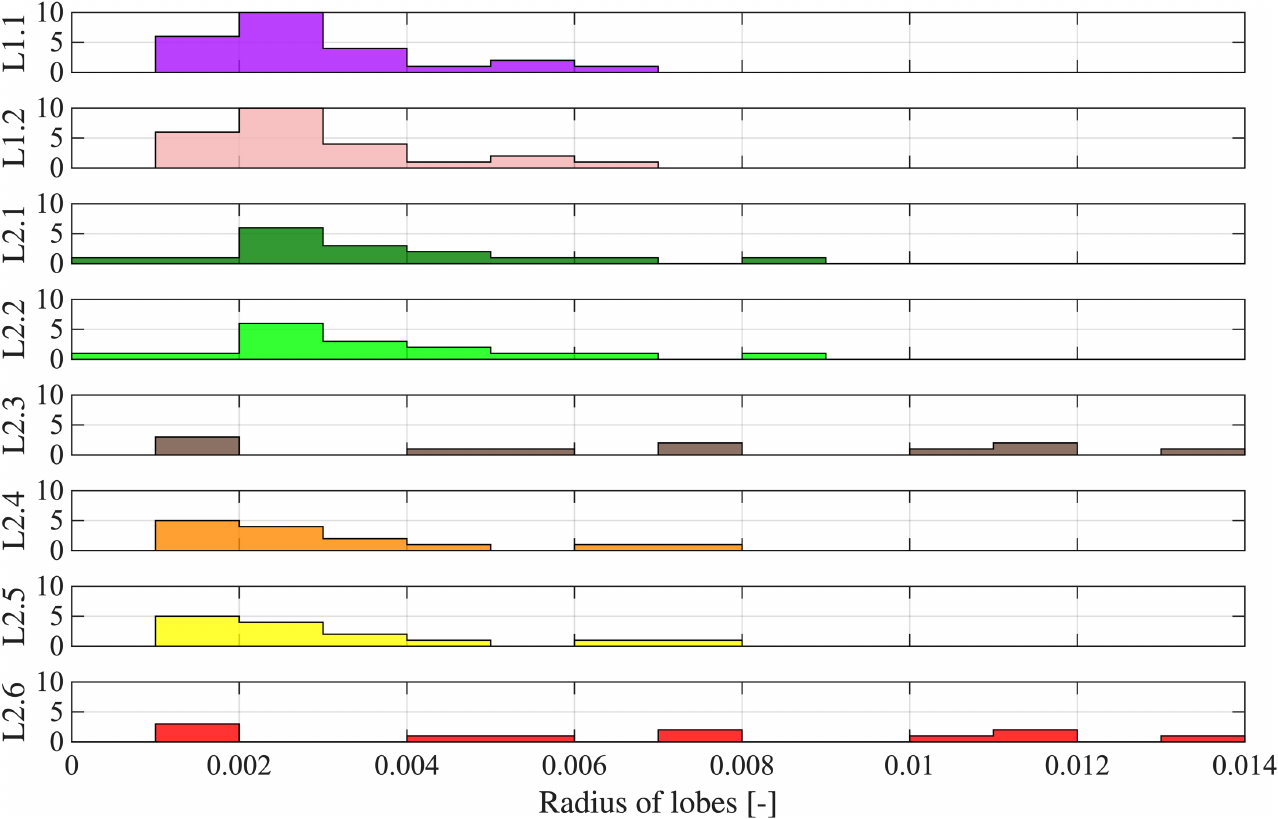}
    \caption{Histogram of lobe radii for each lobe sequence.}
    \label{fig:radius_of_lobe_hist}
\end{figure}
\begin{figure*}[!p]
    \centering
    \begin{minipage}{0.45\textwidth}
        \centering
        \includegraphics[width=\columnwidth]{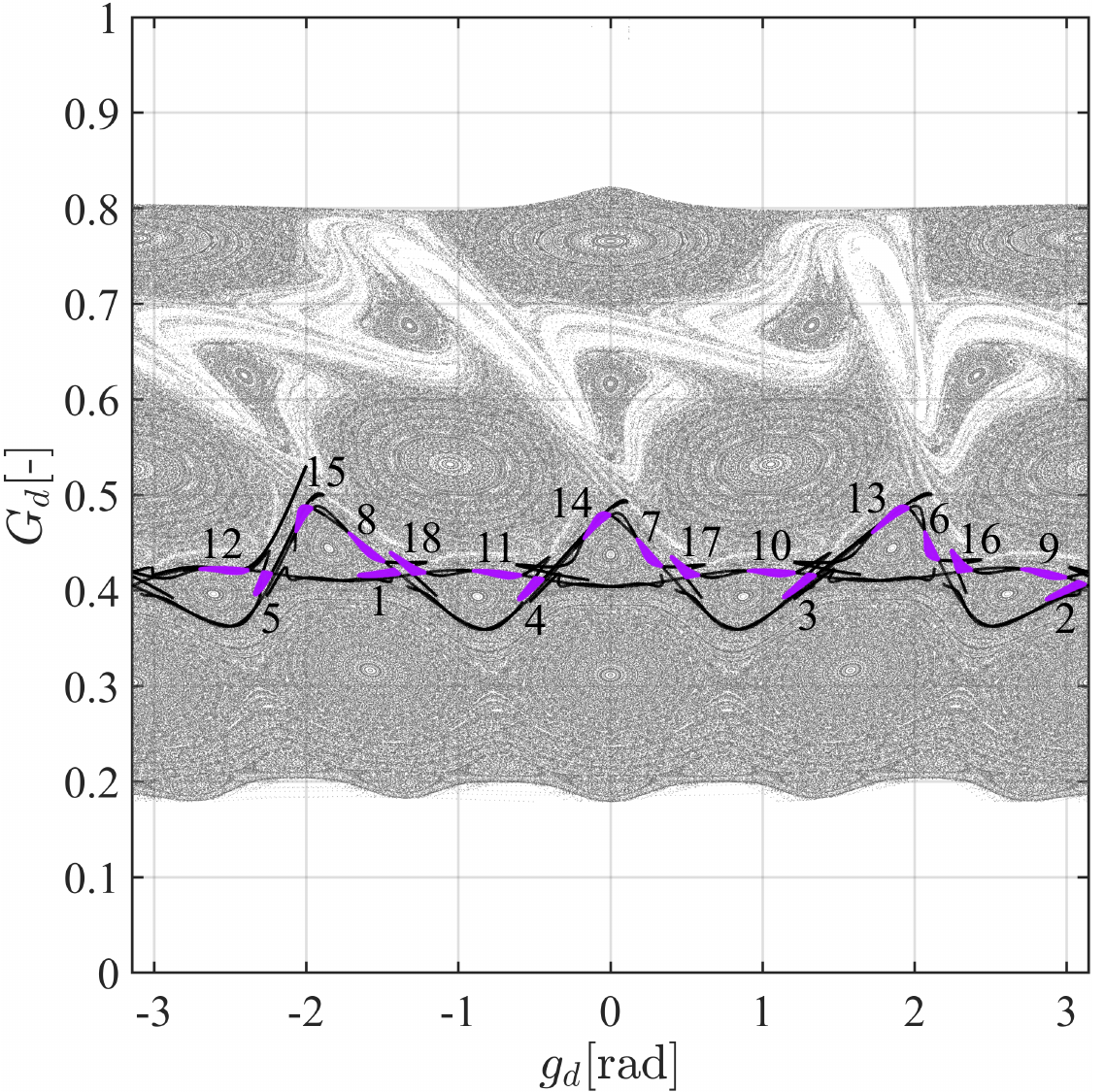}
        \subcaption{L1.1} \label{fig:lobe_1_1}
    \end{minipage}%
    \hspace{10mm}
    \begin{minipage}{0.45\textwidth}
        \centering
        \includegraphics[width=\columnwidth]{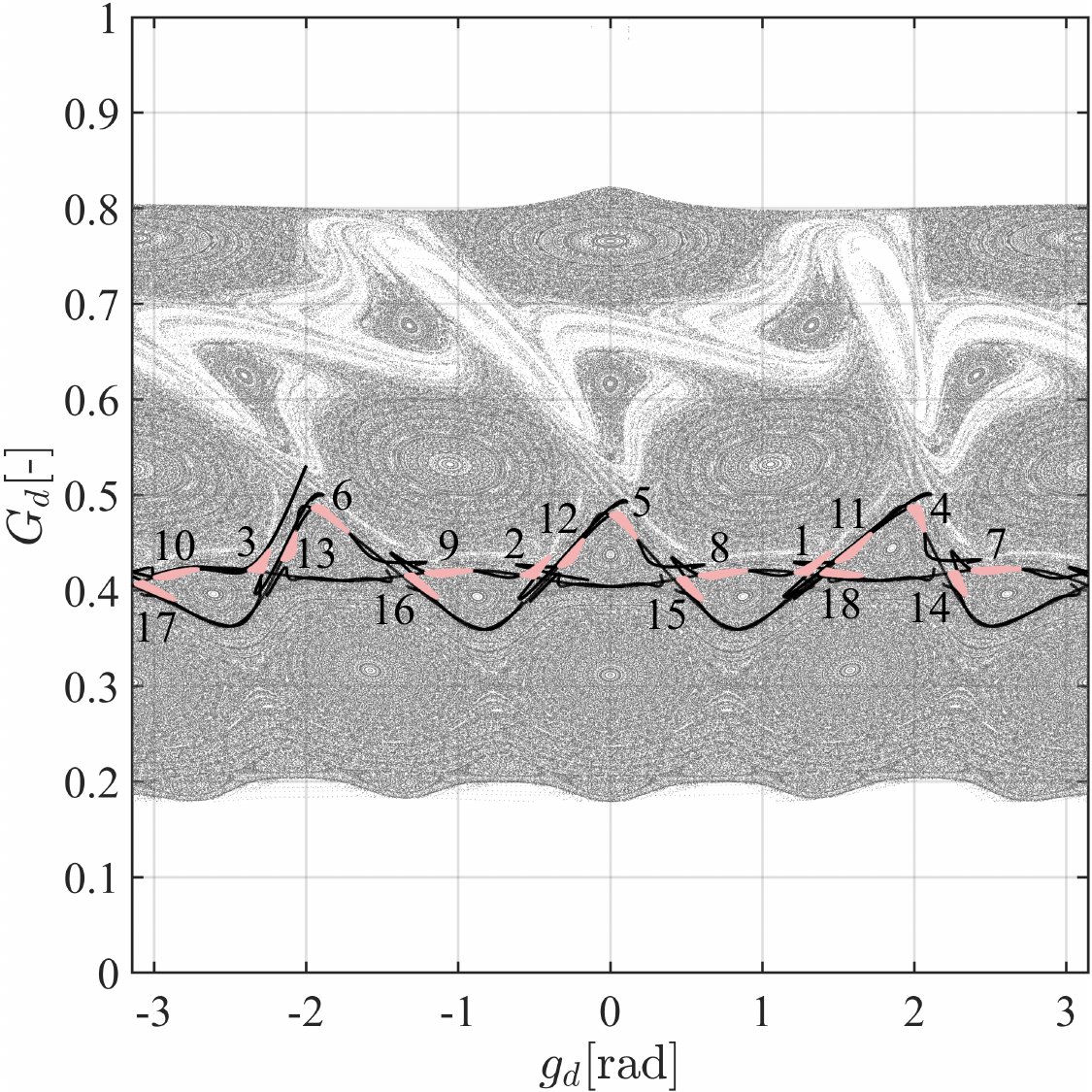}
        \subcaption{L1.2} \label{fig:lobe_1_2}
    \end{minipage}\\
    \vspace{10mm}
    \begin{minipage}{0.45\textwidth}
        \centering
        \includegraphics[width=\columnwidth]{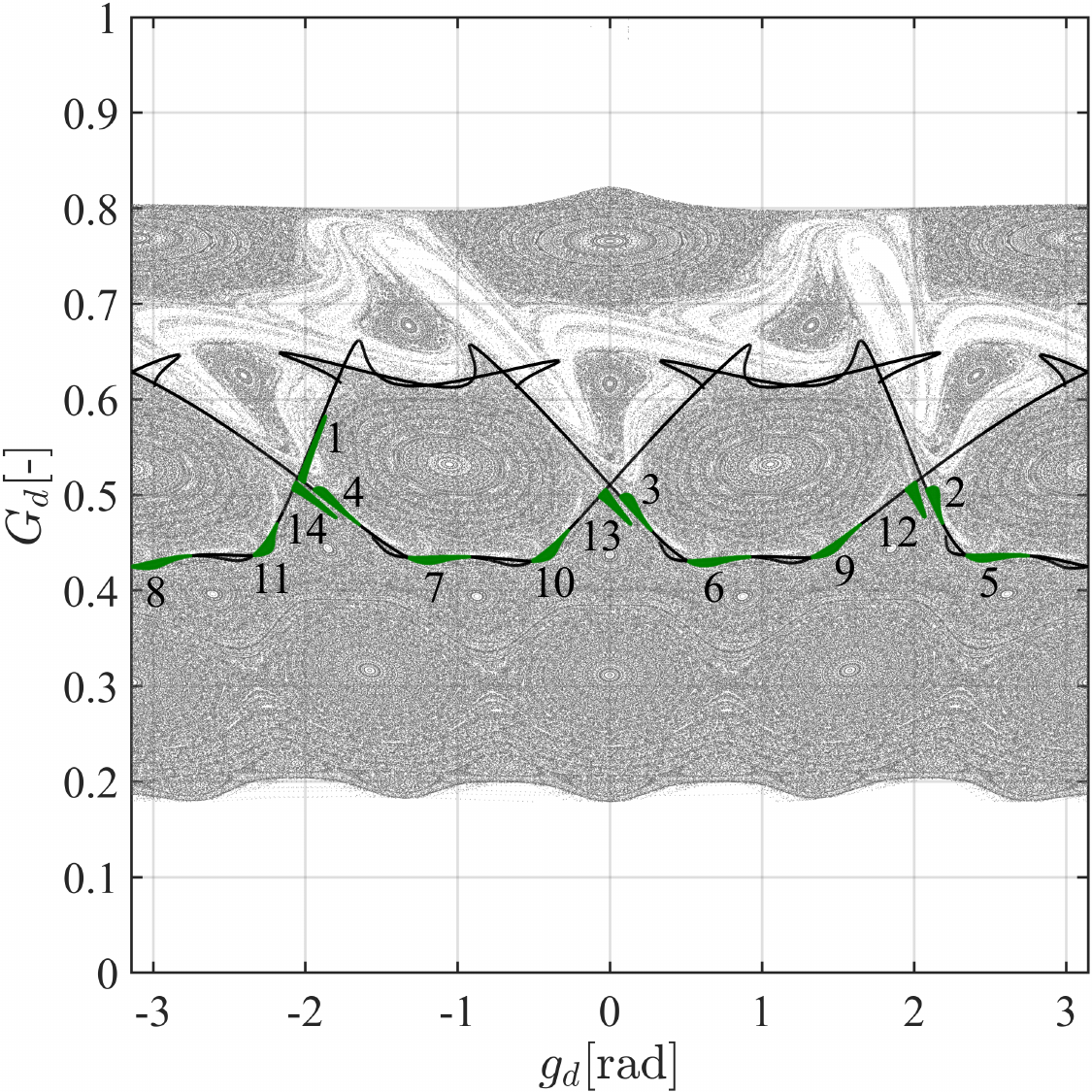}
        \subcaption{L2.1} \label{fig:lobe_2_1}
    \end{minipage}%
    \hspace{10mm}
    \begin{minipage}{0.45\textwidth}
        \centering
        \includegraphics[width=\columnwidth]{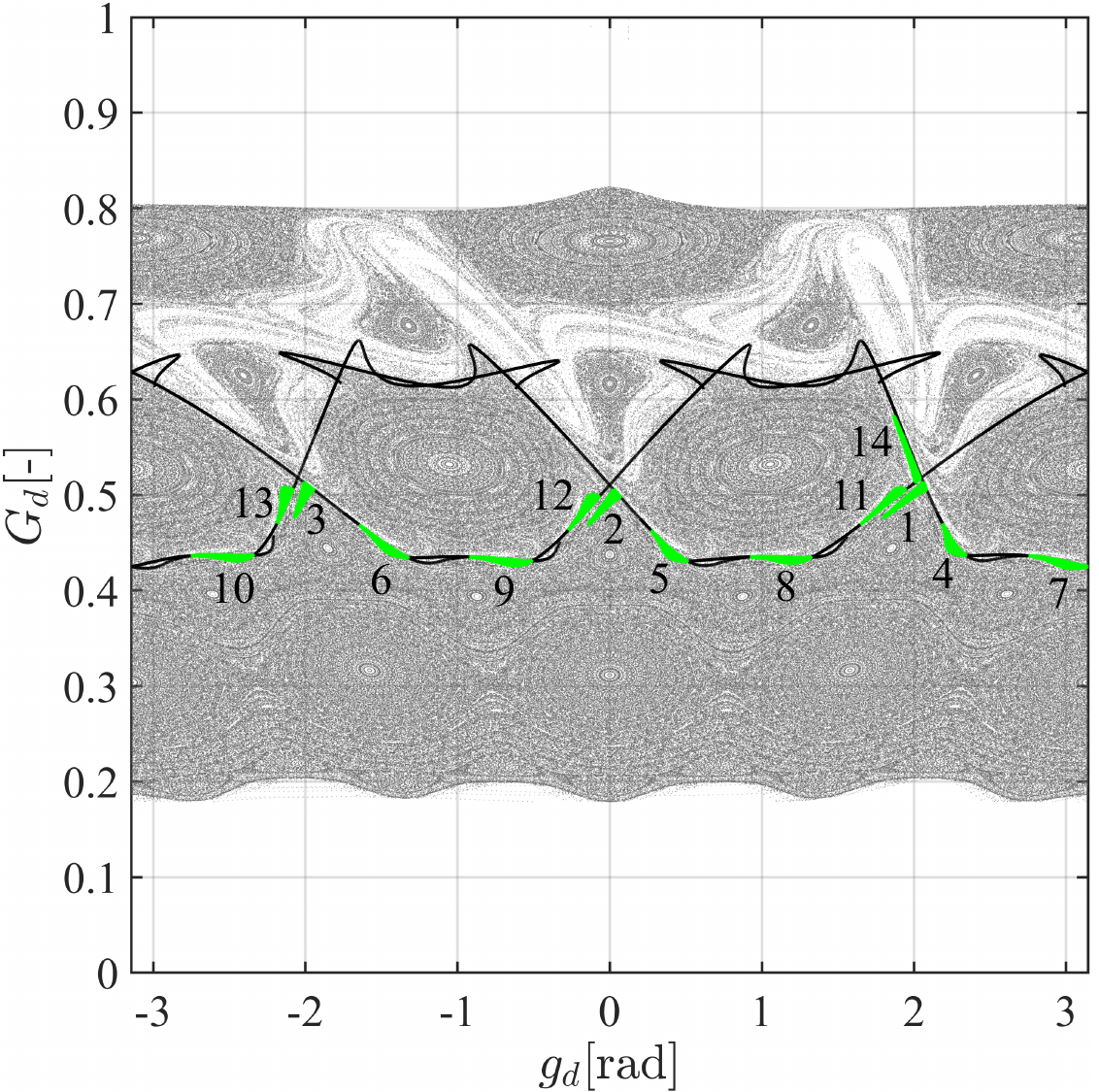}
        \subcaption{L2.2} \label{fig:lobe_2_2}
    \end{minipage}%
    \caption{Effective lobe sequences in the planar Earth--Moon CR3BP.} \label{fig:lobe_sequence_1}
\end{figure*}

Figure~\ref{fig:radius_of_lobe_hist} illustrates the distribution of lobe radii in each lobe sequence.
The name of a lobe sequence, \textit{LX.Y}, indicates the \textit{Y}-th lobe sequence corresponding to the \textit{X}-th periodic orbit.
Specifically, the first and second orbits are the $7$:$2$ and $3$:$1$ unstable resonant orbits, respectively.
Note that lobes with $r_L^\ast < 0.001$ are not thoroughly investigated because of numerical difficulties in detecting finer lobes.
Figure~\ref{fig:lobe_sequence_1} shows the effective lobes when $r_L^\ast = 0.002$ for the selected lobe sequences.
The numbers in this figure indicate the order of transfers in the lobe sequence.

\begin{figure*}[!p]
    \centering
    \begin{minipage}{0.45\textwidth}
        \centering\setcounter{caption@flags}{4}
        \includegraphics[width=\columnwidth]{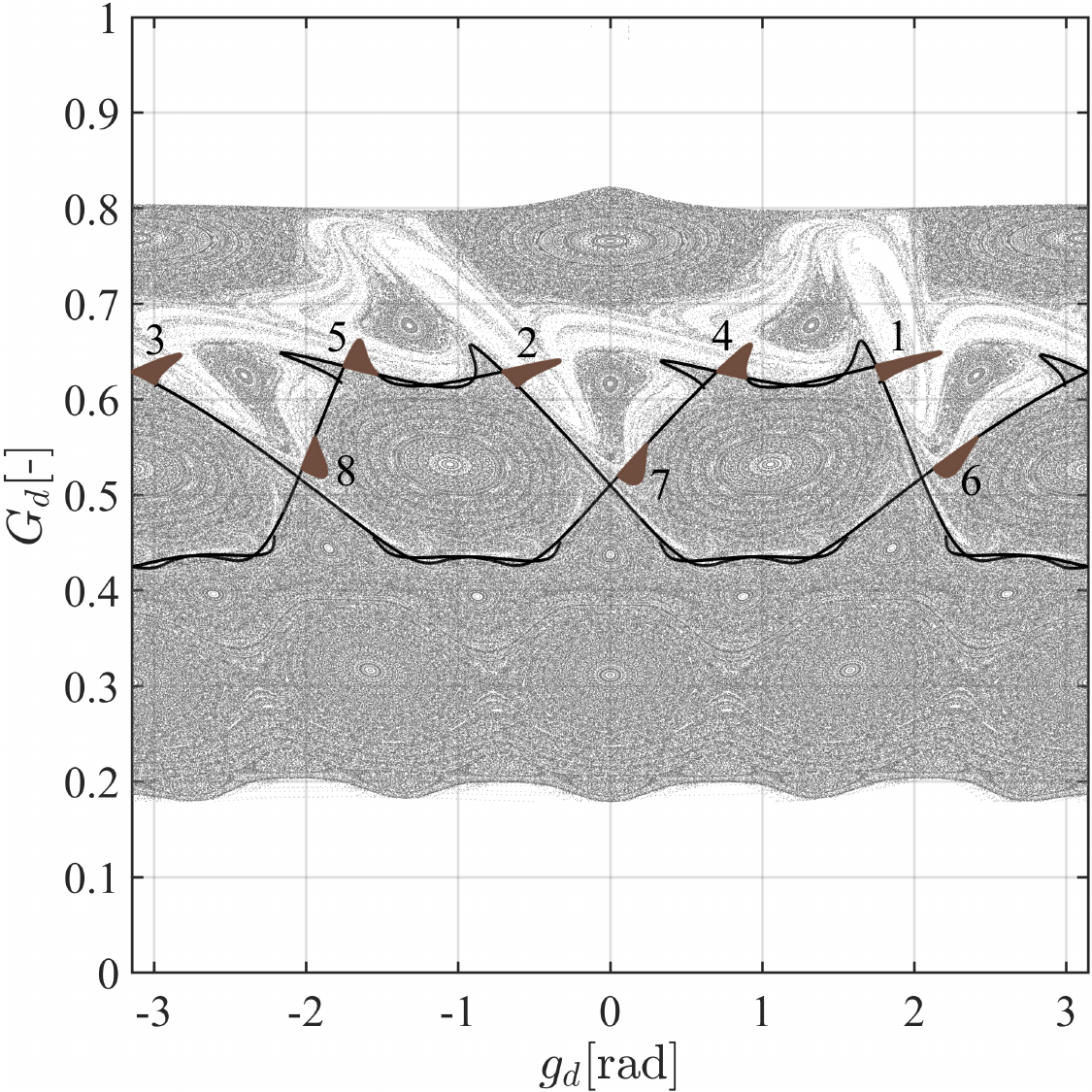}
        \subcaption{L2.3} \label{fig:lobe_2_3}
    \end{minipage}%
    \hspace{10mm}
    \begin{minipage}{0.45\textwidth}
        \centering
        \includegraphics[width=\columnwidth]{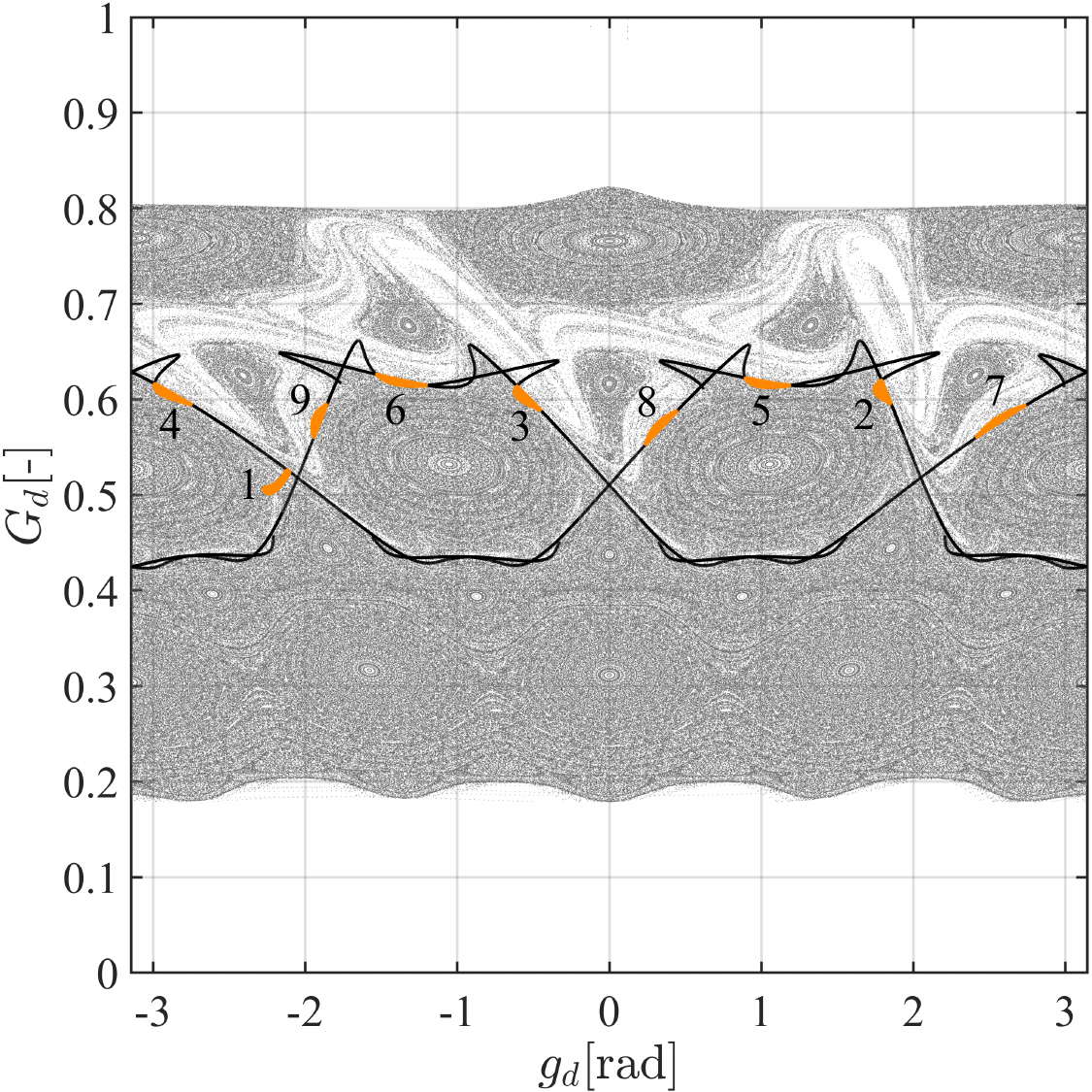}
        \subcaption{L2.4} \label{fig:lobe_2_4}
    \end{minipage}\\
    \vspace{10mm}
    \begin{minipage}{0.45\textwidth}
        \centering
        \includegraphics[width=\columnwidth]{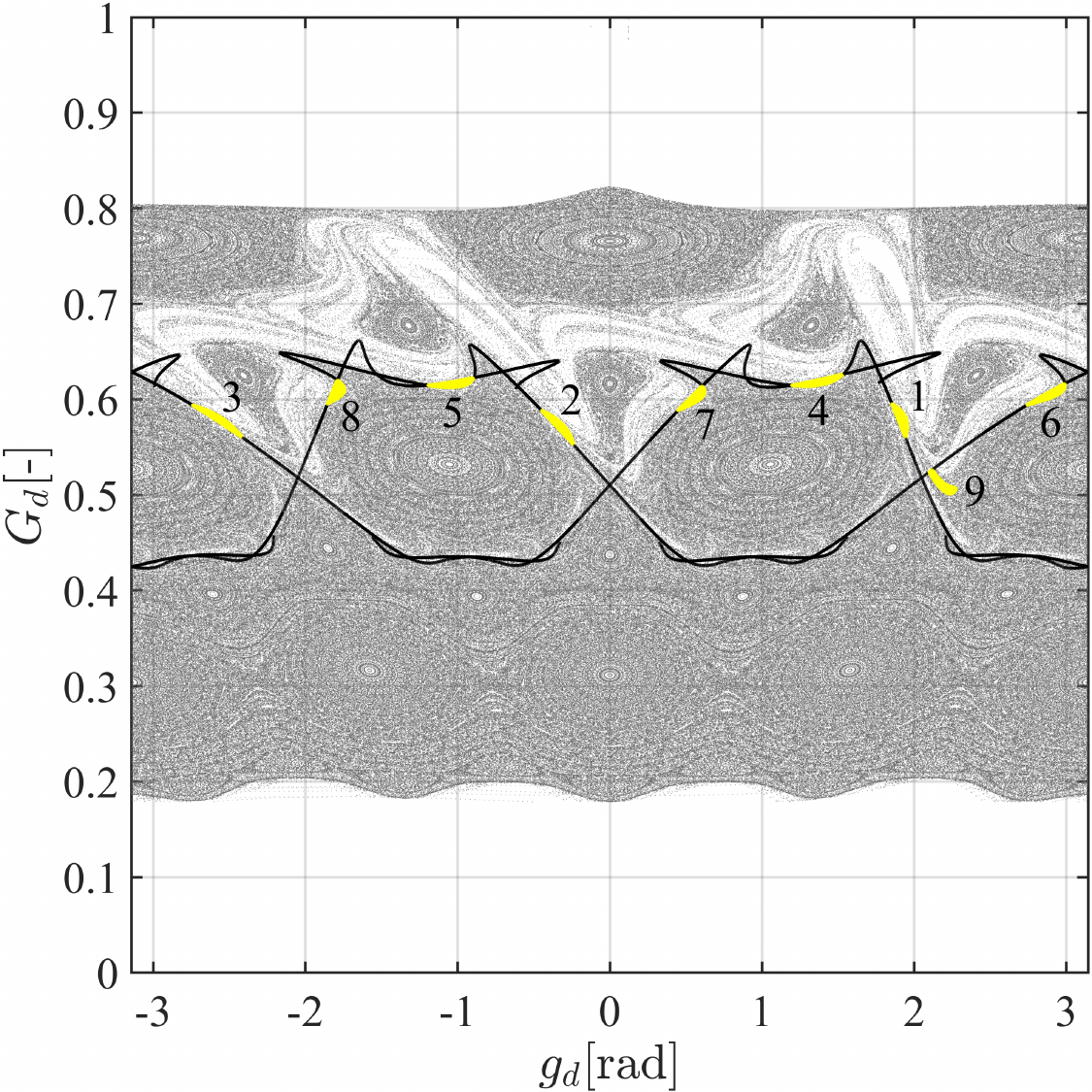}
        \subcaption{L2.5} \label{fig:lobe_2_5}
    \end{minipage}%
    \hspace{10mm}
    \begin{minipage}{0.45\textwidth}
        \centering
        \includegraphics[width=\columnwidth]{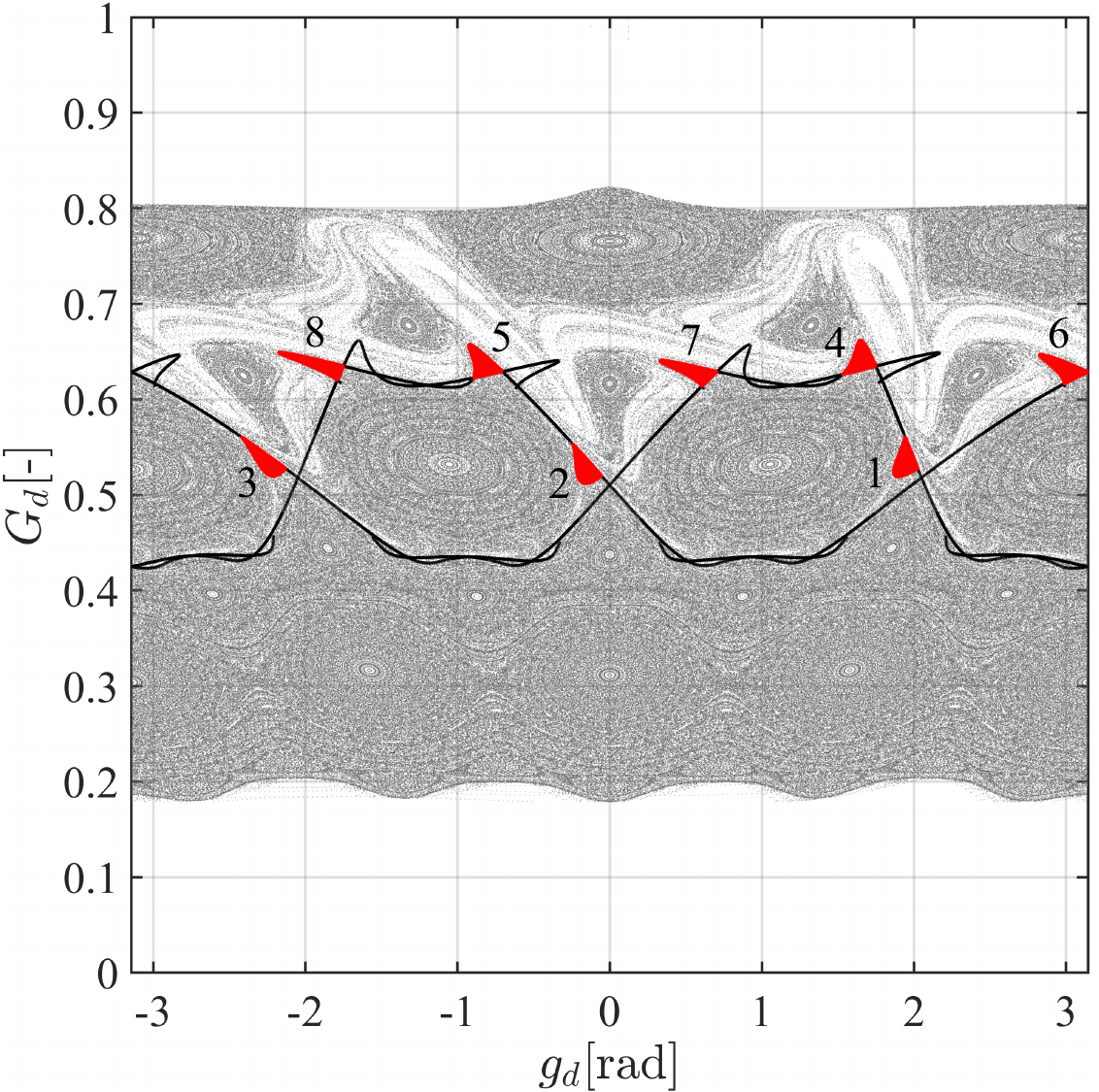}
        \subcaption{L2.6} \label{fig:lobe_2_6}
    \end{minipage}%
    \caption{Effective lobe sequences in the planar Earth--Moon CR3BP. (Continued)} \label{fig:lobe_sequence_2}
\end{figure*}%

%% file: 4_trajectory_design.tex
\section{Trajectory design based on lobe dynamics}\label{sec:trajectory_design}
This section describes the proposed method for designing low-energy chaotic transfers based on lobe dynamics.
Effective lobe sequences are utilized to predict the motion of chaotic orbits within lobes.
The selected chaotic orbits are connected by impulsive maneuvers to construct a low-energy transfer.
These maneuvers help to overcome the partial barriers~\cite{meiss2015thirty} formed by cantori~\cite{mackay1984transport} or the boundaries of resonance regions in the periapsis Poincar{\'e} map of the planar CR3BP.
In the following sections except Section~\ref{sec:impact_rL}, the threshold radius for effective lobes is fixed at $r_L^\ast = 0.002$ as a representative example.

Indeed, natural trajectories passing through different lobe sequences exist due to the chaotic dynamics, as demonstrated by Odashima \textit{et al}.~\cite{odashima2016design} using a gridded search within a lobe associated with the $3$:$1$ resonant orbit.
The proposed method, however, connects different lobe sequences with small maneuvers in order to simplify the low-energy transfer design process.
This simplification enables choosing more resonant orbits for incorporating lobe sequences.

The main idea of the proposed method is to utilize a weighted, directed graph to search for the fuel-optimal transfer path.
After discretizing the solution space, such a graph-based framework is suitable for representing potential impulsive transfer paths.
The graph structure aids in solving combinatorial optimization over a set of natural arcs, such as periodic orbits and their manifolds~\cite{tsirogiannis2012graph,trumbauer2014heuristic}, or a set of motion primitives~\cite{smith2023motion}.
Similarly, Oshima~\cite{oshima2025graph} represents the solution space using a set of periapsis-to-periapsis arcs for low-energy transfers.
This study constructs periapsis-to-periapsis arcs in a simpler manner and focuses on the optimal combination of effective lobe sequences.

\subsection{Overview of the design method}
\begin{figure}[!t]
    \centering
    \includegraphics[width=0.25\linewidth]{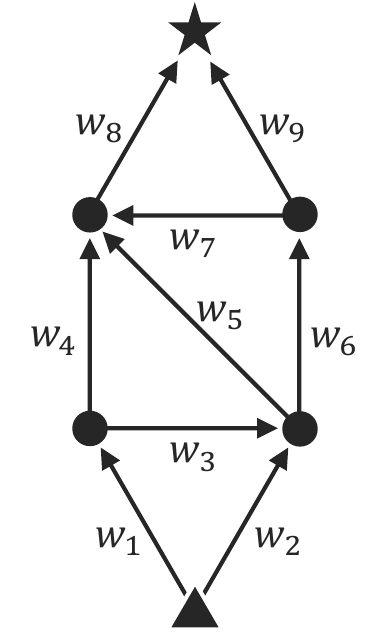}
    \caption{A schematic diagram of the weighted, directed graph for impulsive transfer design.}
    \label{fig:graph_ex}
\end{figure}

Transfer design with effective lobe sequences requires determining the combination of the lobe sequences and the timing for a spacecraft to enter and leave them to minimize the transfer cost.
To this end, the proposed method leverages a weighted, directed graph with three components: nodes (discrete state points), directed edges (transfer paths), and weights (transfer costs).
In the graph, nodes stand for potential start and goal points as well as for the centroids of the effective lobes in Fig.~\ref{fig:lobe_sequence_1}.
The edges connecting these nodes indicate allowable transfer paths, which are predesigned based on their dynamical geometry.
The weight of each edge is assigned as the corresponding transfer cost, such as $\Delta V$ or the transfer time.
Figure~\ref{fig:graph_ex} illustrates a schematic diagram of the graph for the proposed method.
The triangle, dots, and star indicate the start node, the nodes corresponding to effective lobes, and the goal node, respectively.
The directed arrows show predesigned edges, and the edge weights are denoted by $w_k$ ($k = 1,\,2,\,\cdots,\,9$).
The detailed procedure for designing transfer trajectories corresponding to the edges is examined using a test problem in the following subsections.

The optimal transfer is determined by exhaustive search, which computes the total weight for all possible transfers from the start nodes to the goal nodes.
The combinatorial optimization problem can be formulated as follows:
\begin{align}
    &\mathrm{minimize}      & &J = \sum_{k} w_k\label{eq:opt_J}\\
    &\mathrm{subject\,\,to} & &w_k < w^\ast\label{eq:opt_const_w},\\
    &                       & &\mathrm{``Use\;lobe\;sequences\;properly,''} \label{eq:opt_const_lobe}
\end{align}
\begin{figure}[!b]
    \centering
    \includegraphics[width=0.4\columnwidth]{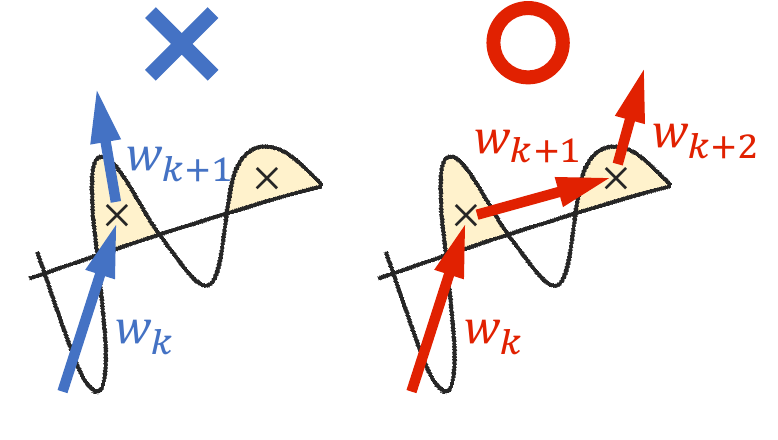}
    \caption{Illustration of the constraint Eq.~\eqref{eq:opt_const_lobe}.}
    \label{fig:opt_const_lobe}
\end{figure}
where $J$ is the sum of $w_k$ along a transfer path from one of the start nodes to one of the goal nodes, and $w^\ast$ is a threshold weight.
The number of the combinations is reduced by considering only edges whose weight satisfies Eq.~\eqref{eq:opt_const_w}.
The second constraint of Eq.~\eqref{eq:opt_const_lobe} implies that if a spacecraft goes through one of the lobe sequences, at least two adjacent lobes in the lobe sequence must be included in the transfer path (see Fig.~\ref{fig:opt_const_lobe}).
This constraint requires the full exploitation of lobe dynamics during transfers, which demonstrates the effectiveness of the approach.
Note that this condition can be neglected for mission design.

\subsection{Test problem settings}\label{sec:test_problem}
\begin{figure}[!b]
    \centering
    \includegraphics[width=0.5\columnwidth]{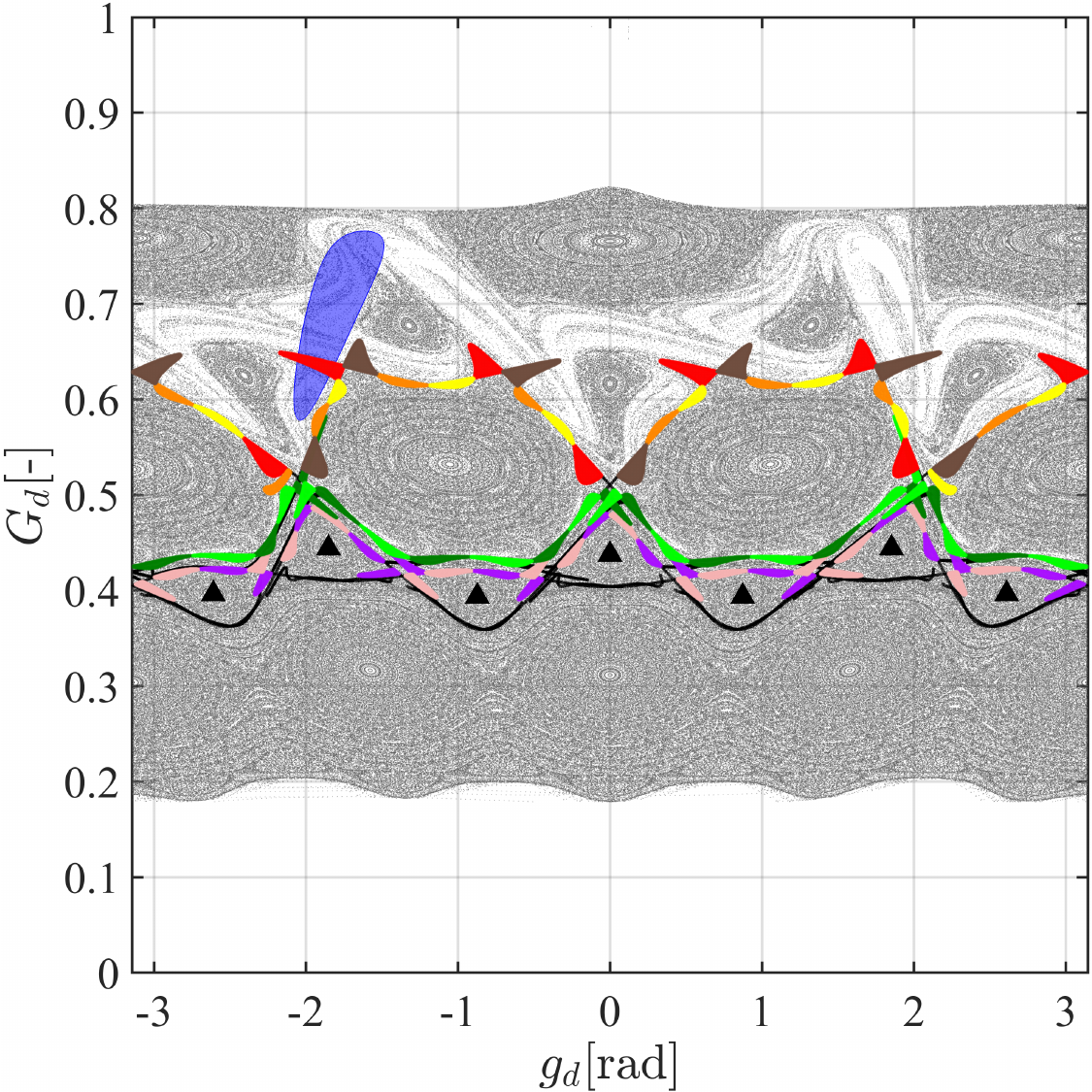}
    \caption{Problem settings for the test problem.} \label{fig:poincare_map_test}
\end{figure}
\begin{figure}[!t]
    \centering
    \includegraphics[width=0.45\linewidth]{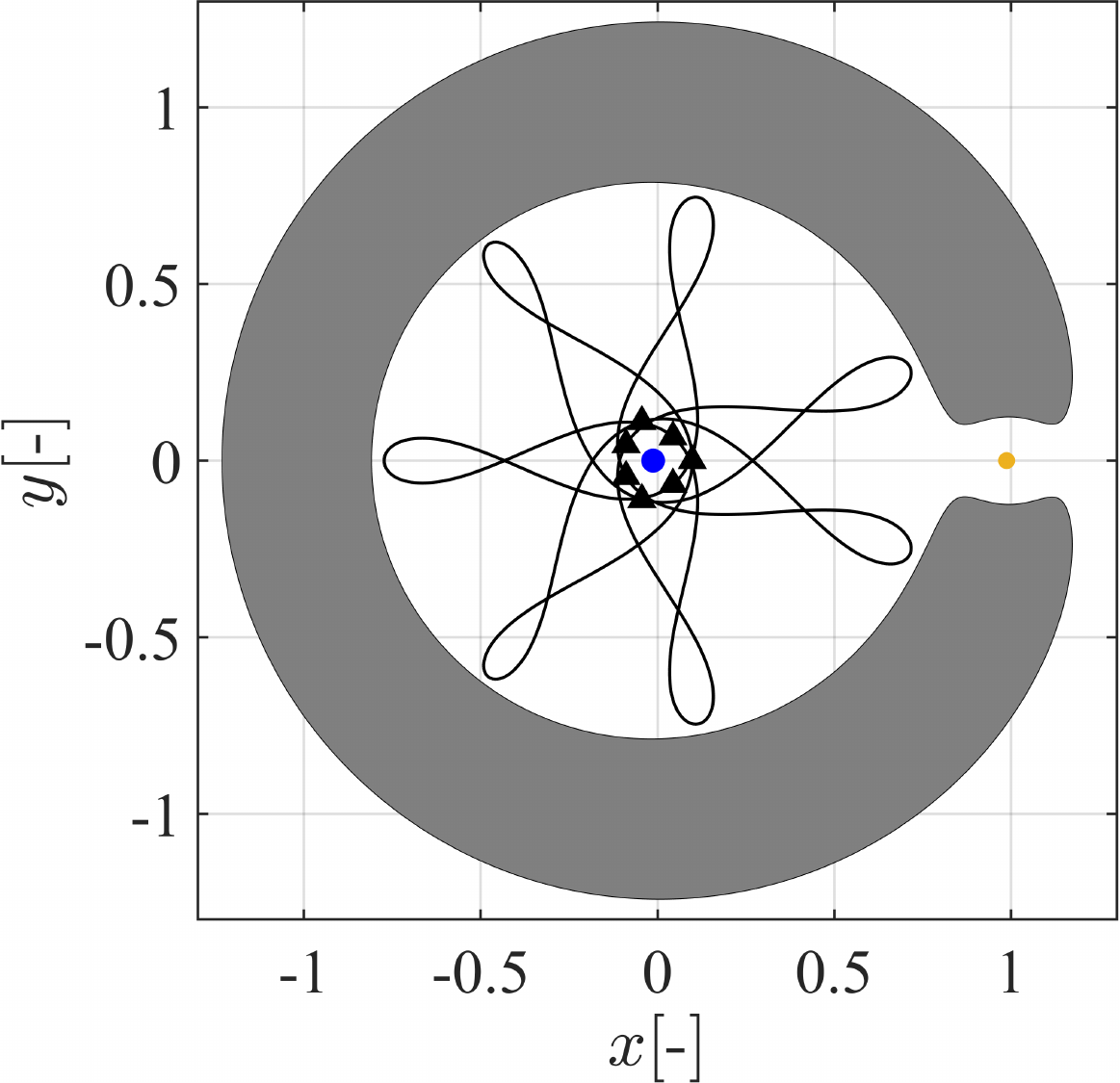}
    \caption{The $7$:$2$ stable resonant orbit when $C_J = 3.16$.}
    \label{fig:resonant_orbit_72}
\end{figure}

\begin{figure}[!t]
    \centering
    \includegraphics[width=0.45\linewidth]{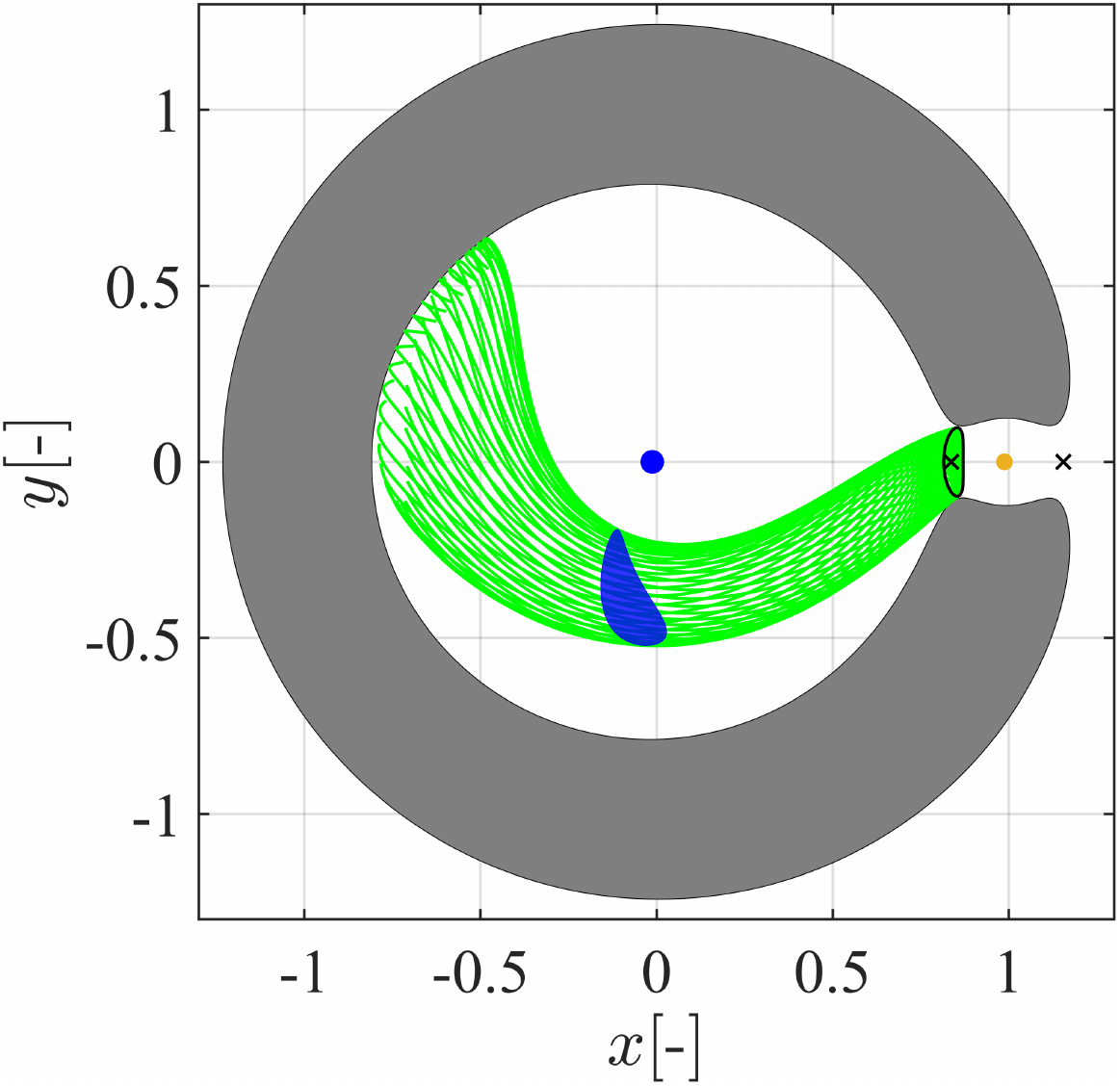}
    \caption{The left half of the stable manifold of the $L_1$ Lyapunov orbit when $C_J = 3.16$.}
    \label{fig:L1_stable}
\end{figure}

%
A test problem for the fuel-optimal transfer is considered to investigate the design of each transfer arc and its corresponding weight.
In this problem, the departure orbit and destination are defined within the periapsis Poincar{\'e} map, as summarized in Fig.~\ref{fig:poincare_map_test}.
The departure orbit is the $7$:$2$ stable resonant orbit in Fig.~\ref{fig:resonant_orbit_72}, and the triangles in Figs.~\ref{fig:poincare_map_test} and~\ref{fig:resonant_orbit_72} indicate its periapses.
It is noteworthy that the altitudes of these periapses to the Earth are from $27,279$~km to $37,746$~km, implying that the starting points exist roughly around the geosynchronous Earth orbit (GEO)~\cite{vallado2001fundamentals} since the altitude of the GEO is about $35,780$~km.
The destination is defined as the left half of the stable manifold of the $L_1$ Lyapunov orbit, as shown in Fig.~\ref{fig:L1_stable}.
The black line is the $L_1$ Lyapunov orbit, and the green lines represent its stable manifold.
The blue regions in Figs.~\ref{fig:poincare_map_test} and~\ref{fig:L1_stable} illustrate the first intersection to the periapsis Poincar{\'e} map.
These blue regions can be regarded as the exit to the moon realm, as a spacecraft within them is delivered to the Moon or possibly to deep space via tube dynamics.
The other colored regions in Fig.~\ref{fig:poincare_map_test} represent the selected effective lobe sequences corresponding to those in Fig.~\ref{fig:lobe_sequence_1}.
Thus, this test problem aims to design a low-energy, fuel-optimal transfer for escaping from the Earth.

\begin{figure}[!t]
    \centering
    \includegraphics[width=0.5\columnwidth]{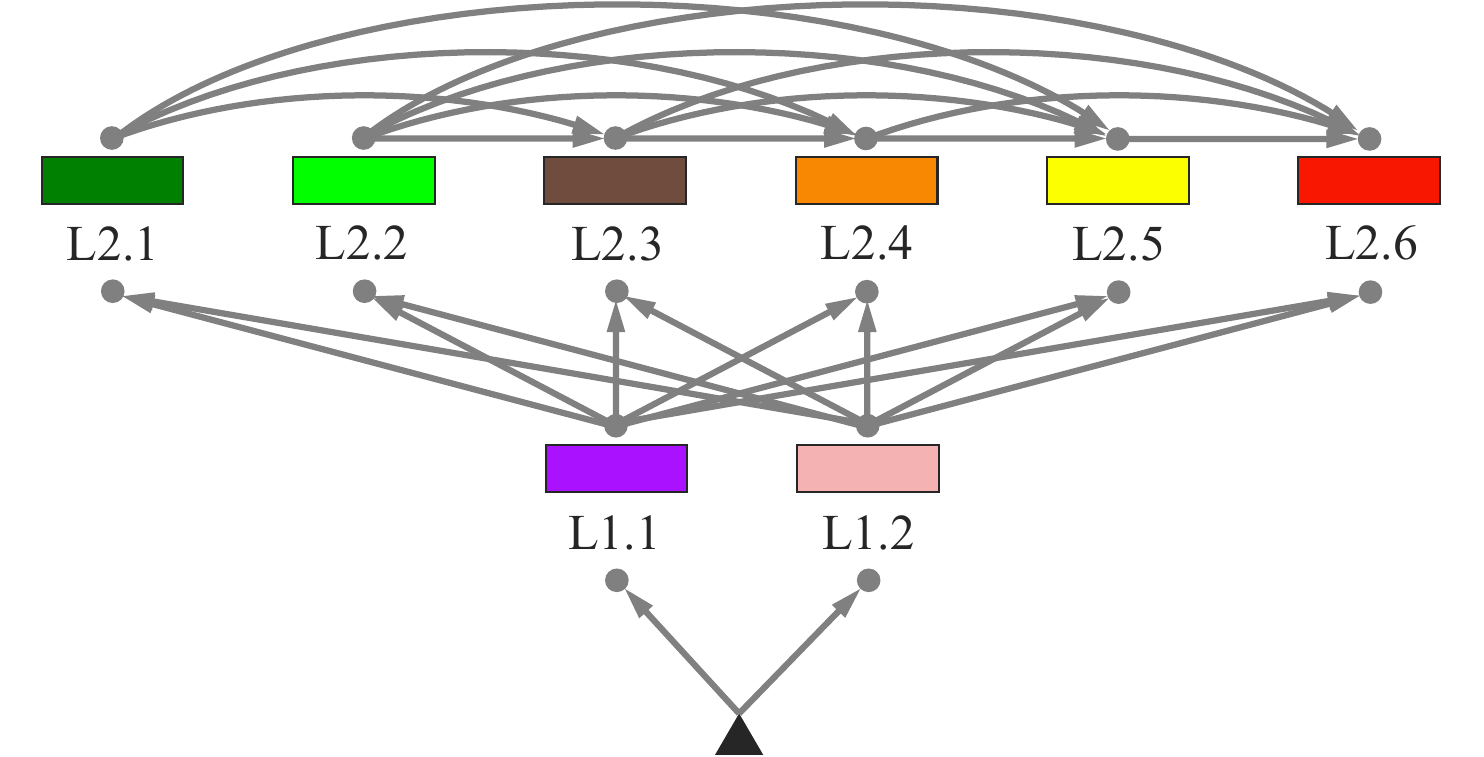}
    \caption{Design of transfer paths for the test problem.}
    \label{fig:order_of_transfer}
\end{figure}

In a graph for this test problem, nodes stand for the periapses of the departure orbit and the centroids of the selected effective lobes.
The goal node is defined as the centroid of the eighth effective lobe in L2.6 in Fig.~\ref{fig:lobe_2_6} because this lobe overlaps the blue region of the destination in Fig.~\ref{fig:poincare_map_test}.
Edges connecting these nodes are constructed as shown in Fig.~\ref{fig:order_of_transfer}.
These edges are designed so that a spacecraft can monotonically increase $G_d$.
The name and color in Fig.~\ref{fig:order_of_transfer} correspond to those of the lobe sequences in Fig.~\ref{fig:lobe_sequence_1}, and the triangle shows the start nodes.
The arrows indicate possible transfer paths.
For example, the arrow from L1.1 to L2.2 means that the graph includes transfer paths from each effective lobe in L1.1 to each effective lobe in L2.2.

\begin{figure}[!b]
    \centering
    \includegraphics[width=0.4\columnwidth]{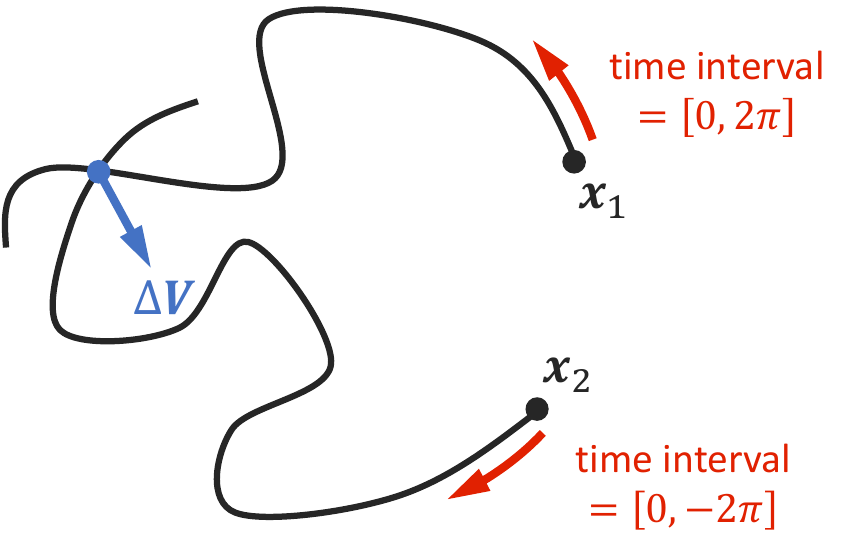}
    \caption{Illustration of the way of building an impulsive transfer arc.}
    \label{fig:impulsive_transfer}
\end{figure}%
In the proposed method, it is essential to define the way of designing transfer arcs corresponding to the edges and evaluating their weights.
For the edges from the start nodes to the lobes or from one lobe sequence to another, i.e., the edges for controlled transfers, transfer arcs are simply designed by so-called targeting~\cite{shinbrot1990using}, as shown in Fig.~\ref{fig:impulsive_transfer}.
When making a transfer from $\bm{x}_1$ to $\bm{x}_2$, for example, a crossing point in the position space is detected by propagating $\bm{x}_1$ forward over time interval $[0,\,\,2\pi]$ and $\bm{x}_2$ backward over time interval $[0,\,\,-2\pi]$.
An impulsive $\Delta \bm{V}$ is applied at this crossing point to adjust the direction of the velocity and keep the Jacobi constant.
The weight for this transfer arc is defined as the magnitude of this $\Delta \bm{V}$.
For the edges between lobes in the same lobe sequence, i.e., the edges for transfers based on natural dynamics, two cases are considered in this test problem:
\begin{itemize}
    \item \textit{Case 1: Without targeting}. Transfer arcs are determined by propagating the initial point $\bm{x}_1$ until it reaches the lobe containing $\bm{x}_2$. The corresponding weight is zero, but the value of $\bm{x}_2$ is updated to the actual final point after the propagation.
    \item \textit{Case 2: With targeting}. Transfer arcs and their weights are determined by the targeting method shown in Fig.~\ref{fig:impulsive_transfer}. The weights are generally small because the transfer arcs are based on lobe dynamics.
\end{itemize}
In all the cases, the value of $w^\ast$ is set as $100$~m/s so that solutions containing infeasible paths are ignored during the optimization.

\subsection{Test-problem results}
\begin{table}[b]
    \centering
    \caption{Number of transfers found by different approaches from Fig.~\ref{fig:order_of_transfer}.}
    \begin{tabular}{lccc}
        \hline
         & Estimated & Without targeting & With targeting\\
        \hline
        Without Eq.~\eqref{eq:opt_const_lobe} & $544$ & $528$ & $544$ \\
        With Eq.~\eqref{eq:opt_const_lobe} & $388$ & $376$ & $388$\\
        \hline
    \end{tabular}
    \label{tab:test_num_traj}
\end{table}
\begin{figure}[!t]
    \centering
    \begin{minipage}{0.45\columnwidth}
        \centering
        \includegraphics[width=\columnwidth]{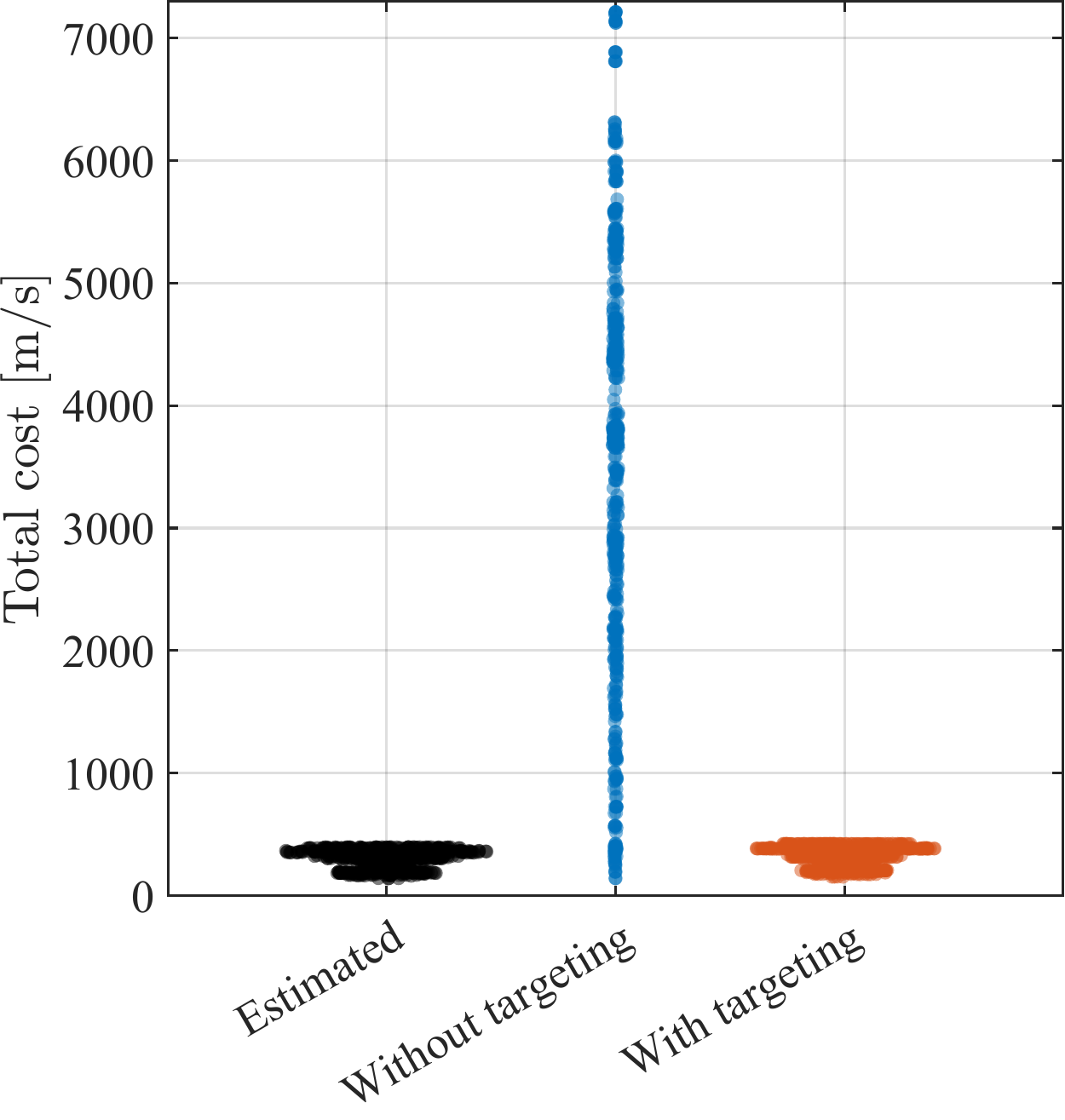}
        \subcaption{Overview: Without Eq.~\eqref{eq:opt_const_lobe}}\label{fig:cf_method}
    \end{minipage}
    \hspace{5mm}
    \begin{minipage}{0.45\columnwidth}
        \centering
        \includegraphics[width=\columnwidth]{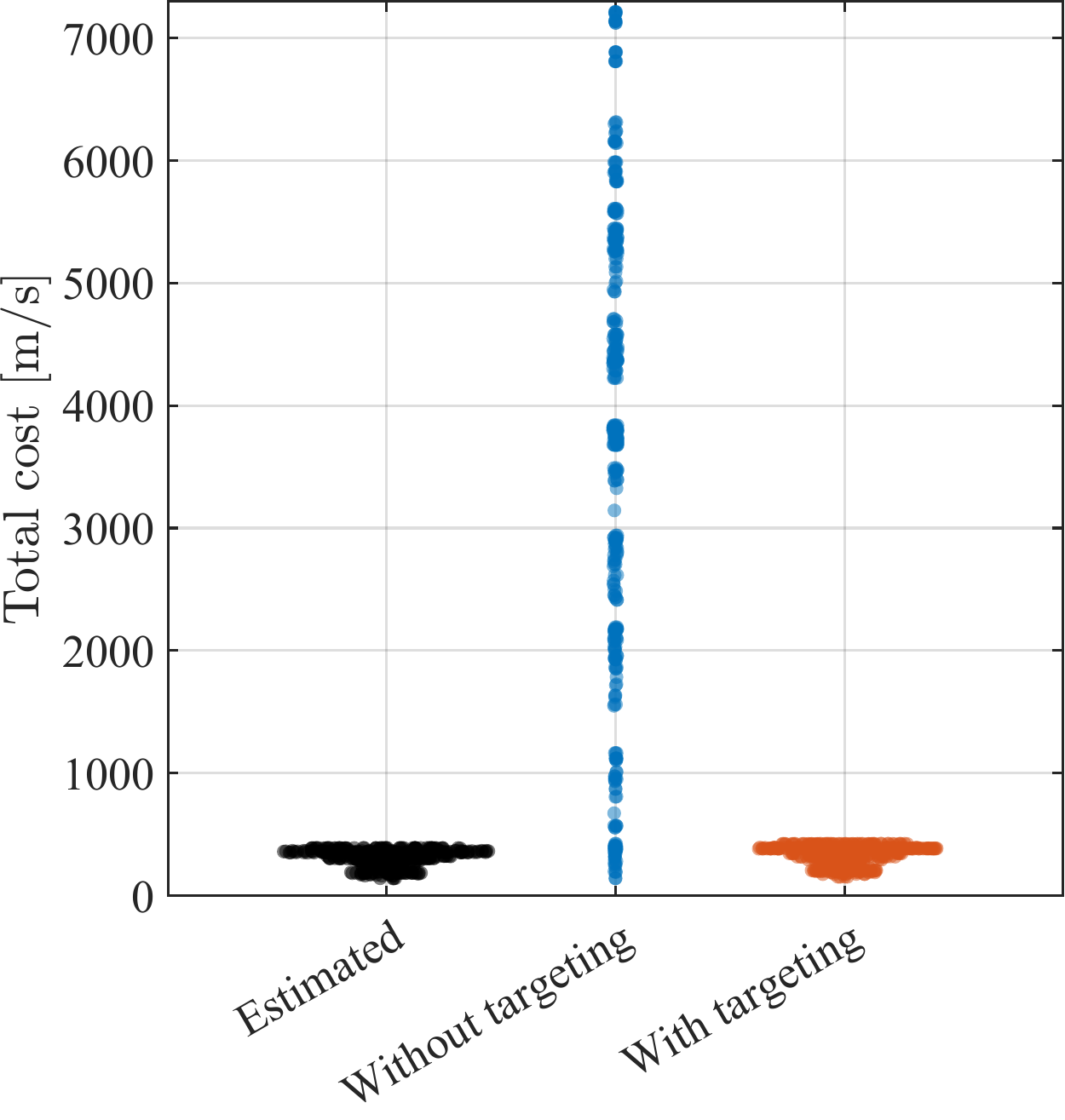}
        \subcaption{Overview: With Eq.~\eqref{eq:opt_const_lobe}}\label{fig:cf_method_lobe}
    \end{minipage}\\
    \vspace{5mm}
    \begin{minipage}{0.45\columnwidth}
        \centering
        \includegraphics[width=\columnwidth]{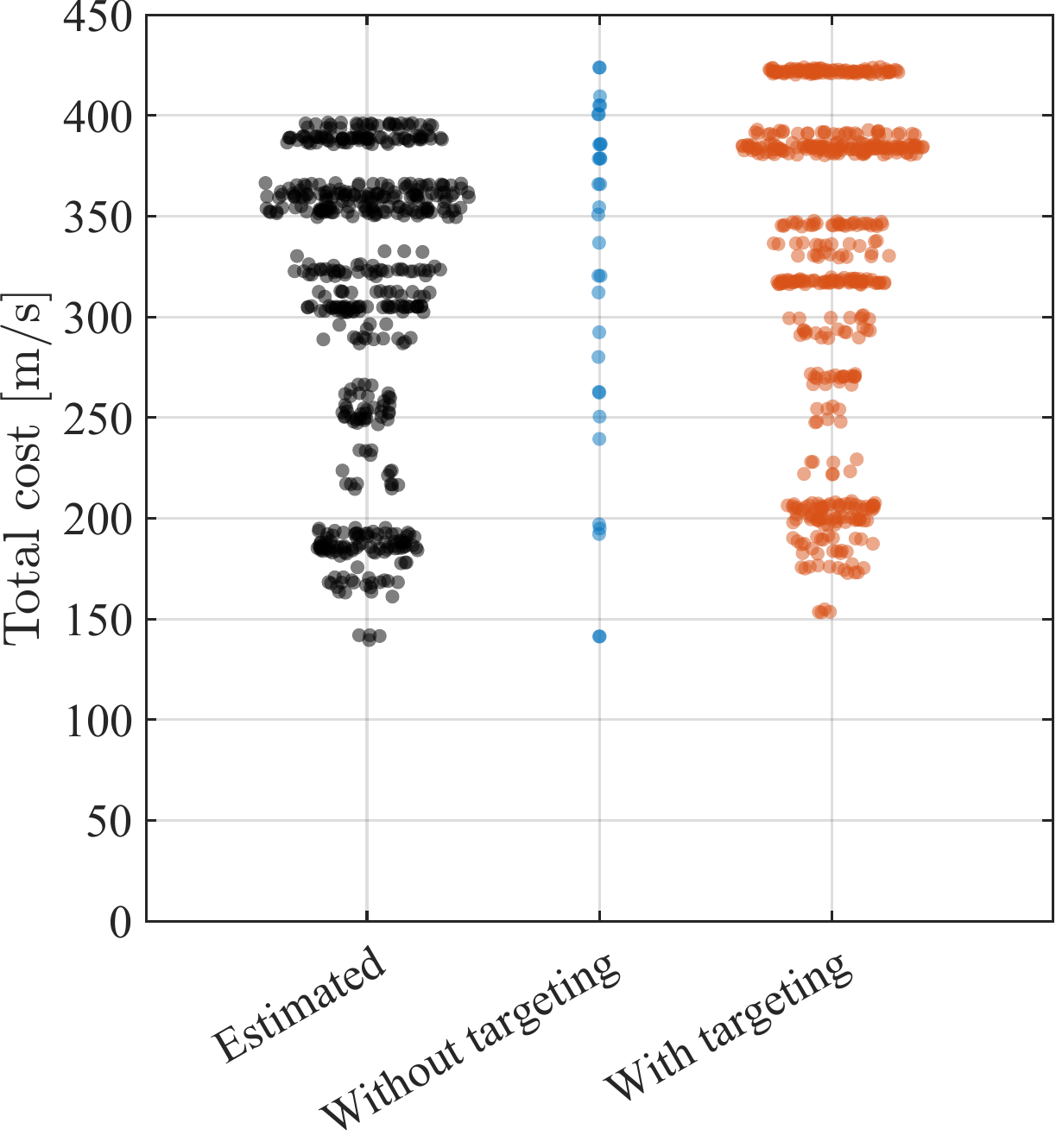}
        \subcaption{Detailed: Without Eq.~\eqref{eq:opt_const_lobe}}\label{fig:cf_method_zoom}
    \end{minipage}
    \hspace{5mm}
    \begin{minipage}{0.45\columnwidth}
        \centering
        \includegraphics[width=\columnwidth]{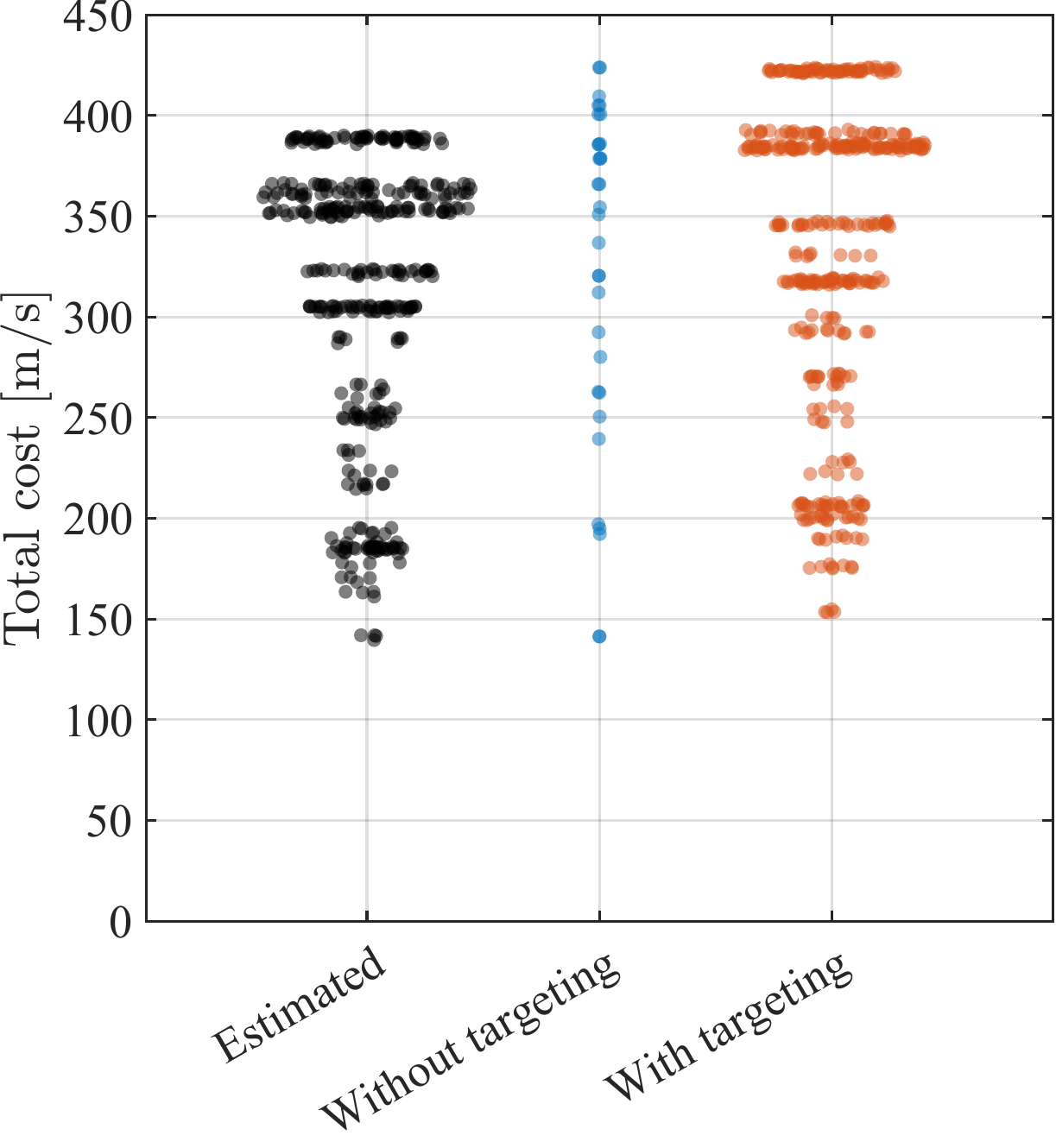}
        \subcaption{Detailed: With Eq.~\eqref{eq:opt_const_lobe}}\label{fig:cf_method_lobe_zoom}
    \end{minipage}%
    \caption{Swarm charts for the transfer costs of all potential paths with different approaches.}
    \label{fig:cf_method_all}
\end{figure}
Test results for all paths in both cases are summarized in Table~\ref{tab:test_num_traj} and Fig.~\ref{fig:cf_method_all}.
The effect of the constraint of Eq.~\eqref{eq:opt_const_lobe} is also examined.
The label ``Estimated'' indicates that the total costs are estimated by assuming that the edge weights for transfers based on lobe dynamics are precisely zero.
The label ``Without targeting'' denotes that the total costs are determined by the graph in Case 1.
The last label ``With targeting'' represents that the total costs are calculated from the graph in Case 2.
% ------
Table~\ref{tab:test_num_traj} indicates the number of transfers found by the three approaches.
In Case 1 (without targeting), some transfer paths do not yield the corresponding transfer trajectories because controlled transfer arcs fail to be reconstructed as a result of the propagation.
In contrast, in Case 2 (with targeting), all transfer paths generate transfer trajectories due to the targeting strategy.
The constraint of Eq.~\eqref{eq:opt_const_lobe} reduces the number of transfers found, regardless of the difference between the approaches.
% ------
Next, Fig.~\ref{fig:cf_method_all} shows swarm charts of the total transfer costs determined by the three approaches.
A swarm chart displays each data point individually, clearly illustrating the distribution of discrete data.
Figures~\ref{fig:cf_method_zoom} and~\ref{fig:cf_method_lobe_zoom} highlight low-cost transfers in Figs.~\ref{fig:cf_method} and~\ref{fig:cf_method_lobe}, respectively.
Figure~\ref{fig:cf_method_all} illustrates that transfer costs in Case 1 (without targeting) are significantly different from the estimated values.
In Case 2 (with targeting), transfer costs have a distribution similar to the estimated values.
Thus, the targeting strategy is practical for realizing transfer trajectories from the graph structure in the proposed method.
The comparison between Fig.~\ref{fig:cf_method} and Fig.~\ref{fig:cf_method_lobe}, and between Fig.~\ref{fig:cf_method_zoom} and Fig.~\ref{fig:cf_method_lobe_zoom} reveals that the number of transfers found is reduced by the constraint of Eq.~\eqref{eq:opt_const_lobe}, but their distribution does not change significantly.

\begin{figure}[!b]
    \centering
    \begin{minipage}{0.45\columnwidth}
        \centering
        \includegraphics[width=\columnwidth]{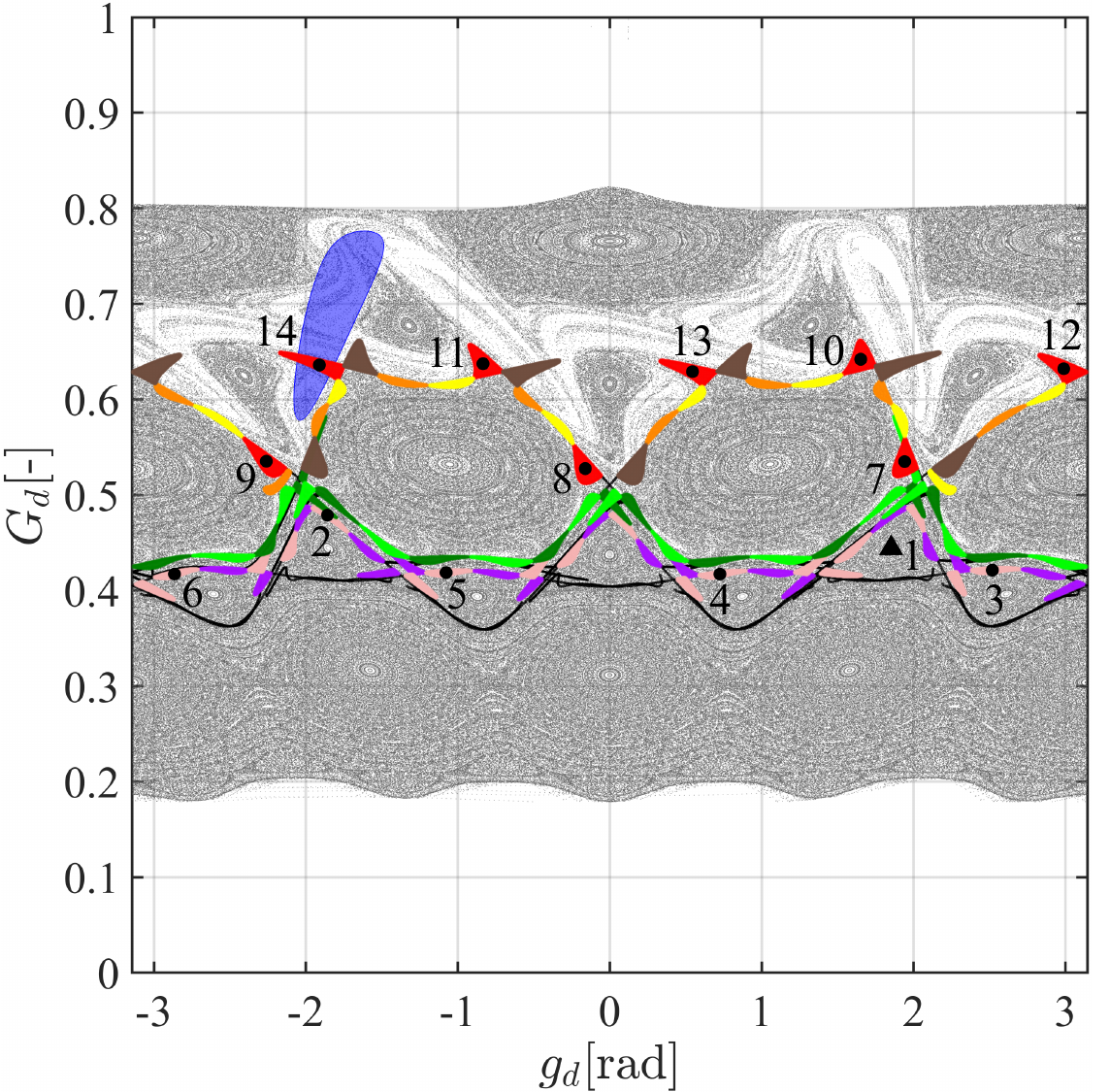}
        \subcaption{Without targeting (Estimated total $\Delta V = 139.5308$~m/s)}\label{fig:optimal_path_1}
    \end{minipage}%
    \hspace{5mm}
    \begin{minipage}{0.45\columnwidth}
        \centering
        \includegraphics[width=\columnwidth]{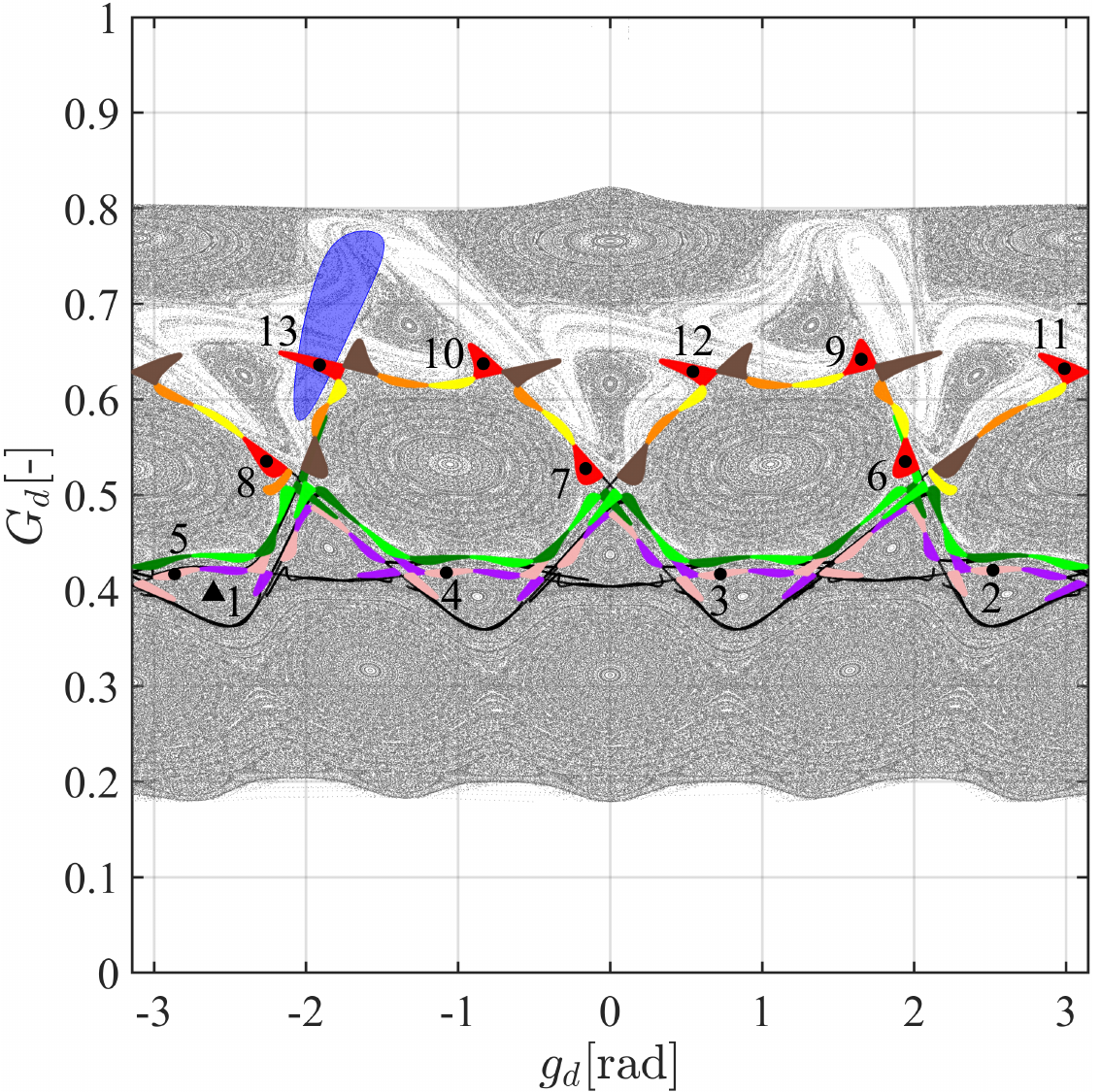}
        \subcaption{With targeting (Total $\Delta V = 153.2523$~m/s)}\label{fig:optimal_path_2}
    \end{minipage}%
    \caption{Optimal transfer paths in the periapsis Poincar{\'e} map for Cases 1 and 2.}
    \label{fig:optimal_path}
\end{figure}

\begin{figure}[!t]
    \centering
    \begin{minipage}{0.45\columnwidth}
        \centering
        \includegraphics[width=\columnwidth]{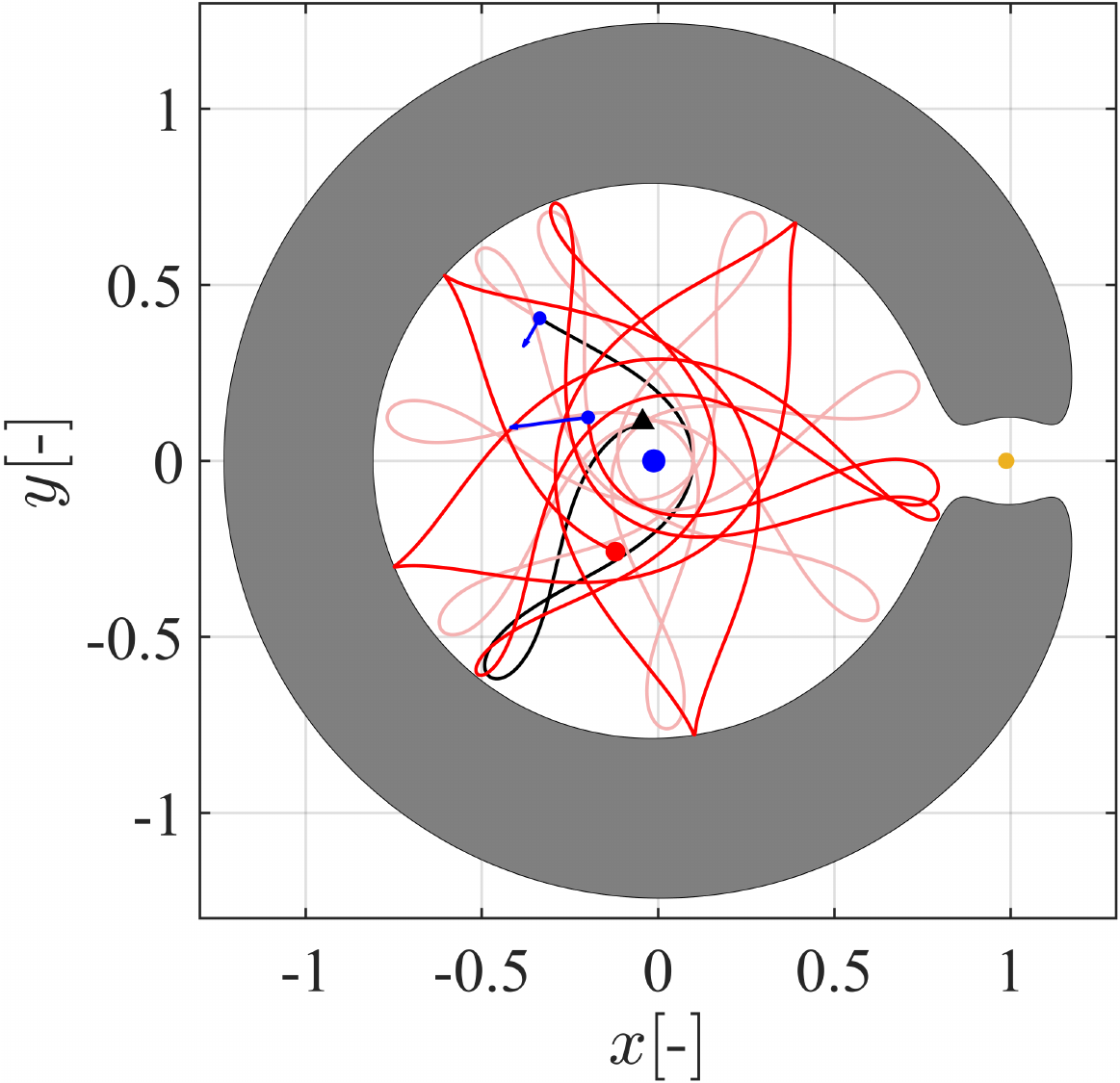}
        \subcaption{Without targeting (Total $\Delta V = 250.4744$~m/s)}\label{fig:optimal_traj_1}
    \end{minipage}%
    \hspace{5mm}
    \begin{minipage}{0.45\columnwidth}
        \centering
        \includegraphics[width=\columnwidth]{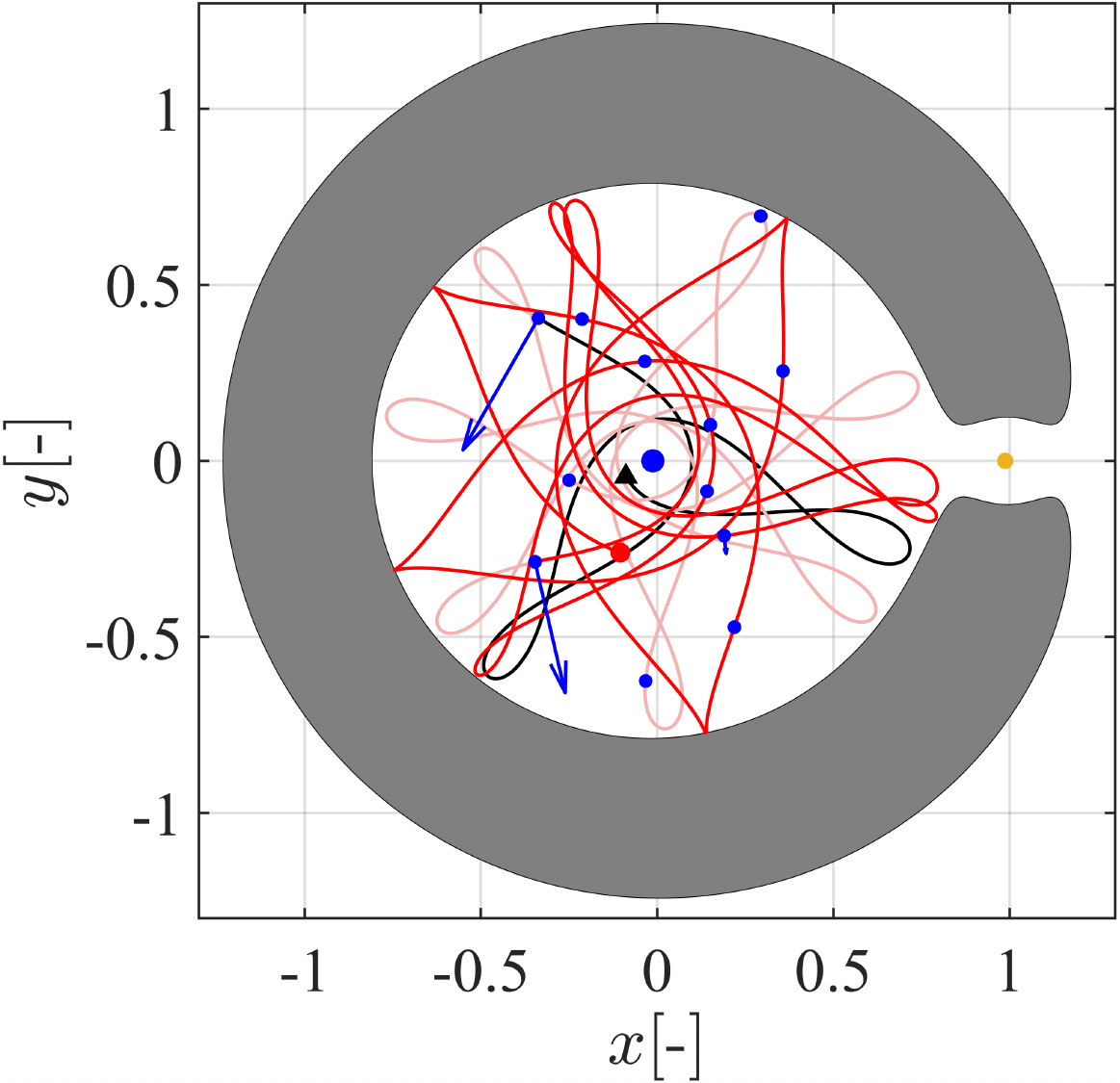}
        \subcaption{With targeting (Total $\Delta V = 153.2523$~m/s)}\label{fig:optimal_traj_2}
    \end{minipage}%
    \caption{Optimal transfer trajectories for Cases 1 and 2.}
    \label{fig:optimal_traj}
\end{figure}

For both Cases 1 and 2, the optimal transfer paths are shown in Fig.~\ref{fig:optimal_path}.
The black triangle and dots represent the start point and the centroids of the selected lobes, respectively.
The numbers indicate the order of transfer.
Figure~\ref{fig:optimal_traj} depicts the corresponding optimal transfer trajectories.
The black lines show the segments of the departure orbit, blue arrows indicate $\Delta \bm{V}$ vectors, and the other colored lines represent natural trajectories within effective lobes (their color corresponds to that of the lobe sequences in Fig.~\ref{fig:lobe_sequence_1}).
The size of the blue arrows is adjusted within the same figure and cannot be compared to that in a different figure.
In Case 1, the estimated total cost is small, but the total cost of the realized trajectory becomes larger in Fig.~\ref{fig:optimal_traj_1} due to propagation errors.
These propagation errors are caused by numerical errors in determining the boundaries and centroids of lobes.
In Case 2, the propagation errors are corrected by small maneuvers with $|\Delta \bm{V}| < 9$~m/s, and as a result, the desired transfer trajectory is constructed with a small total cost.

In summary, the obtained results suggest that the constraint of Eq.~\eqref{eq:opt_const_lobe} does not drastically change total transfer costs and that the approach of Case 2 is preferable for determining transfer arcs and the corresponding weights.

\subsection{Impact of the threshold radius $r_L^\ast$}\label{sec:impact_rL}
\begin{figure}[!tb]
    \centering
    \includegraphics[width=0.45\columnwidth]{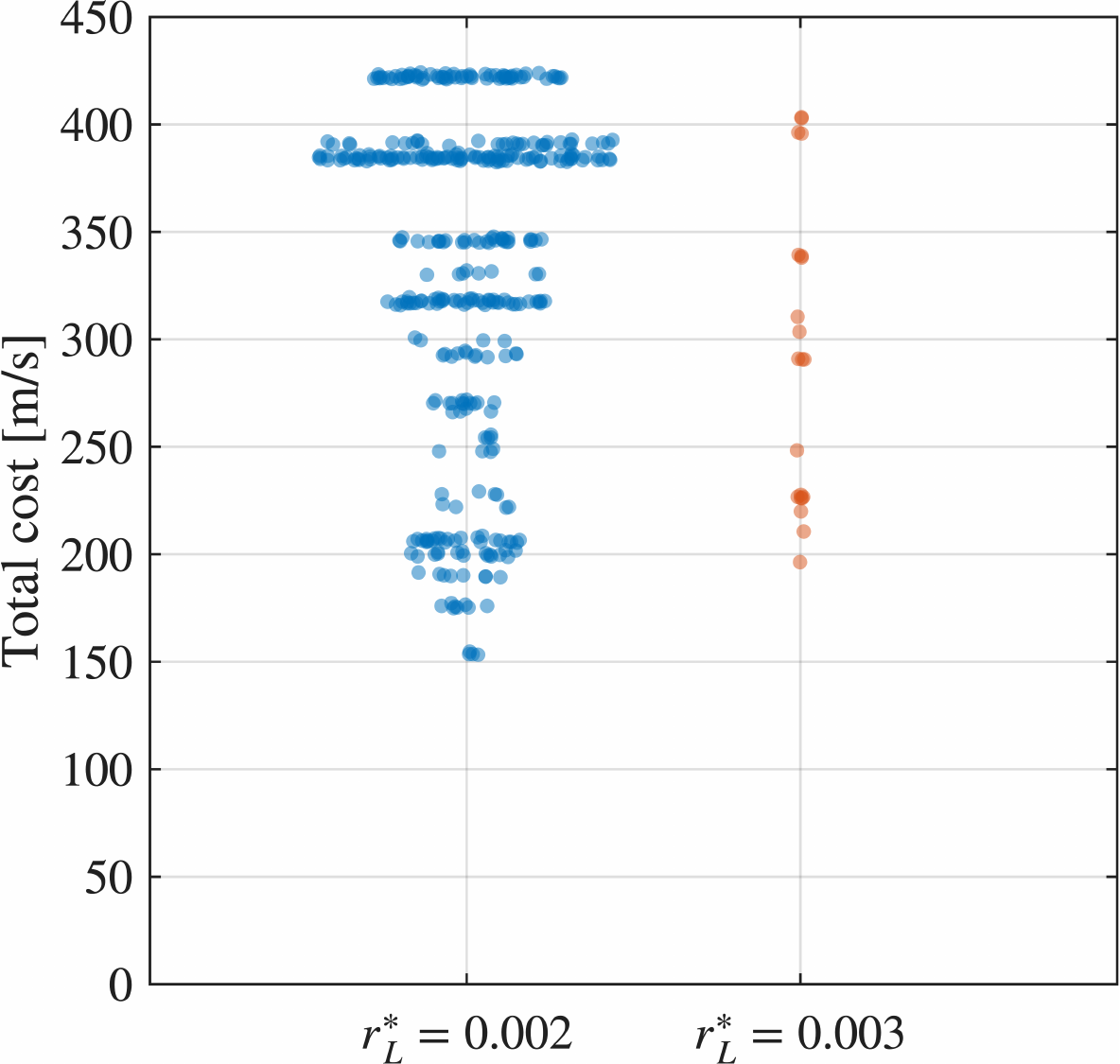}
    \caption{Swarm charts for the transfer costs of all potential paths for $r_L^\ast = 0.002$ and $0.003$.}
    \label{fig:cf_graph_TCM_lobe}
\end{figure}

\begin{figure}[!tb]
    \centering
    \includegraphics[width=0.45\columnwidth]{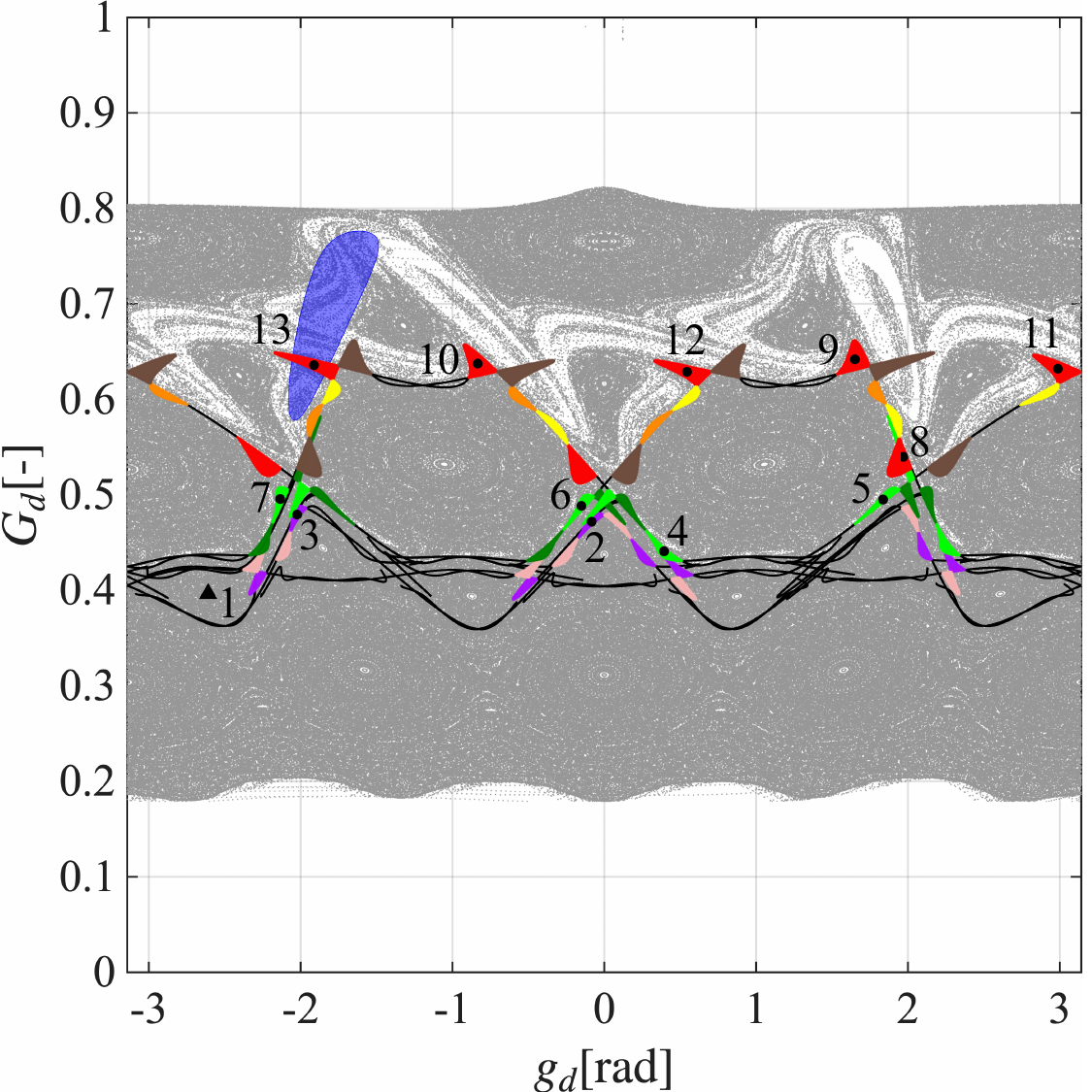}
    \caption{Optimal transfer path in the periapsis Poincar{\'e} map when $r_L^\ast = 0.003$ (Total $\Delta V = 196.2929$~m/s).}
    \label{fig:optimal_path_TCM_3}
\end{figure}

The impact of the threshold radius $r_L^\ast$ on the proposed method is examined based on the test-problem results.
The threshold radius $r_L^\ast$ is an important design parameter because it determines the set of effective lobes and affects potential transfer paths.
To this end, the test problem with the targeting strategy is solved for $r_L^\ast = 0.003$.
Note that the threshold weight $w^\ast$ is set to $110$~m/s, as no feasible transfer exists when $w^\ast = 100$~m/s.
Figure~\ref{fig:cf_graph_TCM_lobe} shows swarm charts of the total transfer costs for $r_L^\ast = 0.002$ and $0.003$.
As expected, a larger $r_L^\ast$ results in a smaller set of effective lobes in the graph, and fewer transfer trajectories are available.
Figure~\ref{fig:optimal_path_TCM_3} indicates the optimal transfer path when $r_L^\ast = 0.003$.
This figure demonstrates that fewer lobes are available, particularly around the $7$:$2$ unstable resonant orbit, resulting in an increase in the total $\Delta V$ by about $43$~m/s.
These results support the selection of $r_L^\ast = 0.002$ as a representative value for the following sections.

%% file: 5_application.tex
\section{Application to Earth--Moon transfer in the CR3BP}\label{sec:application}
The proposed method is applied to a more realistic mission scenario, i.e., Earth--Moon transfers in the planar CR3BP~\cite{hiraiwa2025extending}.
Specifically, the transfer from the low Earth orbit (LEO) to the low lunar orbit (LLO) is considered.

\subsection{Earth--Moon transfer design}
In this study, the LEO is defined by a circular orbit around the Earth with an altitude of $167$~km, and the LLO is a circular orbit around the Moon with an altitude of $100$~km.
These circular orbits are discretized equally into $12$ points.
The angles between the $x$ axis and each point at the LEO and LLO, denoted as $\theta_\mathrm{LEO}$ and $\theta_\mathrm{LLO}$, are given by $\theta_\mathrm{LEO},\,\theta_\mathrm{LLO} = k\pi/6$ ($k = 0, 1, \cdots, 11$).
As intermediate points of transfer, the centroids of the effective lobes in Fig.~\ref{fig:lobe_sequence_1} are used, similar to the test problem in Section~\ref{sec:test_problem}.
The $\Delta \bm{V}$s are performed to connect transfer arcs between the start points at the LEO, the effective lobes, and the goal points at the LLO.
The first transfer arc includes the $\Delta \bm{V}$ for departure from the LEO, $\Delta \bm{V}_\mathrm{E}$, which is applied tangentially to set the Jacobi constant to $C_J = 3.16$.
The last transfer arc contains the $\Delta \bm{V}$ for insertion into the LLO, $\Delta \bm{V}_\mathrm{M}$, which is applied tangentially to change $C_J$ to the value at the goal point.
In this problem, tube dynamics is not explicitly utilized, but effective lobes and the LLO are connected through tubes implicitly.

\begin{figure}[!b]
    \centering
    \includegraphics[width=0.5\columnwidth]{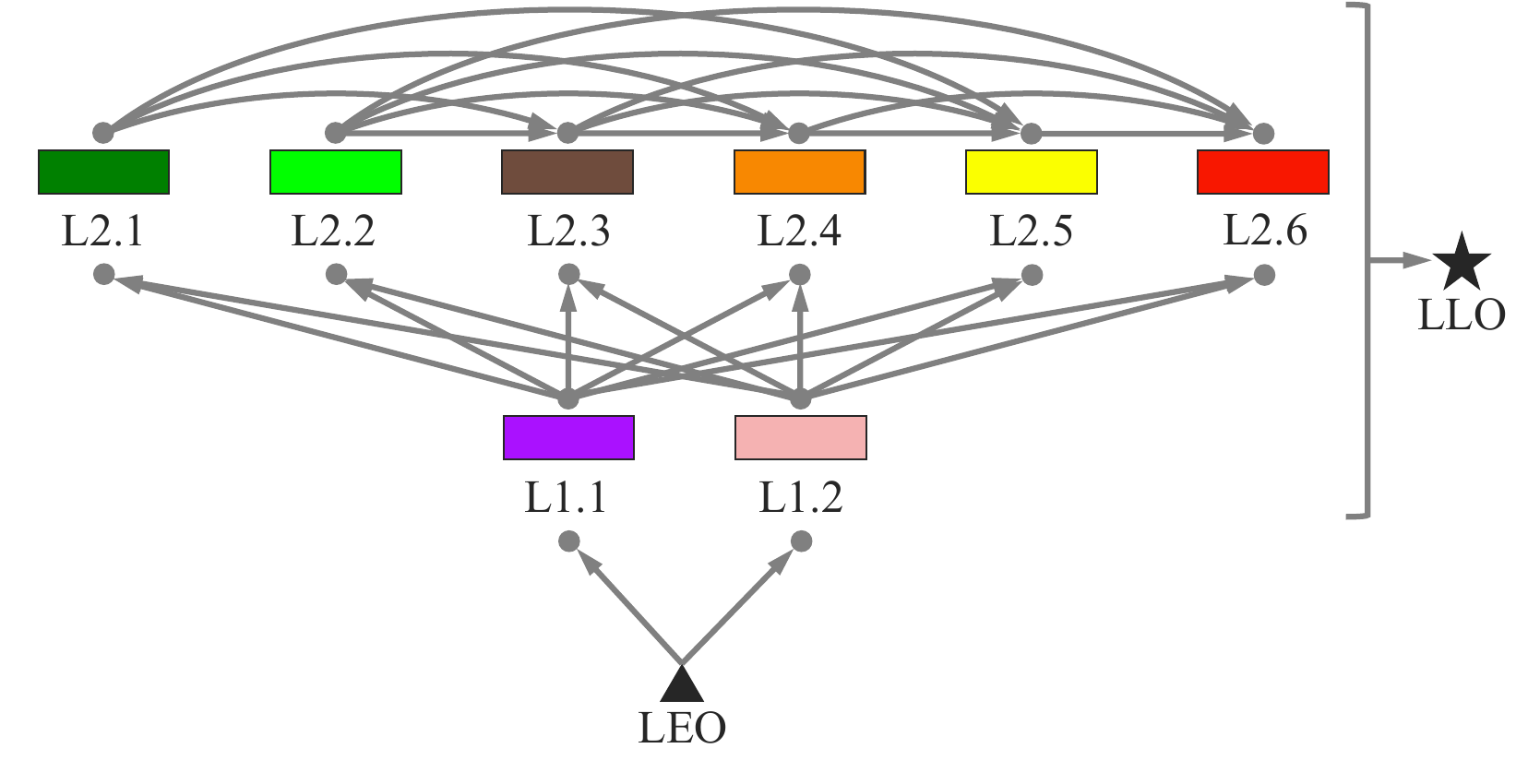}
    \caption{Design of transfer paths for the Earth--Moon transfer.}
    \label{fig:order_of_transfer_EM}
\end{figure}

To build a graph for this problem, the nodes corresponding to the start points, centroids of the effective lobes, and goal points are connected by edges, as shown in Fig.~\ref{fig:order_of_transfer_EM}.
The arrows indicate possible transfer paths, and a spacecraft can move from any selected effective lobe sequence to the LLO.
Transfer arcs and their weights are determined by the targeting strategy shown in Fig.~\ref{fig:impulsive_transfer}.
The $|\Delta \bm{V}_\mathrm{E}|$ and $|\Delta \bm{V}_\mathrm{M}|$ are added to the cost function $J$ in Eq.~\eqref{eq:opt_J} when the total transfer cost is evaluated.
The threshold weight $w^\ast$ is set to $400$~m/s based on the graph structure in this case.

\begin{figure}[!b]
    \centering
    \includegraphics[width=\columnwidth]{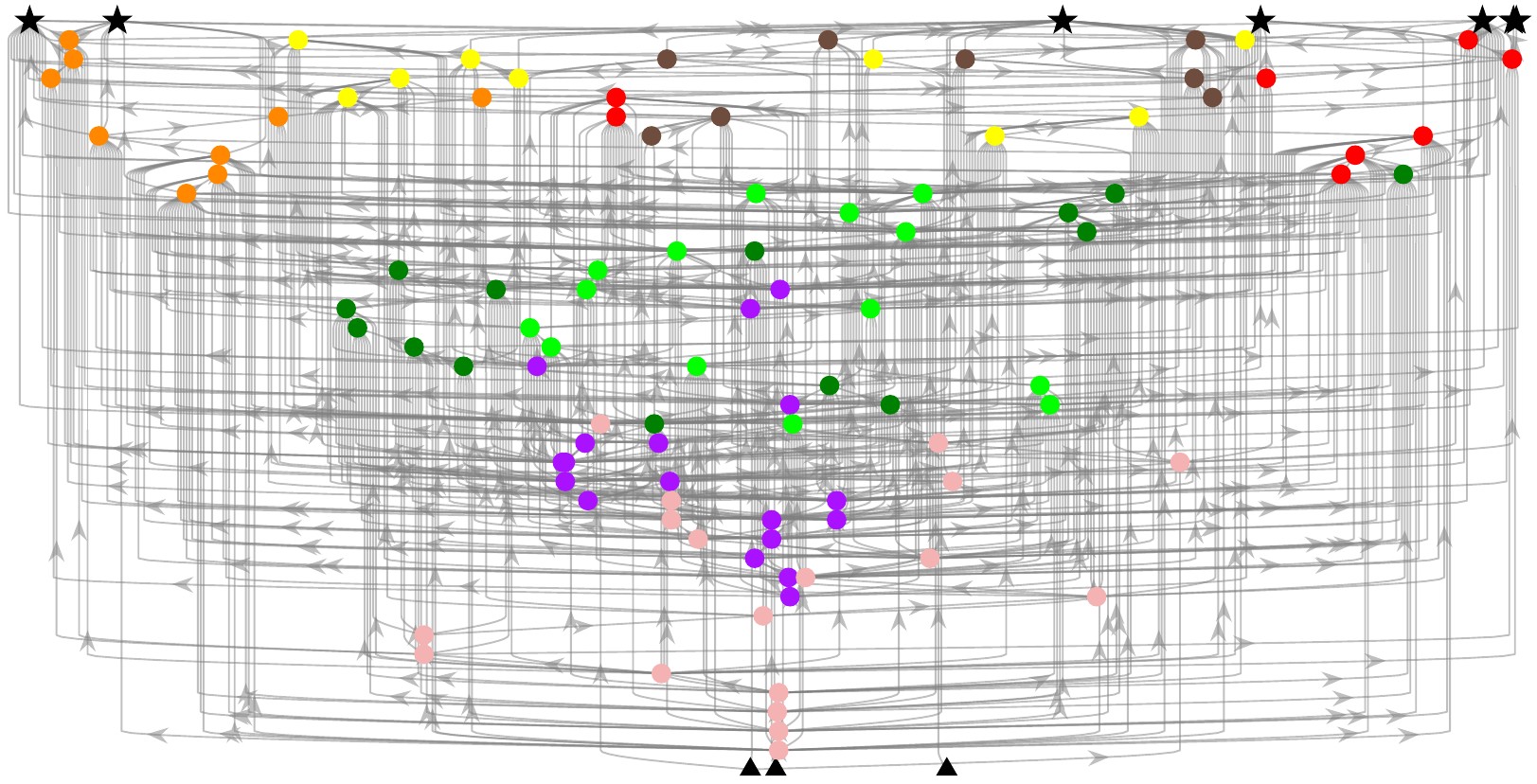}
    \caption{Graph representing possible transfer paths for the Earth--Moon transfer when $w^\ast = 400$~m/s.}
    \label{fig:graph_EM}
\end{figure}%
As a result, the graph for the optimization is obtained as shown in Fig.~\ref{fig:graph_EM}.
The triangles and stars represent the start and goal points, respectively.
The colored dots indicate the centroids of the selected effective lobes, and their color corresponds to that of the lobe sequences in Fig.~\ref{fig:lobe_sequence_1}.
The gray arrows represent possible transfer paths whose weight satisfies $w < w^\ast$.
This graph is utilized to solve the optimization problem by exhaustive search.

\subsection{CR3BP results}
\begin{figure}[!b]
    \centering
    \begin{minipage}{0.45\columnwidth}
        \centering
        \includegraphics[width=\columnwidth]{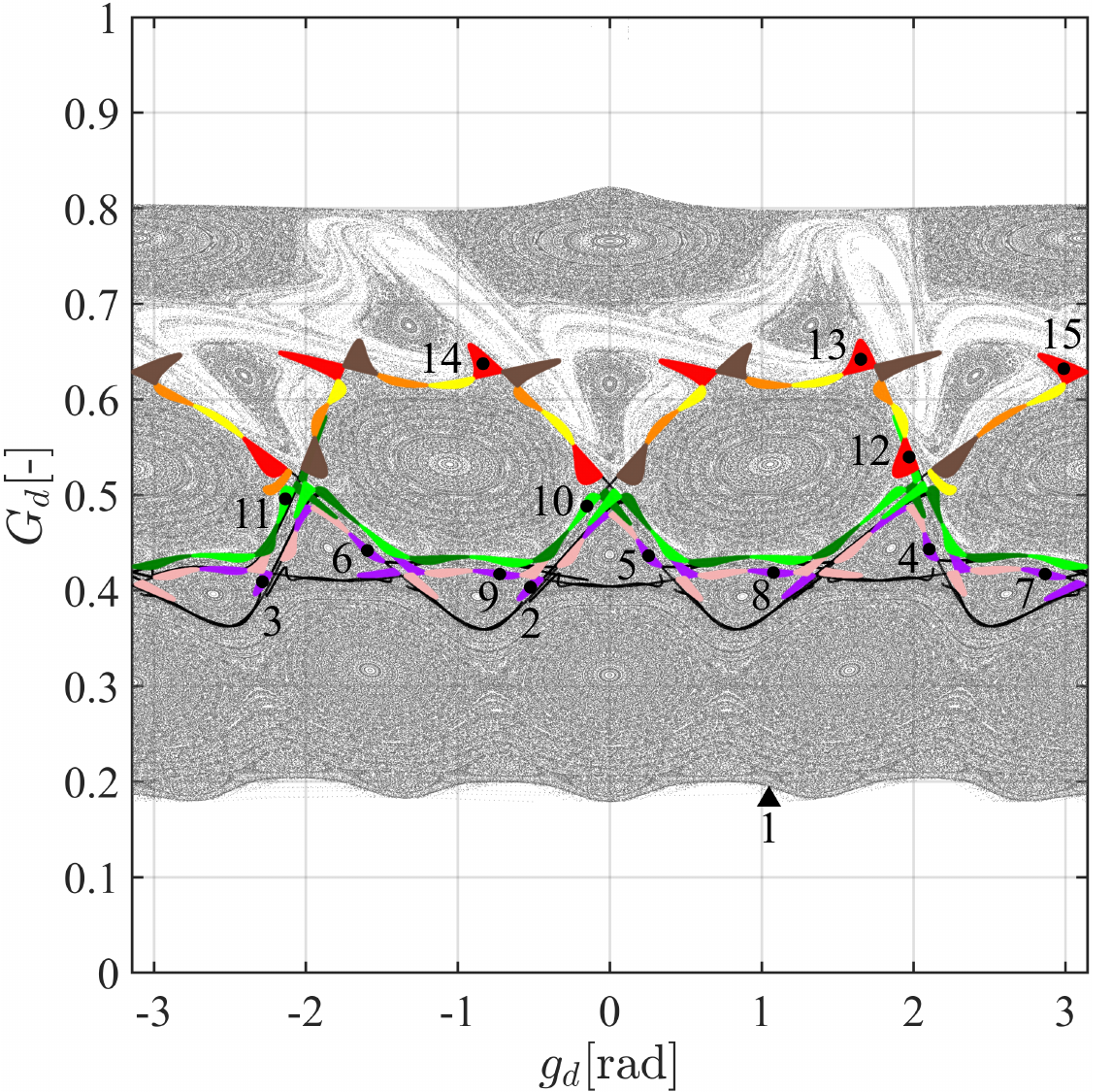}
        \subcaption{Optimal transfer path.}\label{fig:optimal_EM_path}
    \end{minipage}
    \hspace{5mm}
    \begin{minipage}{0.45\columnwidth}
        \centering
        \includegraphics[width=\columnwidth]{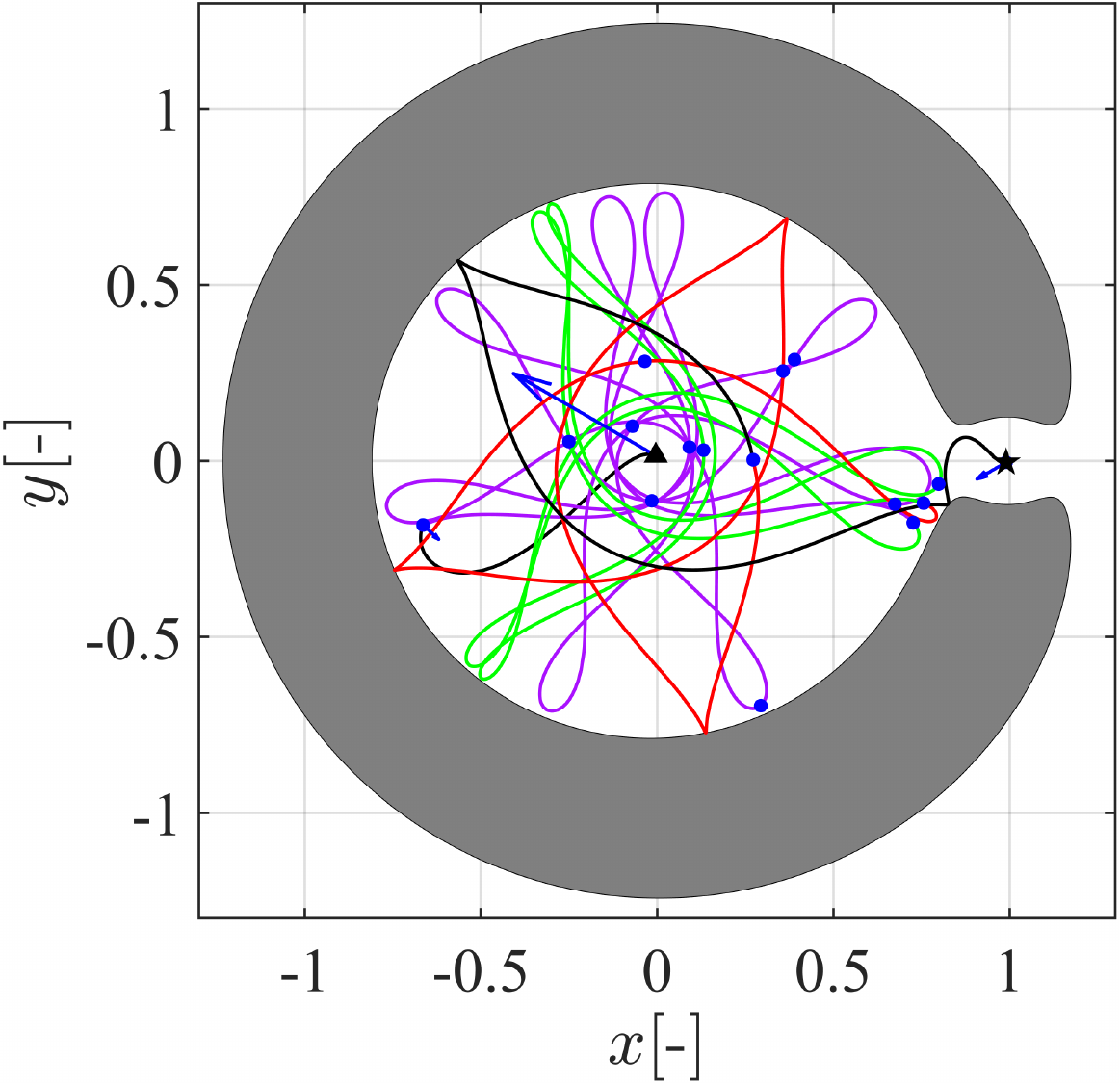}
        \subcaption{Optimal transfer trajectory.}\label{fig:optimal_EM_traj}
    \end{minipage}
    \caption{Optimal solution for the Earth--Moon transfer.}
    \label{fig:optimal_EM}
\end{figure}%
The proposed method yields the optimal Earth--Moon transfer with a total $\Delta V$ of $4274.6742$~m/s and transfer time of $191.8994$~days, as shown in Fig.~\ref{fig:optimal_EM}.
Figure~\ref{fig:optimal_EM_path}, similar to Fig.~\ref{fig:optimal_path}, shows the optimal transfer path in the periapsis Poincar{\'e} map.
Figure~\ref{fig:optimal_EM_traj}, similar to Fig.~\ref{fig:optimal_traj}, illustrates the corresponding optimal transfer trajectory.
The black trajectories are the transfer arcs departing from the LEO and arriving at the LLO.
As a result of the optimization, the start and goal points are set as $\theta_\mathrm{LEO} = \pi/3$ and $\theta_\mathrm{LLO} = 5\pi/3$, respectively.
The four largest $\Delta \bm{V}$s contribute to the majority of the total $\Delta V$.
Specifically, $|\Delta \bm{V}_\mathrm{E}| = 3102.2013$~m/s is applied to depart the LEO, $|\Delta \bm{V}| = 398.9009$~m/s is added to move into the purple lobe sequence, $|\Delta \bm{V}| = 102.6439$~m/s is added to escape from the red lobe sequence toward the LLO, and $|\Delta \bm{V}_\mathrm{M}| = 636.4582$~m/s is finally performed to arrive at the LLO.
The other small $\Delta \bm{V}$s satisfying $|\Delta \bm{V}| < 22$~m/s are used to adjust the transfer trajectory.
This result implies that further investigation may be necessary to improve the design of transfer arcs between lobe sequences and circular orbits, which is left for future work.

The obtained results demonstrate that the proposed method can systematically incorporate multiple lobe dynamics into trajectory design by constructing a graph that represents possible transfer paths.
This method combines chaotic orbits within effective lobes based on the targeting strategy.
The Earth-Moon transfers are then realized by connecting trajectories within lobes to lunar transfer arcs through tubes.

\subsection{Role of lobe dynamics with respect to WSB}
\begin{figure}[!t]
    \centering
    \begin{minipage}{0.45\columnwidth}
        \centering
        \includegraphics[width=\columnwidth]{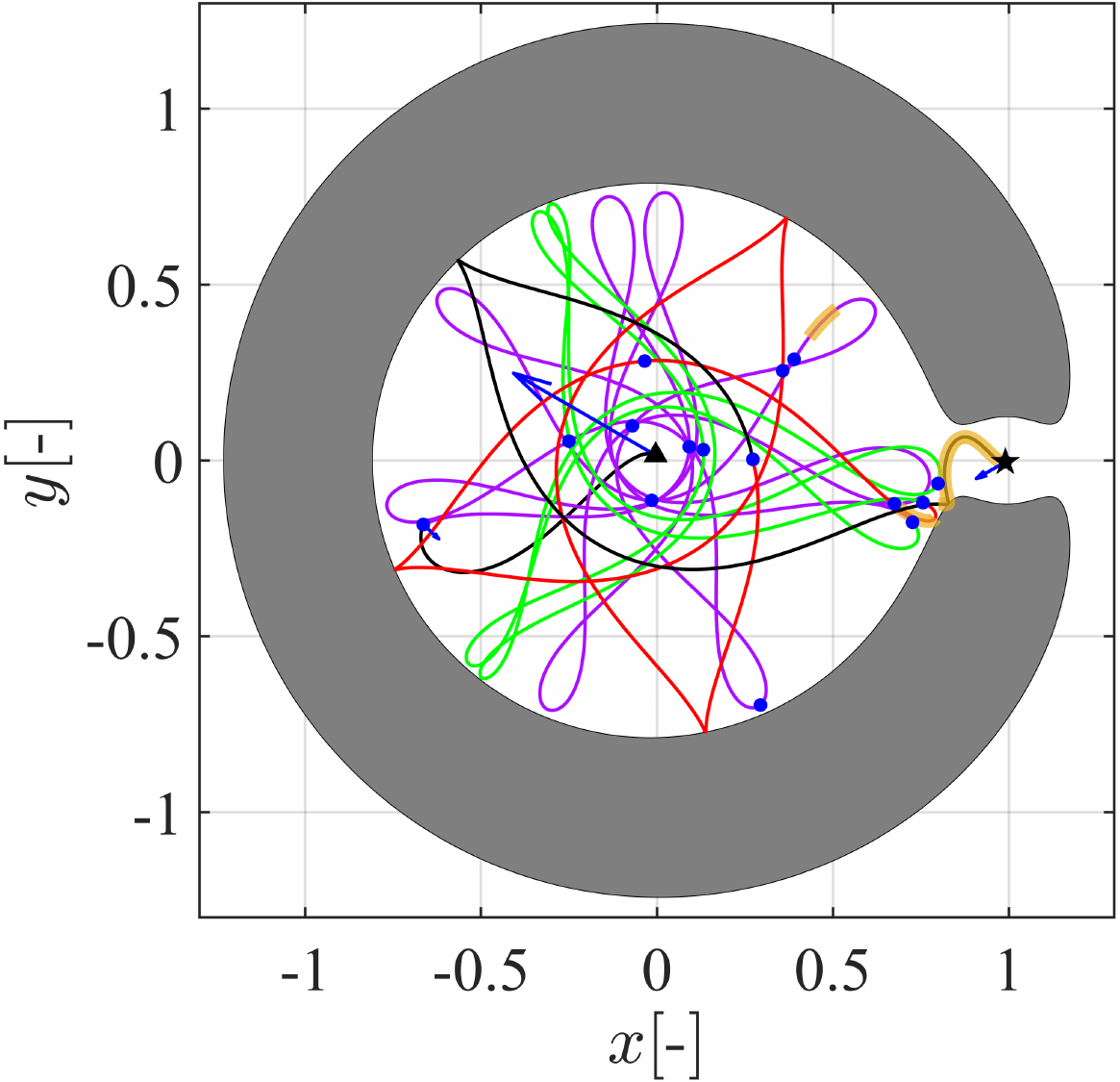}
        \subcaption{In the $x$-$y$ plane.\\ \textcolor{white}{blank}}\label{fig:optimal_EM_WSB_traj}
    \end{minipage}
    \hspace{5mm}
    \begin{minipage}{0.45\columnwidth}
        \centering
        \includegraphics[width=\columnwidth]{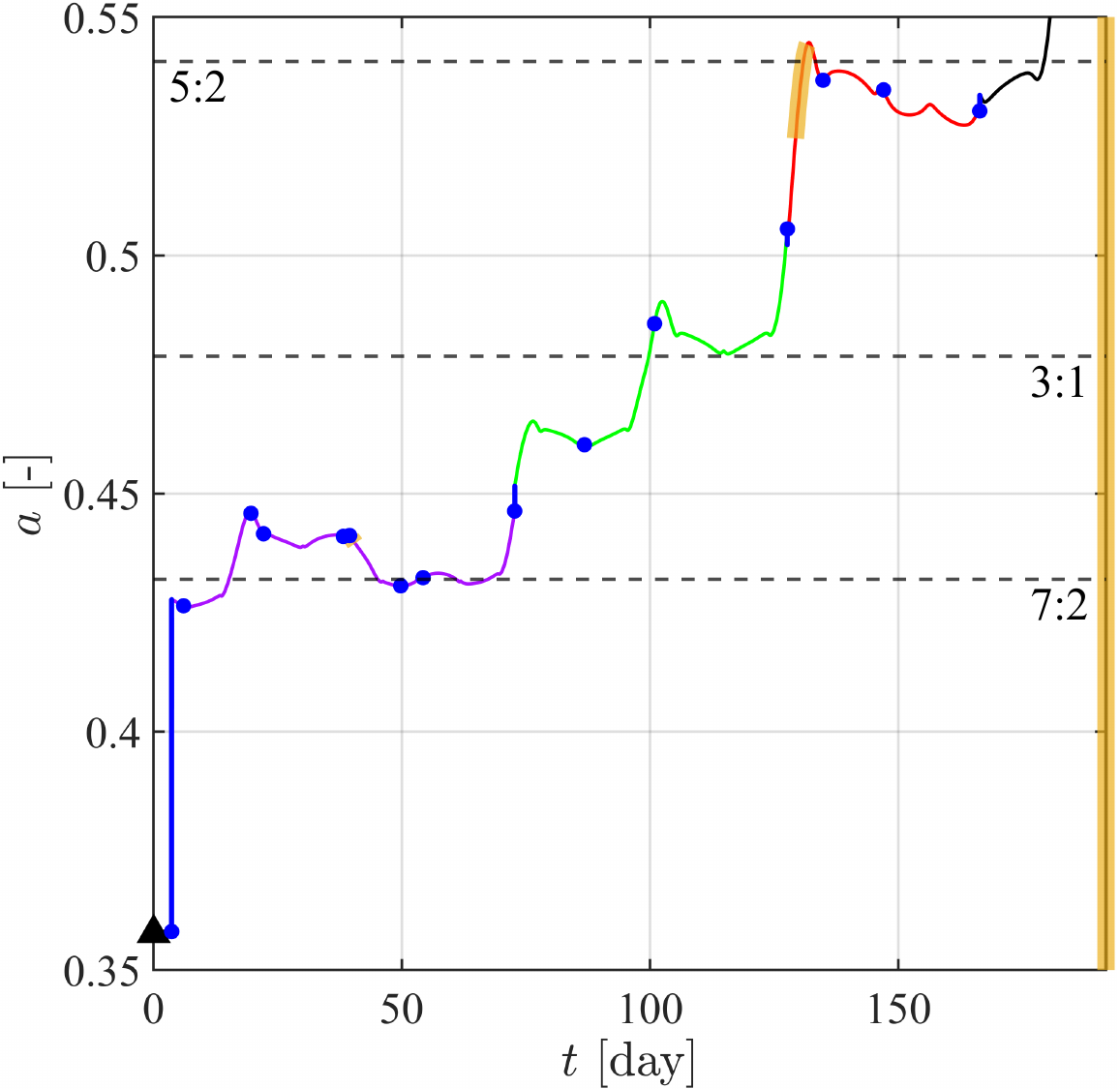}
        \subcaption{Time history of the semi-major axis.}\label{fig:optimal_EM_WSB_a}
    \end{minipage}
    \caption{Optimal trajectory in Fig.~\ref{fig:optimal_EM_traj} and its semi-major axis with extended WSB.}
    \label{fig:optimal_EM_WSB}
\end{figure}

This subsection discusses the difference between the concept of lobe dynamics and extended WSB~\cite{belbruno2008resonance}, both of which are related to resonance transitions.
The optimal trajectory in Fig.~\ref{fig:optimal_EM_traj} is a good example for investigating this difference.
Figure~\ref{fig:optimal_EM_WSB} shows extended WSBs along this trajectory by yellow thick lines.
Specifically, in Fig.~\ref{fig:optimal_EM_WSB_traj}, some transfer arcs within effective lobe sequences pass through extended WSBs, i.e., experience weak capture at the Moon.
The time history of the semi-major axis along the optimal trajectory is shown in Fig.~\ref{fig:optimal_EM_WSB_a}.
The dashed lines indicate the values from Eq.~\eqref{eq:resonant_a} for the corresponding resonance ratios.
The blue lines indicate impulsive $\Delta \bm{V}$s, and the other colored lines represent the transfer arcs within the lobe sequences of the corresponding colors.

Although the WSB is helpful for constructing ballistic lunar transfers, Fig.~\ref{fig:optimal_EM_WSB_a} demonstrates that weak capture does not always cause resonance transition.
Thus, traversing extended WSBs may not be a sufficient condition to perform resonance transitions, as discussed in Ref.~\cite{belbruno2008resonance}.

On the other hand, analyzing lobe dynamics in Fig.~\ref{fig:lobe_sequence_1} aids in roughly predicting changes in resonance ratios, as shown in Fig.~\ref{fig:optimal_EM_WSB_a}.
Based on this analysis, the proposed method can efficiently connect lobe sequences to increase/\hspace{0pt}decrease resonance ratios.
Thus, lobe dynamics is advantageous for designing transfer trajectories.

%% file: 6_extension.tex
\section{Extension of the CR3BP results into the BCR4BP}\label{sec:extension}
In this section, the optimal trajectory obtained in Section~\ref{sec:application} is extended to the BCR4BP to demonstrate the effectiveness of the proposed method~\cite{hiraiwa2025extending}.
Ballistic lunar transfers, such as the trajectory shown in Fig.~\ref{fig:optimal_EM_traj}, can be categorized into two types: exterior and interior.
Exterior transfers are trajectories that arrive at the Moon from the Earth--Moon $L_2$ side and are often designed based on WSBs~\cite{belbruno1993sun} or Poincar{\'e} maps~\cite{koon2001low,scheuerle2024energy}.
Interior transfers arrive at the Moon from the Earth--Moon $L_1$ side.
This section focuses only on interior Earth--Moon transfers because the graph constructed in Section~\ref{sec:application} does not include paths for exterior lunar transfers.

\subsection{Optimization of the Earth--Moon transfer trajectory}
\begin{figure}[!t]
    \centering
    \includegraphics[width=0.45\linewidth]{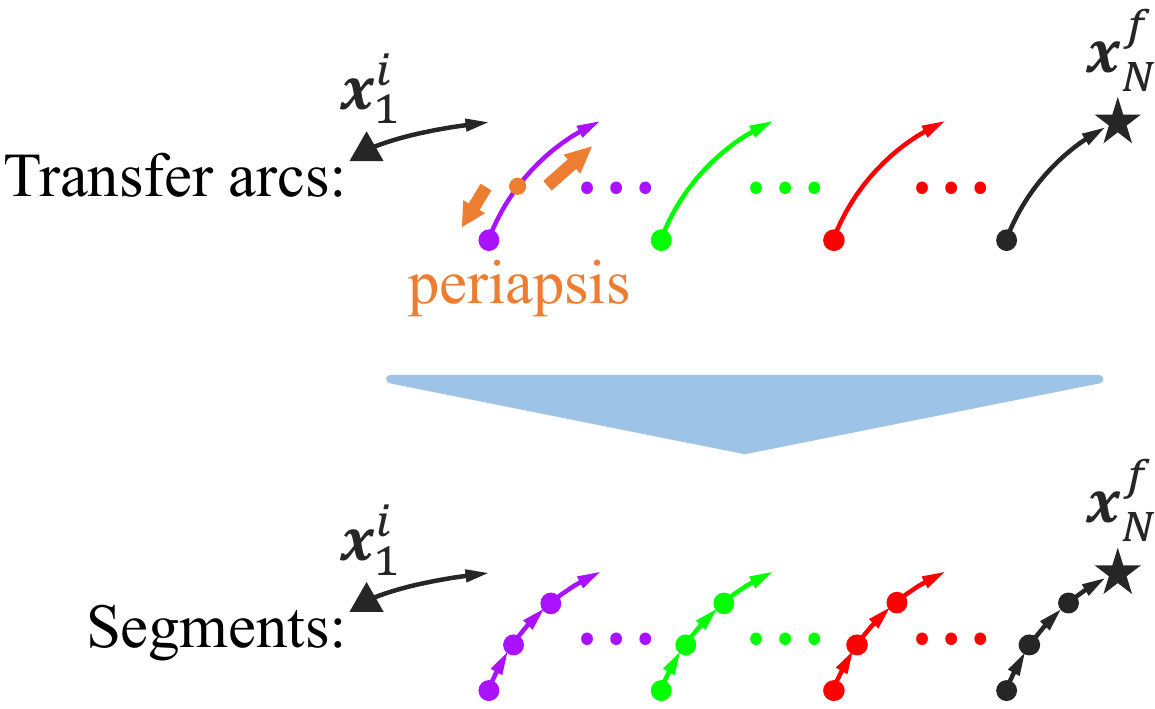}
    \caption{Illustration of the segments to construct the initial guess.}
    \label{fig:segments_MS}
\end{figure}
The optimal Earth--Moon transfer in the BCR4BP is constructed simply by using the optimal trajectory in the CR3BP as an initial guess and solving a minimum-fuel optimization problem with the multiple-shooting scheme.
As illustrated in Fig.~\ref{fig:segments_MS}, the initial guess is constructed by propagating the centroids of effective lobes forward and backward in time.
Every transfer arc is then subdivided into shorter segments to ensure that the transfer time of each segment is approximately $\Delta t \simeq 1$.
In this case, the total number of segments is determined to be $N = 45$.
The initial points of each segment are propagated using Eq.~\eqref{eq:EoM_BCR4BP} in the optimization.
As the BCR4BP is a non-autonomous but periodic system, the initial phase angle of the Sun, $\theta_s^\ast$, is set as $\theta_s^\ast = k\pi/6$ ($k = 0, 1, \cdots, 11$).
Thus, $12$ initial guesses, the same trajectory segments with different values of $\theta_s^\ast$, are used.
To obtain the minimum-fuel interior Earth--Moon transfer in the BCR4BP, the optimization problem is formulated as follows:
\begin{align}
  &\mathrm{minimize}      & & |\Delta \bm{V}_\mathrm{E}| + \sum_{n = 1}^{N-1}\left|\bm{v}_{n+1}^i-\bm{v}_n^f\right| + |\Delta \bm{V}_\mathrm{M}|\\
  &\mathrm{subject\,\,to} & & \bm{r}_{1}^i = \bm{r}_\mathrm{E}\\
  &                       & & \theta_{s,1}^i = \theta_s^\ast\\
  &                       & & \bm{r}_{n+1}^i = \bm{r}_{n}^f\,\,(n = 1,2,\cdots,N-1)\\
  &                       & & \theta_{s,n+1}^i = \theta_{s,n}^f\,(n = 1,2,\cdots,N-1)\\
  &                       & & \bm{r}_{N}^f = \bm{r}_\mathrm{M}
\end{align}
where the position, velocity, and phase angle of the Sun along the $n$-th segment are denoted by $\bm{r}_n$, $\bm{v}_n$, and $\theta_{s,n}$, respectively ($n = 1,\,2,\,\cdots,\,N$).
The starting position at the LEO is expressed as $\bm{r}_\mathrm{E}$, while the final position at the LLO is expressed as $\bm{r}_\mathrm{M}$.
These positions $\bm{r}_\mathrm{E}$ and $\bm{r}_\mathrm{M}$ are the same as those in the optimal trajectory in Fig.~\ref{fig:optimal_EM_traj}.
The initial and final values of each segment are expressed by $(\bm{\cdot})^i$ and $(\bm{\cdot})^f$, respectively.
In this problem, $\Delta \bm{V}_\mathrm{E}$ and $\Delta \bm{V}_\mathrm{M}$ can be applied along any direction.
The optimization problem is solved using the \textit{fmincon} function in MATLAB with the SQP algorithm.

\begin{table}[!t]
  \centering
  \caption{Results of the optimization in the BCR4BP}
  \begin{tabular}{ccc}
      \hline
      $\theta_s^\ast$ & Total $\Delta V$~[m/s] & Transfer time~[day] \\
      \hline
      $0$            & $3832.6088$      & $193.2512$ \\
      $\pi/6$        & -                & - \\
      $\pi/3$        & -                & - \\
      $\pi/2$        & $5201.0019$      & $191.7576$ \\
      $2\pi/3$       & $3841.6832$      & $193.7585$ \\
      $5\pi/6$       & -                & - \\
      $\pi$          & -                & - \\
      $7\pi/6$       & -                & - \\
      $4\pi/3$       & -                & - \\
      $3\pi/2$       & $4934.6452$      & $192.5052$ \\
      $5\pi/3$       & $3848.4479$      & $193.9369$ \\
      $11\pi/6$      & $3837.8299$      & $193.3733$ \\
      \hline
  \end{tabular}
  \label{tab:result_BCR4BP}
\end{table}%
\begin{figure}[!b]
    \centering
    \includegraphics[width=0.5\linewidth]{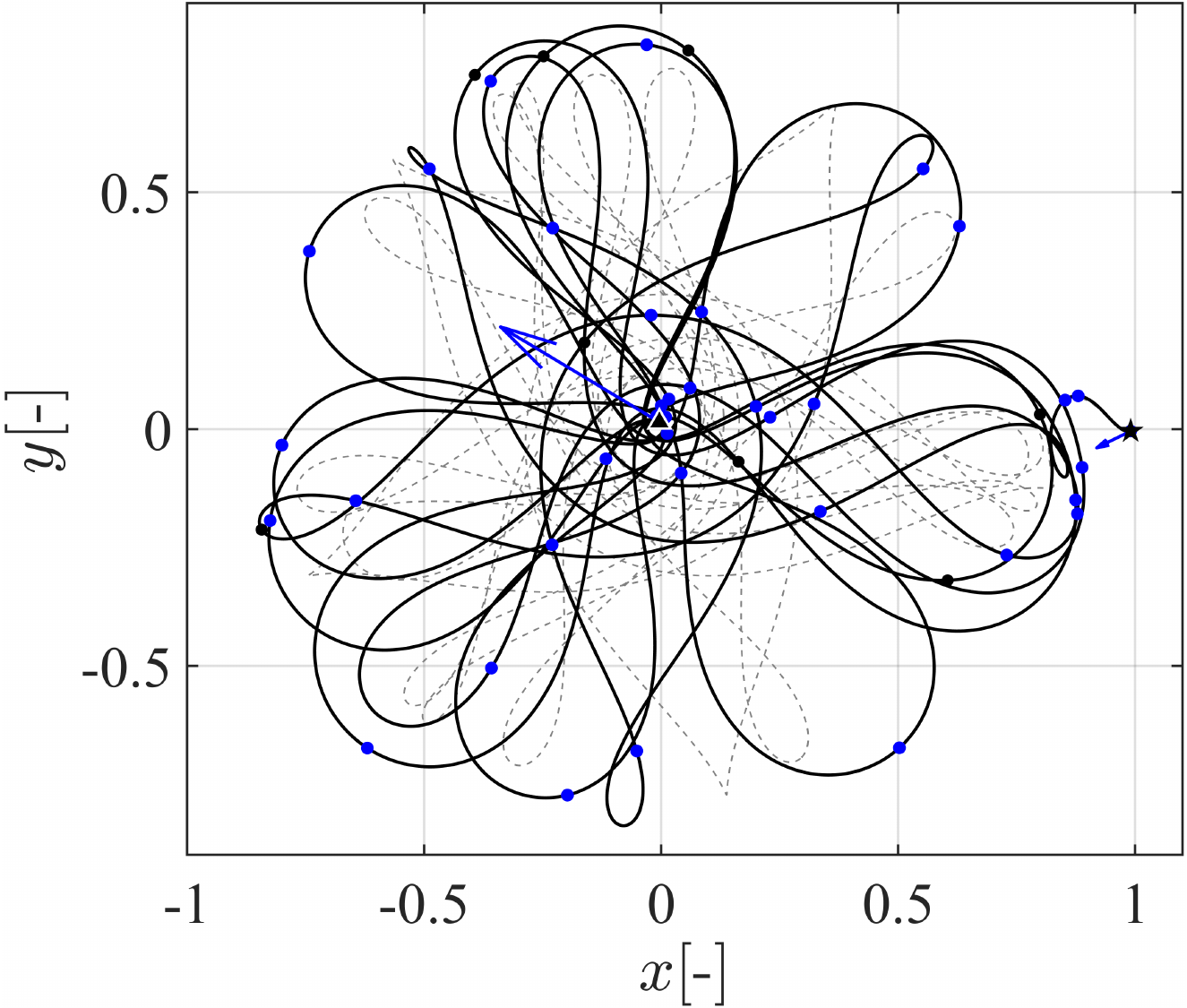}
    \caption{Optimal interior Earth--Moon transfer in the BCR4BP.}
    \label{fig:optimal_EM_traj_BCR4BP}
\end{figure}%
The obtained results are summarized in Table~\ref{tab:result_BCR4BP}.
This table uses a ``-'' to denote that the optimization with the corresponding $\theta_s^\ast$ does not converge after $50000$ iterations.
The failure of convergence for some values of $\theta_s^\ast$ is due to the sensitivity of the initial guess to $\theta_s^\ast$.
These results imply that the convergence may be improved by gradually adding the perturbation of the Sun.
All the converged solutions have a transfer time similar to that of the initial guess ($191.8994$~days), but the total $\Delta V$ varies significantly.
The best solution in this table is the one with $\theta_s^\ast = 0$.
Figure~\ref{fig:optimal_EM_traj_BCR4BP} illustrates the optimal trajectory in the BCR4BP with $\theta_s^\ast = 0$.
The dashed line in this figure represents the optimal trajectory in the CR3BP (Fig.~\ref{fig:optimal_EM_traj}), while the solid line indicates the optimal trajectory in the BCR4BP.
The triangle and star in this figure correspond to the start and goal points, respectively.
Similar to Fig.~\ref{fig:optimal_EM_traj}, blue arrows indicate $\Delta \bm{V}$ vectors, and the length of these arrows corresponds to the relative magnitude of the $\Delta \bm{V}$s.
Unlike the optimal transfer in the CR3BP, the majority of the total $\Delta V$ is contributed solely by the two largest $\Delta \bm{V}$s.
Specifically, $|\Delta \bm{V}_\mathrm{E}| = 3120.8662$~m/s is applied to depart the LEO, and $|\Delta \bm{V}_\mathrm{M}| = 638.5176$~m/s is applied to arrive at the LLO.
The other small $\Delta \bm{V}$s satisfying $|\Delta \bm{V}| < 21$~m/s are used to adjust the transfer trajectory.

\subsection{Comparison of the results with the literature}
The results obtained in the BCR4BP are compared with those of the previous studies~\cite{sweetser1991estimate,topputo2013optimal,pernicka1994search,yagasaki2004computation,yagasaki2004sun,topputo2005earth,mengali2005optimization,mingotti2011ways,DaSilvaFernandes2011sun}.
The known results of impulsive interior transfers from the LEO with an altitude of $167$~km to the LLO with an altitude of $100$~km are summarized in Table~\ref{tab:result_reference} based on Topputo~\cite{topputo2013optimal}.
The previous studies have employed various dynamical models for the transfer design, making direct comparisons difficult.
However, the results in the literature are valuable for estimating Pareto-optimal solutions for this transfer problem.
Specifically, Sweetser~\cite{sweetser1991estimate} estimated the theoretical minimum total $\Delta V$ ($3726$~m/s) for the transfer between the given LEO and LLO.
Topputo~\cite{topputo2013optimal} calculated the Hohmann transfer between the LEO and LLO, with a total $\Delta V$ of $3954$~m/s and transfer time of $5$~days.
\begin{table}[!p]
    \centering
    \caption{List of known Earth--Moon interior transfers between the LEO with an altitude of $167$~km and the LLO with an altitude of $100$~km.}
    \begin{tabular}{lcc}
        \hline
        References & Total $\Delta V$~[m/s] & Transfer time~[day] \\
        \hline
        Hohmann transfer~\cite{topputo2013optimal}         & $3954$ & $5$ \\[3mm]
        Sweetser~\cite{sweetser1991estimate}               & $3726$ & - \\[3mm]
        Pernicka \textit{et al.}~\cite{pernicka1994search} & $3824$ & $292$ \\[3mm]
        Yagasaki~\cite{yagasaki2004computation}            & $3925$ & $31$ \\
                                                           & $3947$ & $14$ \\
                                                           & $3951$ & $4$ \\[3mm]
        Yagasaki~\cite{yagasaki2004sun}                    & $3941$ & $14$ \\
                                                           & $3949$ & $5$ \\[3mm]
        Topputo \textit{et al.}~\cite{topputo2005earth}    & $3895$ & $256$ \\
                                                           & $3900$ & $194$ \\[3mm]
        Mengali and Quarta~\cite{mengali2005optimization}  & $3861$ & $85$ \\
                                                           & $3920$ & $68$ \\
                                                           & $3950$ & $14$ \\
                                                           & $4005$ & $3$ \\[3mm]
        Mingotti \textit{et al.}~\cite{mingotti2011ways}   & $3896$ & $31$ \\
                                                           & $3917$ & $30$ \\
                                                           & $3936$ & $14$ \\[3mm]
        Da Silva Fernandes                                 & $3850$ & $58$ \\
        and Marinho~\cite{DaSilvaFernandes2011sun}         & $3902$ & $32$ \\
                                                           & $3943$ & $14$ \\
                                                           & $3950$ & $5$ \\[3mm]
        Topputo~\cite{topputo2013optimal}                  & $3893$ & $31$ \\
                                                           & $3937$ & $14$ \\
                                                           & $3945$ & $5$ \\
        \hline
    \end{tabular}
    \label{tab:result_reference}
\end{table}%

\begin{figure}[!t]
    \centering
    \includegraphics[width=\linewidth]{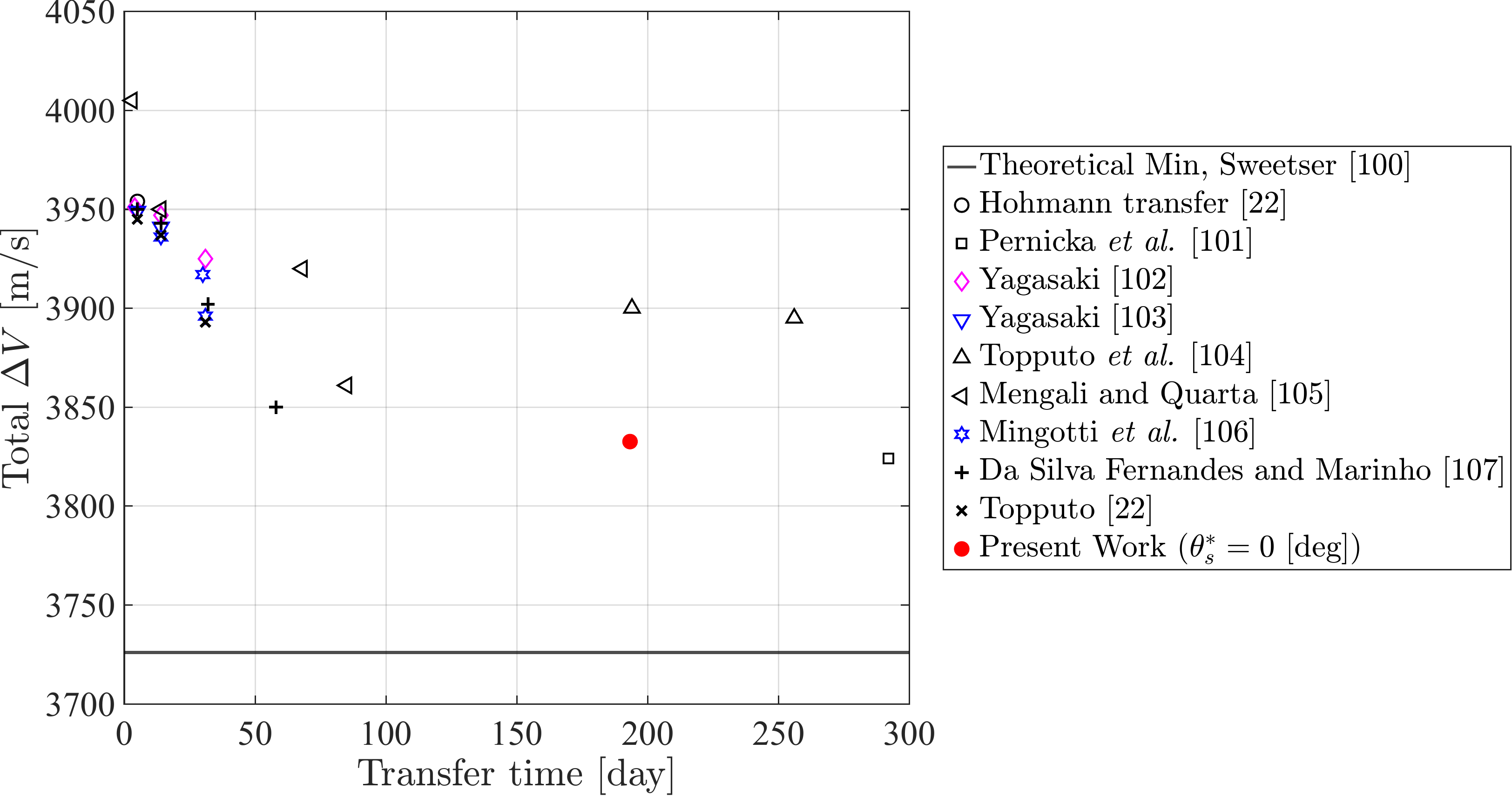}
    \caption{Comparison of impulsive interior Earth--Moon transfers with respect to the total $\Delta V$ and transfer time.}
    \label{fig:comparison_results}
\end{figure}

The result of the comparison is summarized in Fig.~\ref{fig:comparison_results}.
This figure illustrates that the proposed method can identify one of the Pareto-optimal solutions with a medium transfer time.
The obtained transfer with $\theta_s^\ast = 0$ may be helpful, especially for missions with strict fuel constraints, such as CubeSat missions.
Note that it may be possible to find better solutions by adjusting the initial LEO point, the final LLO point, and the initial phase angle of the Sun.

\subsection{Analysis of the optimal Earth--Moon transfer}
To investigate transfer structures in the BCR4BP, this subsection analyzes the optimal trajectory in the BCR4BP with $\theta_s^\ast = 0$.
Although analyzing dynamical structures in the higher-dimensional system of the BCR4BP is challenging, a numerical investigation is conducted along the optimal trajectory.

\begin{figure}[!t]
    \centering
    \begin{minipage}{0.45\columnwidth}
        \centering\includegraphics[width=\columnwidth]{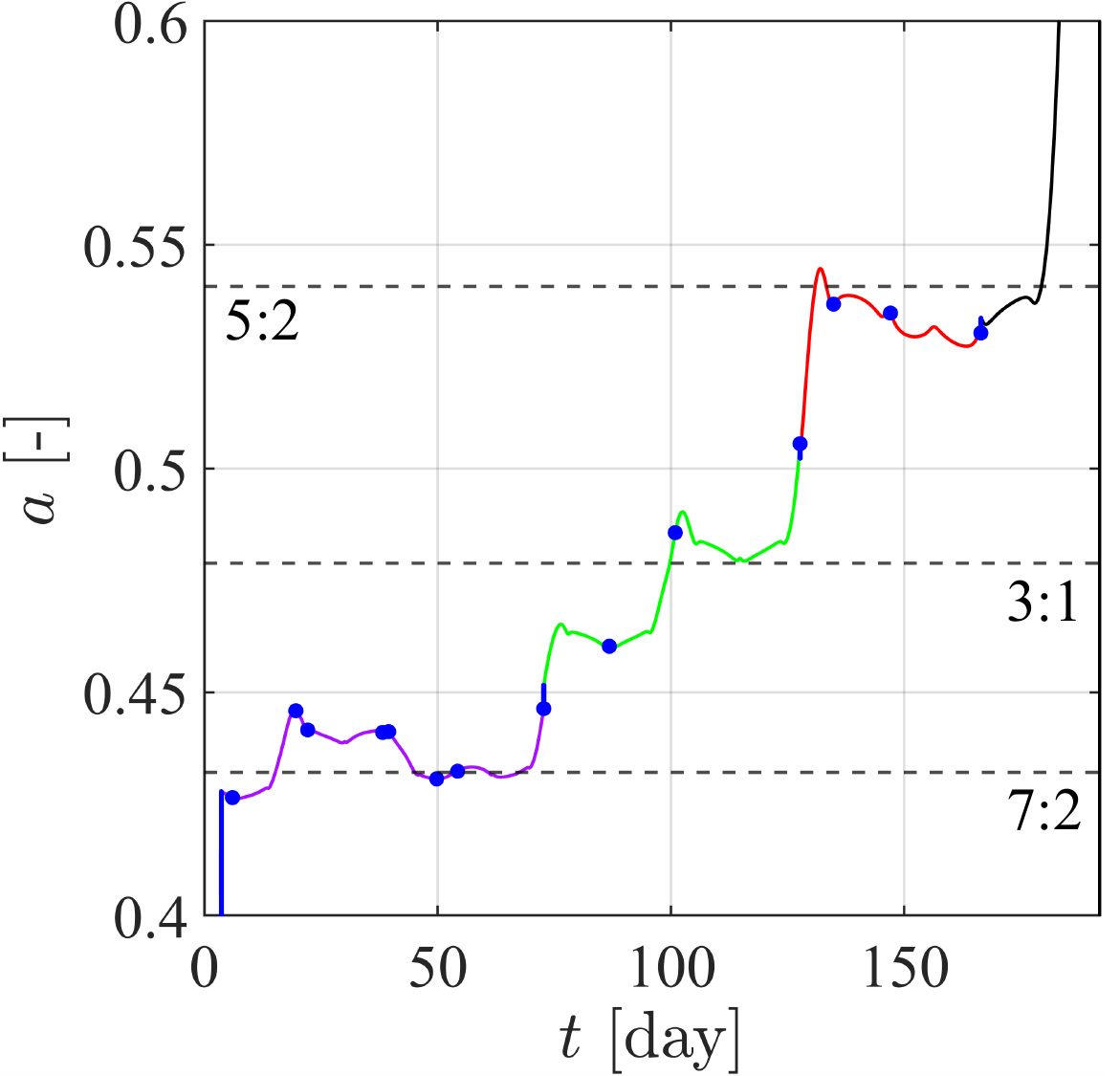}
        \subcaption{In the CR3BP.}
        \label{fig:optimal_EM_BCR4BP_a_CR3BP}
    \end{minipage}
    \hspace{5mm}
    \begin{minipage}{0.45\columnwidth}
        \centering\includegraphics[width=\columnwidth]{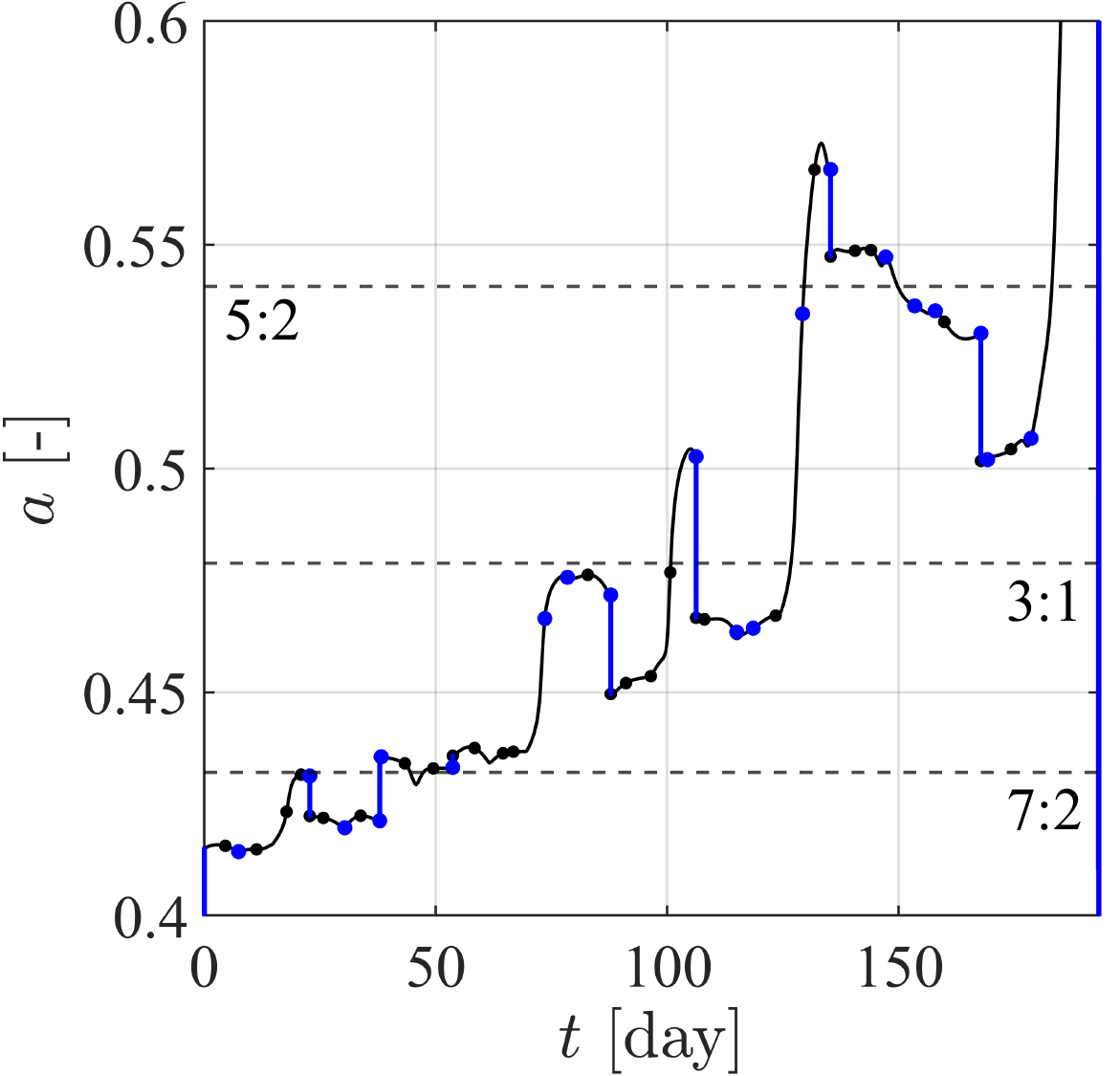}
        \subcaption{In the BCR4BP.}
        \label{fig:optimal_EM_BCR4BP_a_BCR4BP}
    \end{minipage}
    \caption{Time history of the semi-major axis along the optimal trajectory.}
    \label{fig:optimal_EM_BCR4BP_a}
\end{figure}

First, the existence of lobe dynamics in the BCR4BP is numerically investigated based on mean-motion resonances.
The time history of the semi-major axis along the optimal trajectory in the BCR4BP is shown in Fig.~\ref{fig:optimal_EM_BCR4BP_a}.
The dashed lines in Fig.~\ref{fig:optimal_EM_BCR4BP_a} indicate the values from Eq.~\eqref{eq:resonant_a} for the corresponding resonance ratios, and the blue lines indicate impulsive $\Delta \bm{V}$s.
Note that Fig.~\ref{fig:optimal_EM_BCR4BP_a_CR3BP} is the same as Fig.~\ref{fig:optimal_EM_WSB_a}.
Similar to Fig.~\ref{fig:optimal_EM_BCR4BP_a_CR3BP}, some parts of the semi-major axis in Fig.~\ref{fig:optimal_EM_BCR4BP_a_BCR4BP} are roughly constant.
This result suggests that mean-motion resonances may also be important for transfers in the BCR4BP; therefore, the analysis of lobe dynamics in the BCR4BP is a promising direction for future work.
Interestingly, small impulsive maneuvers with $\Delta V < 21$~m/s can significantly alter the semi-major axis.

\begin{figure}[!t]
    \centering
    \begin{minipage}{0.45\columnwidth}
        \centering\includegraphics[width=\columnwidth]{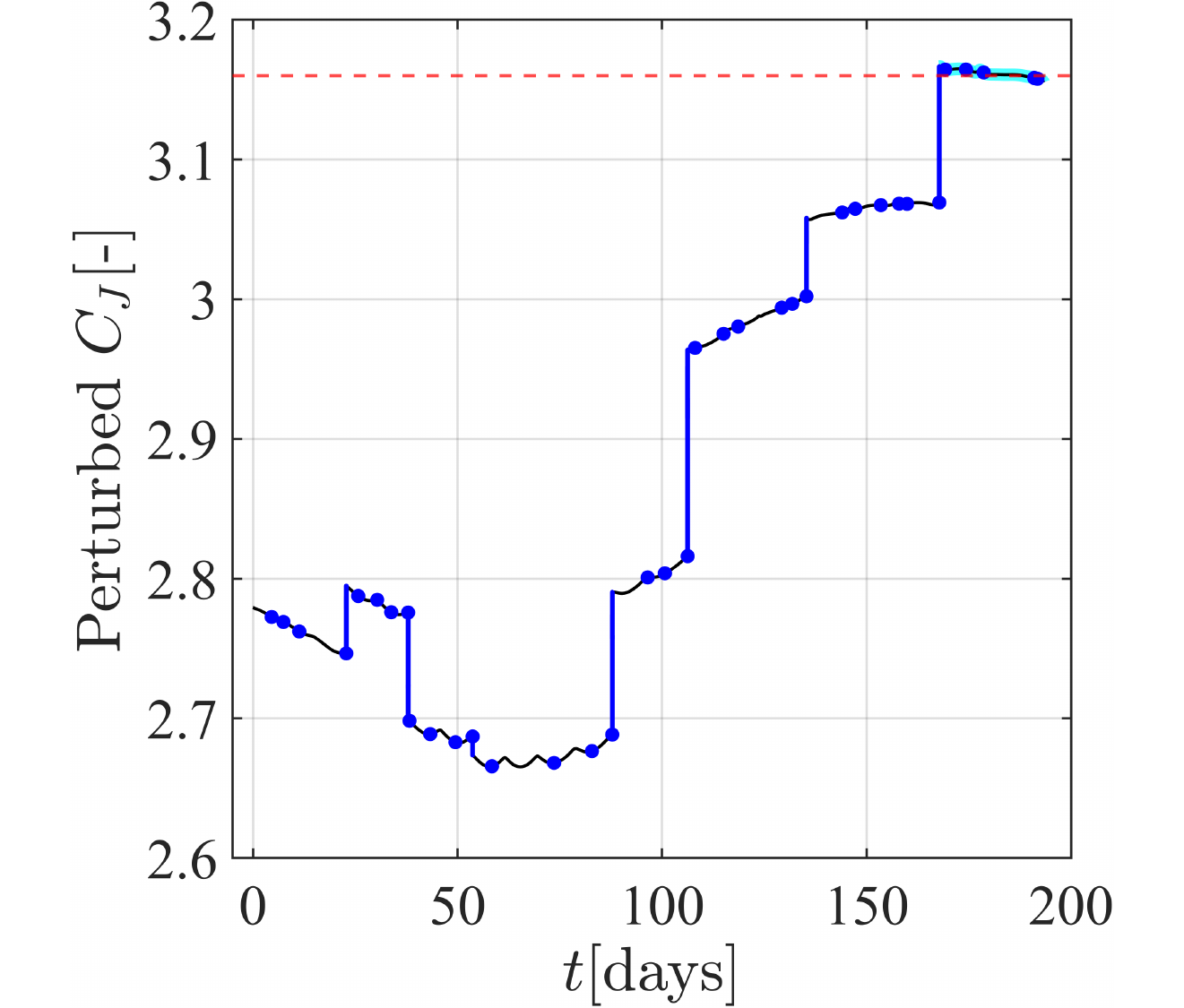}
        \subcaption{Jacobi constant.}
        \label{fig:optimal_EM_traj_BCR4BP_CJ_CJ}
    \end{minipage}
    \hspace{5mm}
    \begin{minipage}{0.45\columnwidth}
        \centering\includegraphics[width=\columnwidth]{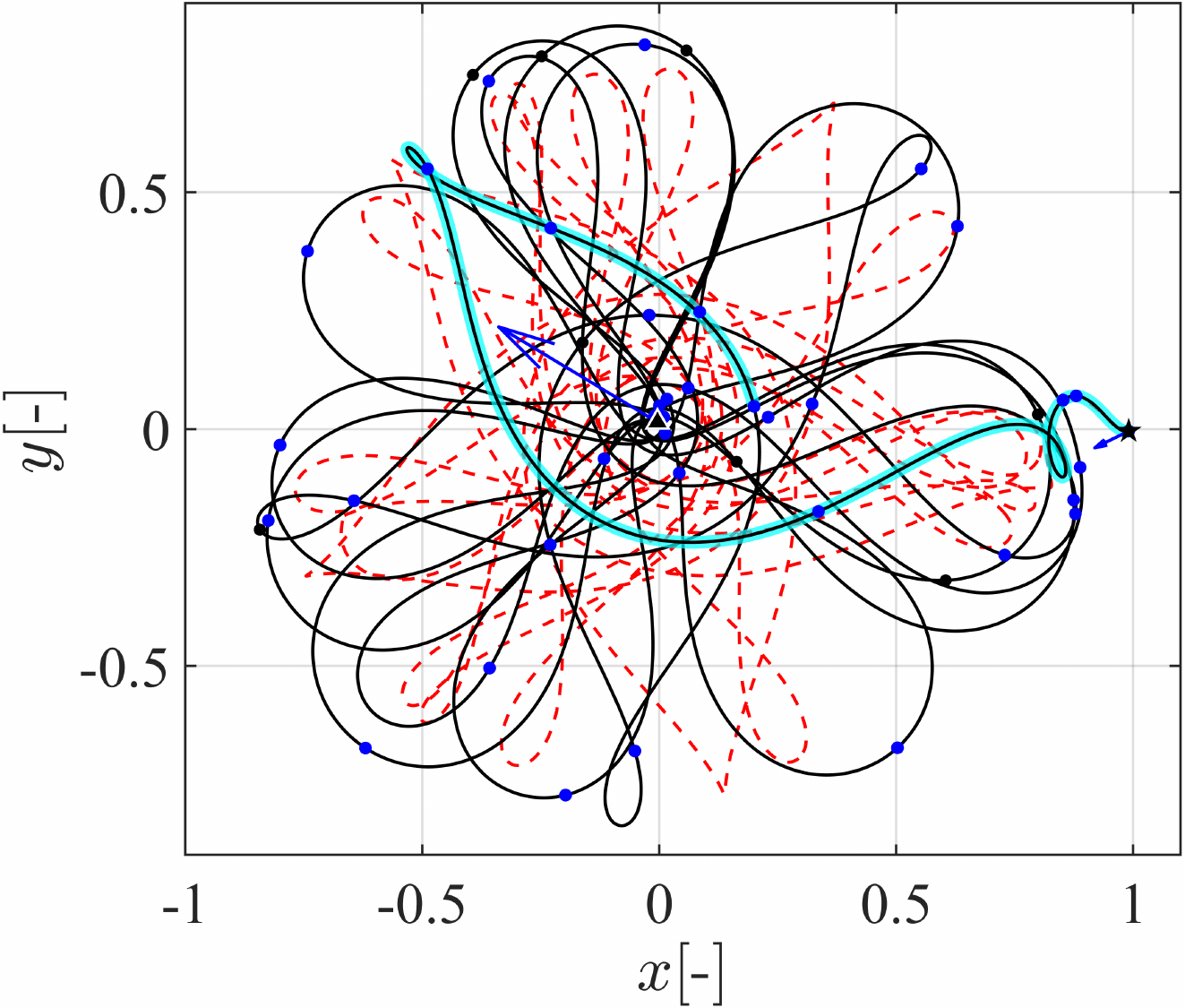}
        \subcaption{Optimal trajectory in Fig.~\ref{fig:optimal_EM_traj_BCR4BP}.}
        \label{fig:optimal_EM_traj_BCR4BP_CJ_traj}
    \end{minipage}
    \caption{Jacobi constant in the Earth--Moon CR3BP along the optimal trajectory in Fig.~\ref{fig:optimal_EM_traj_BCR4BP}.}
    \label{fig:optimal_EM_traj_BCR4BP_CJ}
\end{figure}

Next, the existence of tube dynamics in the BCR4BP is examined along the optimal trajectory.
Figure~\ref{fig:optimal_EM_traj_BCR4BP_CJ} indicates the Jacobi constant in the Earth--Moon CR3BP along the optimal trajectory in Fig.~\ref{fig:optimal_EM_traj_BCR4BP}, perturbed by the gravitational force of the Sun.
The red dashed lines show the reference in the CR3BP with $C_J = 3.16$, and the blue lines indicate impulsive $\Delta \bm{V}$s.
The cyan thick line in Fig.~\ref{fig:optimal_EM_traj_BCR4BP_CJ_CJ} shows the perturbed Jacobi constant $C_J \simeq 3.16$ (precisely, $3.1551 < C_J < 3.1666$), and the corresponding optimal trajectory is also plotted as the cyan thick line in Fig.~\ref{fig:optimal_EM_traj_BCR4BP_CJ_traj}.
This result implies that the transfer structure to the Moon in the CR3BP with $C_J = 3.16$, i.e., the stable manifold of the $L_1$ Lyapunov orbit in Fig.~\ref{fig:L1_stable}, may remain in the BCR4BP.

For the analysis of transfer structures in Fig.~\ref{fig:optimal_EM_traj_BCR4BP_CJ}, Lagrangian coherent structures (LCSs)~\cite{haller2000lagrangian,haller2001distinguished,haller2015lagrangian} are leveraged.
Haller~\cite{haller2001distinguished} has defined LCSs as local extrema of the finite-time Lyapunov exponent (FTLE).
The FTLE at $\bm{x}$ from $t = 0$ to $t = T$ is defined as
\begin{equation}
    \sigma(\bm{x}, T) = \frac{1}{|T|}\ln\|\Phi(T,0)\|
\end{equation}
where $\Phi(T,0)$ is the state transition matrix from $t = 0$ to $t = T$ along the reference trajectory starting from $\bm{x}$, and $\|\Phi(T,0)\|$ denotes the square root of the largest eigenvalue of the Cauchy--Green strain tensor $\Phi(T,0)^\mathrm{T}\Phi(T,0)$.
In autonomous systems, LCSs correspond to the stable and unstable manifolds associated with periodic orbits; in non-autonomous systems, LCSs can represent manifold-like lines or surfaces that mediate phase space transport.
Previous studies have demonstrated that LCSs can be detected based on the FTLE in the CR3BP~\cite{short2011lagrangian,perez2012detecting} and in the elliptic restricted three-body problem~\cite{gawlik2009lagrangian}.
Reference~\cite{short2014lagrangian} has investigated LCSs in the CR3BP, BCR4BP, and the Sun--Earth--Moon ephemeris model by analyzing the FTLE fields.

\begin{figure}[!b]
    \centering
    \begin{minipage}{0.45\columnwidth}
        \centering\includegraphics[width=\columnwidth]{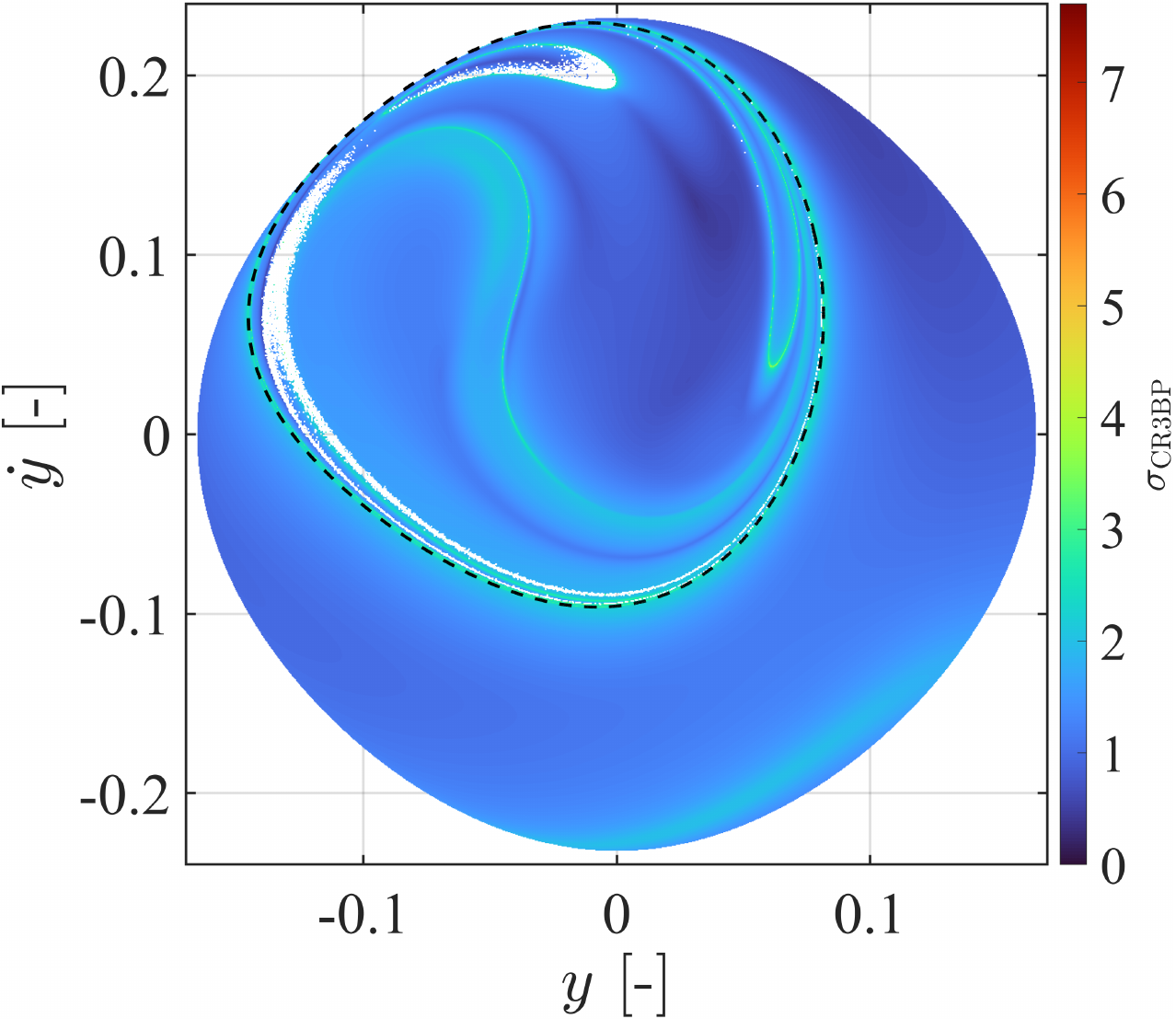}
        \subcaption{In the CR3BP.}
        \label{fig:FTLE_CR3BP_manifold}
    \end{minipage}
    \hspace{5mm}
    \begin{minipage}{0.45\columnwidth}
        \centering\includegraphics[width=\columnwidth]{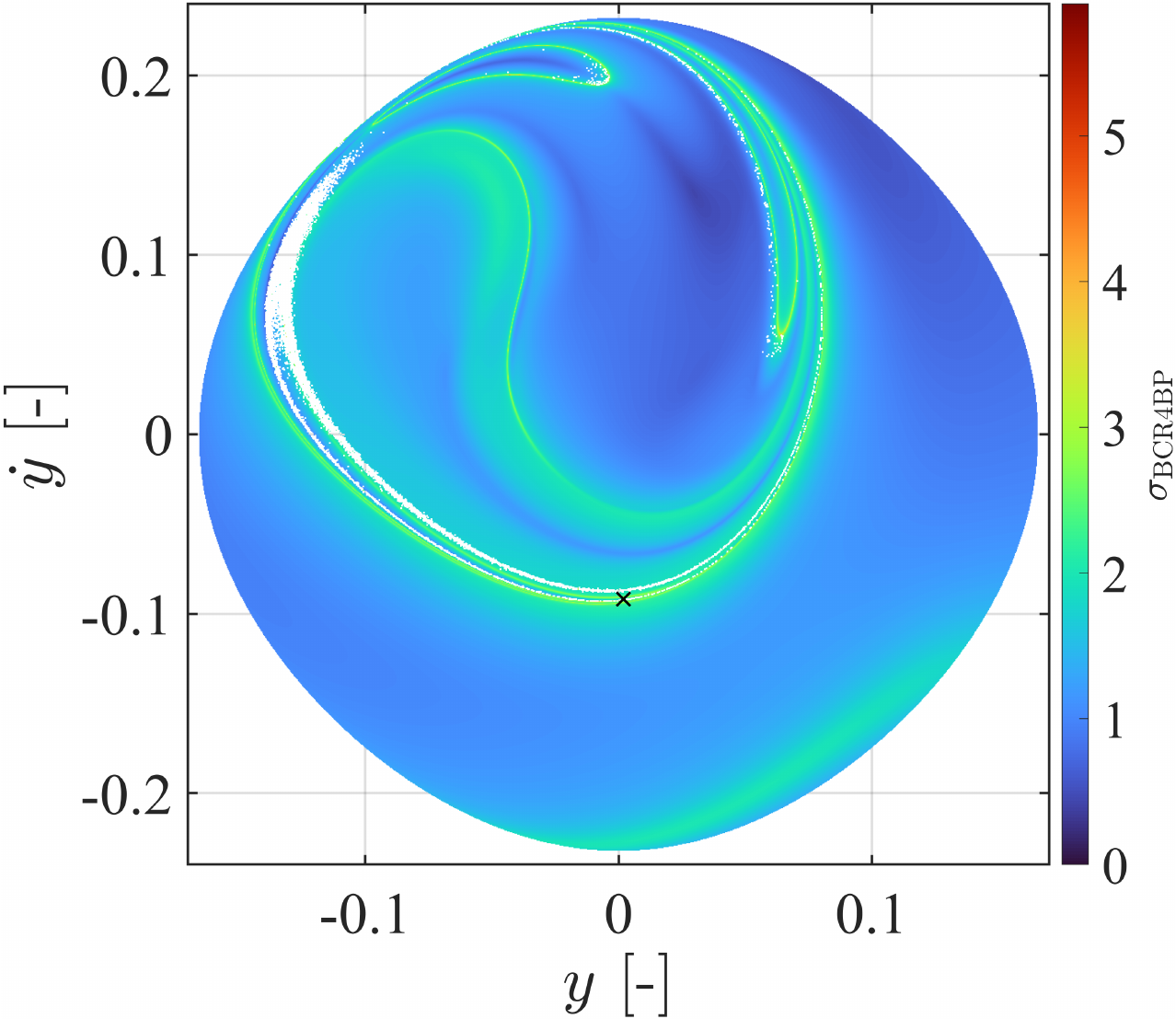}
        \subcaption{In the BCR4BP.}
        \label{fig:FTLE_BCR4BP}
    \end{minipage}
    \caption{Forward-time FTLE fields at $x = 0.8$ in the CR3BP and BCR4BP.}
    \label{fig:FTLE}
\end{figure}%
Similar to Ref.~\cite{short2014lagrangian}, the FTLE field on the surface of section at $x = 0.8$ is examined under the dynamics of the BCR4BP.
For the calculation of FTLE fields, a grid of $1000\times1000$ points is set within the region of $-0.1658 \leq y \leq 0.1658$ and $-0.2325 \leq \dot{y} \leq 0.2325$.
The forward-time FTLE fields in the CR3BP and BCR4BP (with $\theta_s^\ast = 297.1825$~deg) are calculated with $T = 14$~days, as shown in Fig.~\ref{fig:FTLE}.
In Fig.~\ref{fig:FTLE_CR3BP_manifold}, the dotted line denotes the stable manifold of the $L_1$ Lyapunov orbit in Fig.~\ref{fig:L1_stable} at $x = 0.8$.
It is confirmed that the ridges in the FTLE field in the CR3BP match this stable manifold for this time interval.
Figure~\ref{fig:FTLE_BCR4BP} shows the FTLE field in the BCR4BP.
The cross represents the optimal trajectory in Fig.~\ref{fig:optimal_EM_traj_BCR4BP} at $x = 0.8$, and thus this trajectory is located slightly inside of the ridges of the FTLE field in the BCR4BP.
Note that the phase angle of the Sun is $\theta_s = 297.1825$~deg at this point.
This result reveals that the optimal trajectory passes through the region divided by the LCS in the BCR4BP for the transfer to the Moon, and this LCS in the BCR4BP can be understood as the perturbed stable manifold of the $L_1$ Lyapunov orbit with $C_J = 3.16$ in the CR3BP.

%% file: 7_conclusion.tex
\section{Conclusions and future work}\label{sec:conclusion}
This paper proposes and validates a new method for systematically designing low-energy transfers by combining multiple lobe dynamics for spacecraft trajectory design.
The numerical difficulty of detecting lobe sequences, i.e., transport structures by lobe dynamics, is avoided by focusing on effective lobes.
The concept of effective lobes is also valuable for trajectory design, as its definition implies that the designed trajectory is robust to small errors.
Effective lobes are utilized as intermediate transfer points between start and goal orbits.
The combination of effective lobe sequences is determined by exploiting a weighted, directed graph that represents possible transfer paths.
In the graph, the transfer arcs between each point are determined by the targeting strategy.
This graph aids in solving the combinatorial optimization to find the desired transfers.
Thus, the proposed method constructs low-energy transfers by connecting chaotic orbits via effective lobes and offers a new type of transfer option.

The numerical results in Sections~\ref{sec:trajectory_design} and \ref{sec:application} demonstrate a way to combine lobe and tube dynamics.
In other words, lobe dynamics can be utilized to identify transport structures between a celestial body and tubes of libration point orbits.
Thus, the proposed framework advances the geometrical approach to trajectory design.

This paper also illustrates the application of the proposed method to the design of fuel-optimal transfers from a low Earth orbit to a low lunar orbit.
In Section~\ref{sec:extension}, the optimization of the Earth--Moon transfer trajectory in the BCR4BP enables comparison of the fuel consumption and transfer time with previous studies~\cite{sweetser1991estimate,topputo2013optimal,pernicka1994search,yagasaki2004computation,yagasaki2004sun,topputo2005earth,mengali2005optimization,mingotti2011ways,DaSilvaFernandes2011sun}.
The effectiveness of the proposed method is verified by demonstrating that it can help identify one of the Pareto-optimal solutions.
Note that the proposed framework can easily be applied to other transfer problems by replacing the start and goal points with preferred ones.

A natural progression of this study is to extend the proposed method by leveraging lobe dynamics in a higher-dimensional phase space, such as the spatial CR3BP or the BCR4BP.
It is necessary to develop a method to understand the higher-dimensional phase space using at most three-dimensional figures for trajectory design.
Another direction of future work is to verify some aspects of the proposed method.
The robustness of transfer trajectories designed by the proposed method can be explored in mission scenarios with small uncertainties.
It is also essential to improve the method for determining transfer arcs, especially for connecting trajectories with different Jacobi constants.